\documentclass[12pt,Bold,letterpaper]{mcgilletdclassmine}
\usepackage[pdftex,final]{graphicx}
\usepackage{graphicx}
\usepackage{latexsym, amssymb, epsfig, epic, eepic}
\usepackage{amsmath,amsfonts}
\usepackage[dvips,letterpaper,left=3.8cm,top=3.2cm,bottom=2.5cm,right=2.5cm,includefoot]{geometry}
\usepackage{epsfig}
\usepackage{epsf}
\usepackage{xspace}
\usepackage{multirow}
\usepackage{rotating}
\usepackage{subfig}
\input{epsf.sty}
\usepackage{verbatim} 
\usepackage{enumitem}
\usepackage{wrapfig}
\usepackage{url}
\usepackage{longtable}

\def\be{\begin{equation}}
\def\ee{\end{equation}}
\def\ba{\begin{eqnarray}}
\def\ea{\end{eqnarray}}

\newcommand{\roughly}[1]{\mathrel{\raise.3ex\hbox{$#1$\kern-0.85em
\lower1ex\hbox{$\sim$}}}}

\def\2pi{\left(2\pi\right)}

\def\beq{\begin{equation}}
\def\eeq{\end{equation}}
\def\bg{\begin{eqnarray}}
\def\nd{\end{eqnarray}}
\def\bea{\begin{eqnarray}}
\def\eea{\end{eqnarray}}

\def\D3{\overline{\mbox{D3}}}

\SetTitle{\huge{Towards New Classes\\
of Flux Compactifications}}%
\SetAuthor{Paul Franche}%
\SetDegreeType{Doctor of Philosophy}%
\SetDepartment{Department of Physics}%
\SetUniversity{McGill University}%
\SetUniversityAddr{Montr\'eal,Qu\'ebec}%
\SetThesisDate{August 2012}%
\SetRequirements{A thesis submitted to McGill University in partial fulfillment of the requirements of the degree of Doctor of Philosophy}%
\SetCopyright{Copyright \copyright  paulfranche, 2012}%

\makeindex[keylist]
\makeindex[abbr]
\begin{document}

\maketitle%

\begin{romanPagenumber}{2}%
\SetAbstractEnName{Abstract}%
\SetAbstractEnText{We derive novel solutions of flux compactification with D7-branes on the resolved conifold in type IIB String Theory and later extend this solution to allow for non-zero temperature. At zero temperature, we find that adding D7-branes via the Ouyang embedding contributes to the supersymmetry-breaking (1,2) imaginary-self-dual flux, without generating a bulk cosmological constant. We further find that having  D7-branes and a resolved conifold together give rise to a non-trivial D-term on the D7-branes. This supersymmetry-breaking term vanishes when we take the singular conifold limit, although supersymmetry appears to remain broken. We also lift our construction to F-theory where we show that the type IIB (1,2) flux goes to (2,2) non-primitive flux on the fourfold. 
\\
In the second part of the thesis, we extend these results by taking the non-extremal limit of our geometry to incorporate temperature. In this case, the internal NS-NS and R-R fluxes are no longer expected to be self-dual, but they should also naturally be extensions of the fluxes found above. From the supergravity equations of motion, we compute how the new contributions to the fluxes should enter, due to the squashing of the resolved metric and non-extremality. This provides us with a compelling gravity dual of large $N$ thermal quantum chromodynamics with flavor.}
\AbstractEn%

%A faire si je veux
%\SetDedicationName{\MakeUppercase{Dedication}}%
%\SetDedicationText{This document is dedicated to the graduate students of the McGill University.}%
%\Dedication%

%The results presented in this thesis are based on original research carried out with collaborators keshav Dasgupta, Anke Knauf, James Sully, Mohammed Mia, Fang Chen, Sachindeo Vaidya, and is presented here with their permission.
%\begin{itemize}
 %\item K.~Dasgupta, {\bf P.~Franche}, A.~Knauf, J.~Sully, ``D-terms on the resolved conifold,'' Journal of High Energy Physics {\bf 0904}, 027 (2009). 55 pp. ArXiv:0802.0202 [hep-th].

%Chapter \ref{DtermPaper} blebleble

\TOCHeading{\MakeUppercase{Table of Contents}}%
%\LOTHeading{\MakeUppercase{List of Tables}}%
%\LOFHeading{\MakeUppercase{List of Figures}}%
\tableofcontents %
%\listoftables %
\listoffigures %

\end{romanPagenumber}

%\mainmatter %
% \centering
\chapter{Introduction}
At present, the Standard Model of particle physics and General Relativity together provide compelling explanations for almost all physical observations in extreme regimes, i.e. both at the very small and very large scales respectively. These theories however depend heavily on some fine-tuning of the values of their many parameters and furthermore fail to provide a natural mechanism to allow these particular values to arise. The absence of such mechanisms to explain the hierarchy between the mass of the quark generations and between the gravity and electroweak scales are interesting examples from the Standard Model, while the fine-tuning of the initial conditions for slow-roll inflation is another one from cosmology. 

The hopes that String Theory could encompass the four forces of physics from the Standard Model and General Relativity and simultaneously provide a natural mechanism for these specific theories to arise are well founded. It contains only one free parameter to be fixed, the string scale, while all other parameters are dynamical variables to be determined by the ground state. This fact alone legitimizes the hopes that String Theory could naturally give rise to the observed parameters of the ``real world". 

Unfortunately, two key features of String Theory can, at first sight, appear problematic. First, consistency of the theory requires to have 10 spacetime dimensions. Six extra dimensions must then be decoupled consistently from the four macroscopic dimensions. Second, solutions to String Theory are far from unique. There is a landscape of solutions, each with a different ground state leading to different ``real world" physics. This then makes it very hard to allow for general features of physics to be extracted. Nevertheless, the situation is far from hopeless.

These ``problematic" features of String Theory might actually be exactly what we are looking for. The way we deal with these six extra dimensions is crucial in defining what will be the resulting four dimensional physics. Finding what type of conditions have to be imposed on these extra dimensions is precisely the type of natural mechanism we are looking for. Type IIB String Theory will be of particular interest for us as its D3-branes are natural objects to associate to the spacetime dimensions of our real universe. Flux compactification will then further determine what combination of geometry, fluxes and objects are allowed in the extra dimensions as full String solutions. Understanding solutions that are close to realistic physics is then of primordial importance. Lets see how this can be achieved in two particular approaches of interest to us.

~\newline\\
\begin{flushleft}
{\it String Cosmology}
\end{flushleft}
Inflation provides a compelling explanation for the homogeneity and isotropy of the universe as well as for the observed spectrum of density perturbations. However, many cosmological models involve a range of energy scales large enough for generic Planck-scale corrections to become significant or inevitably hit a regime where their theoretical description breaks down \cite{cosmohitproblem1,cosmohitproblem2}. This makes the study of inflation a good opportunity to explore the high energy physics predicted by String Theory as well as to possibly develop a more fundamental understanding of the microphysics of inflation.

D-brane inflation, where the role of the inflaton field is identified with the position of a mobile probe D3-brane, is a particularly promising framework to look for such models \cite{refinflationreview1,refinflationreview2,refinflationreview3,refinflationreview4,refinflationreview5}, but very few concrete models are actually completely successful from a top-down point of view. The well-known work of S. Kachru et al. (KKLMMT) \cite{KKLMMT} in particular examined the inflaton potential arising from the simple attraction between a mobile $D3$- and a distant $\overline{D3}$-brane. Although the potential naively appears to be flat enough to sustain enough slow-roll inflation, it also shows that generic moduli stabilization gives rise to corrections making the potential too steep\footnote{Note however the recent criticism by Bena et al. \cite{Bena}}.

Subsequently, additional evidence for a possible successful D-brane inflation scenario were recovered in String Theory using a new key ingredient: D7-branes. The work of \cite{BDKMMM} first added 7-branes to the Klebanov-Tseytlin setup \cite{kt} through the Kuperstein embedding \cite{kuperstein} to stabilize the moduli, before they computed their exact contribution to the moduli stabilization potential found in KKLMMT. A following paper \cite{bdkm} then examined the resulting inflationary dynamics. They found that the potential for the mobile D3-brane depends heavily on the position of the D7-branes and that even heavy fine-tuning is, in general, not possible to allow for inflation. Fortunately, they also found that there exists a small region of parameter space that allows for a sufficiently flat potential which can lead to enough inflation at an inflection point. This occurs when the attraction of the inflaton brane towards the tip of the conifold is almost exactly balanced by the attraction towards the 7-brane located far in the conifold. This success however depends on the fine-tuning of the position and embedding of the D7-brane and the details of the flux compactification. Improving our understanding of such types of solution and generalizing them is crucial to bring String Theory closer to the physics of the early universe.

This was our initial motivation for constructing the solution of \cite{DtermPaper} and is presented in chapter \ref{DtermPaper}. We developed a similar solution as in \cite{BDKMMM,bdkm}, but we embedded our D7-branes via the holomorphic Ouyang embedding \cite{ouyang} and we use the resolved conifold geometry, as opposed to the non-supersymmetric Kuperstein embedding and the singular conifold of \cite{bdkm}. Although the D7-branes don't break supersymmetry spontaneously, the non-primitive fluxes induced by the geometry do break it and allow for a D-term potential to appear on the D7-branes. Our hope was that the non-primitive fluxes or the D-term might have helped to give slow-roll dynamics with less fine-tuning then the usual $D3-\overline{D3}$ scenario \cite{KKLMMT,BDKMMM}. This question however was not addressed precisely in \cite{DtermPaper}. Our main focus was to construct a new class of exact supergravity solution that could be used in a top-down approach from String Theory to ``real-world" physics. The results obtained here turned out to be crucial in a different domain: particle physics. In chapters \ref{chapitreintrotemperature} and \ref{actfluxes}, it will become obvious how \cite{DtermPaper} formed a proper starting point towards more realistic embedding the Standard Model in String Theory.

~\newline\\
\begin{flushleft}
{\it Towards Embedding the Standard Model}
\end{flushleft}
As we mentioned earlier, the Standard Model of particle physics provides a theoretical explanation for almost all of particle physics observations, but fails to provide a natural mechanism to explain its specific symmetries and  the values taken by its parameters. String Theory is again an ideal candidate as an extension of the Standard Model as its internal dynamics could provide such a mechanism. 

The primary tool to introduce Standard Model-like theories within String Theory is the AdS/CFT correspondence (also named holography or gauge/gravity duality). It establishes that the gauge theory living on D-branes, induced by the open strings ending on it, is equivalent to the resulting closed string spectrum, which includes gravity. In other words, an open string stretched between two branes is equivalent to a closed string exchange between the branes. This duality was first formulated by Maldacena \cite{MaldacenaoriginalAdSCFT} where he conjectured that the $AdS_5\times S^5$ geometry induced by a stack of $N$ parallel D3-branes is dual to the ${\mathcal N}=4$ $SU(N)$ Super Yang-Mills gauge theory living on the branes. The four gauge fields of the gauge theory are associated to the position of the string endpoints on the brane, while the six scalars correspond to the oscillations of the position of the D3-branes in the transverse space.

Since the gauge and gravity descriptions of the same setup are suggested to be equivalent, one might immediately want to compare their individual predictions. We quickly realize however that the validity of their respective calculations are in opposite regimes. Let us look at the parameters involved. First, the string coupling $g_s$, which describes the couplings between the strings and thus between the gauge fields, can be identified with the Yang-Mills coupling, $g^2_{YM}=4\pi g_s$. Then,  we take $g_s$ to be small in order to suppress the effect of the closed string loops. Moreover, the 't Hooft couplings of the $SU(N)$ gauge theory is given by $\lambda=g^2_{YM}N$ and the $AdS$ radius of the geometry is given by $L^4=4\pi g_sN\alpha'^2$, or $L^4/\alpha'^2=\lambda$. Let us remember that $\lambda$ is the natural expansion parameter for Yang-Mills theories for small $\lambda$. In order for the approximation done when we go from the worldsheet action to supergravity to be valid, the $\alpha'$ parameter must be taken to be small. This in turn tells us, in combination with the previous relationship, that the 't Hooft coupling $\lambda$ of the gauge theory must be large, forcing us into the strong couping regime of the gauge theory. Consequently, if the conjecture is true, the well-understood weak coupling regime of supergravity would then be a window into the less-understood strongly coupled field theories, and vice versa. Notice from the definition of the 't Hooft couplings that $\lambda/g_s\sim N$, which tells us that when $\lambda$ is large and $g_s$ is small, $N$ must then be very large.

Starting from this specific gauge/gravity duality, one might then be interested in exploring more general classes of gauge theories. Indeed, Maldacena's conjecture is for highly supersymmetric flavorless conformal zero-temperature large $N$ gauge theories. This was due to the simplicity of the gravity setup which only included D3-branes with 5-form flux compactified on an $S^5$, without internal fluxes or other types of branes. There is then obvious interest to generalize the AdS/CFT correspondence towards more realistic field theories.

A first generalization was done by Klebanov and Tseytlin (KT) \cite{kt} where $M$ additional parallel D5-branes were added. From the gravity side, this added self-dual 3-form fluxes and changed the internal geometry to a $T^{1,1}$ singular conifold, which is less symmetric than $S^5$. From the gauge theory side, the gauge group was enhanced to $SU(N+M)\times SU(N)$, the previous six scalars are now chiral complex scalar fields and a logarithmic running of the couplings was induced by the new fluxes. Here, the supergravity solution with additional branes and fluxes reduced the amount of supersymmetry but enhanced the gauge group and enriched the internal geometry.

The KT solutions however had an inherent singularity which was resolved by Klebanov and Strassler (KS) \cite{ks}. The running of the fluxes on the gravity side was understood from the gauge theory point of view as a Seiberg-duality cascade. As we move down the RG scale (towards smaller radius in gravity), the symmetry of the gauge flux is progressively broken, which in turn is seen as a reduced amount of fluxes and branes on the gravity side. This mechanism avoids the IR singularity and provides a gravity realisation (deformation of the conifold) of the confinement in the gauge theory (reduction to $SU(M)$ symmetry).

Two further extensions then brought us even closer to the realistic gauge theories. First, P. Ouyang \cite{ouyang} successfully added flavor to the gauge theory by including supersymmetric D7-branes to the supergravity solution of KT. This made the fluxes even more general by allowing the running of the axion-dilaton field, but neglected the backreaction of the branes on the geometry. Second, the extension the KT gravity solution to include temperature, first without fundamental matter \cite{PleinTemp11,PleinTemp12} and later with fundamental matter \cite{PleinTemp21,PleinTemp22,PleinTemp23,PleinTemp24}, attracted a lot of attention. The key tool used here is to modify the geometry to that of a non-extremal black hole which induces a non-zero temperature. The work of \cite{KT-non-ex1,KT-non-ex2} in particular demonstrated how one should deviate from the usual self-dual 3-form fluxes once temperature is incorporated, but could not find a complete answer. These works also focused only on the IR physics and did not address the question of UV completion. This left the questions of Wilson loop divergences and Landau poles unanswered.

Based on some results obtained in \cite{DtermPaper} and presented in chapter \ref{DtermPaper}, the work of \cite{Mia:2009wj,jpsi1} conjectured an explicit KS-type supergravity background that would include both flavor D7-branes and non-zero temperature on a squashed conifold. From this, they further conjectured the precise form the fluxes should take to allow a UV completion, but never computed the precise equations for the fluxes and the geometry. The first part of \cite{OurLastPaper} made significant progress towards solving exactly the full supergravity equations of motion in this generalized background and the later part, presented in chapter \ref{actfluxes}, looks more precisely at the particular equations for the fluxes. 

From the AdS/CFT point of view, this construction brings us much closer to a realistic gravity realisation of the Standard Model's gauge theory than Maldacena's original construction. After introducing fundamental flavor via D7-branes, the gauge theory is no longer conformal and could even be UV complete. Temperature is included via non-extremality and supersymmetry is broken by the generalization of the geometry. If the large $N$ requirements could be relaxed, both for the number of colors and flavors, then we would be as close as we could imagine to a gravity dual of the Standard Model.

~\newline\\
\begin{flushleft}
{\it Overview}
\end{flushleft}

This thesis is organized as follows: Chapter \ref{chapitreintrobase} presents a literature review of the supergravity constructions presented above, up to the Klebanov-Strassler construction and GKP's flux compactification. Chapter \ref{DtermPaper} then presents our results on D7-branes embedding on the resolved conifold, the resulting D-term and the F-theory lift of the fluxes. The second part of the thesis then goes towards the thermal solutions. Chapter \ref{chapitreintrotemperature} presents how temperature is embedded via non-extremality in gravity, reviews previous approaches to a non-extremal Klebanov-Tseytlin solution, and gives some details of the background for which we constructed and solved the supergravity equations. Finally, chapter \ref{actfluxes} presents our calculations for the equations for thermal fluxes. %Although String inflation is one of our initial motivation, since we do not directly address this question here, we will not review it.

\chapter{Towards the Duality Cascade and Flux Compactification}\label{chapitreintrobase}

Finding new solutions to the equations of motion of String Theory with a clear holographic dual is not an easy task. Significant progress has been made since the early work of Maldacena \cite{MaldacenaoriginalAdSCFT}, especially towards flux compactification on conifold geometries.

Here, we will review how new solutions were developed by gradually implementing new ingredients. We will mostly follow the historical development of their findings but with a modern point of view. Starting with the simple Klebanov-Witten construction, we will progressively make our way to the Klebanov-Strassler solution with its duality cascade \cite{ks} and its compact version developed by Giddings, Kachru and Polchinski\cite{GKP}. This forms the basis upon which we will develop our solutions presented later in this thesis. Attention will be given to both the gravity and the gauge theory aspects of the particular solutions and only the features that are of interest to us will be presented.

%%%%%%%%%%%%%%%%%%%%%%%%%%%%%%%%%%%%%%%%%%%%%%%%%%%%%%%%%%%%%%%%%%%%%%%%%%%%%
\section{The Klebanov-Witten Solution}\label{SectionKW}
\begin{figure}[h]
 \begin{center}
 \includegraphics[height=6cm,angle=0]{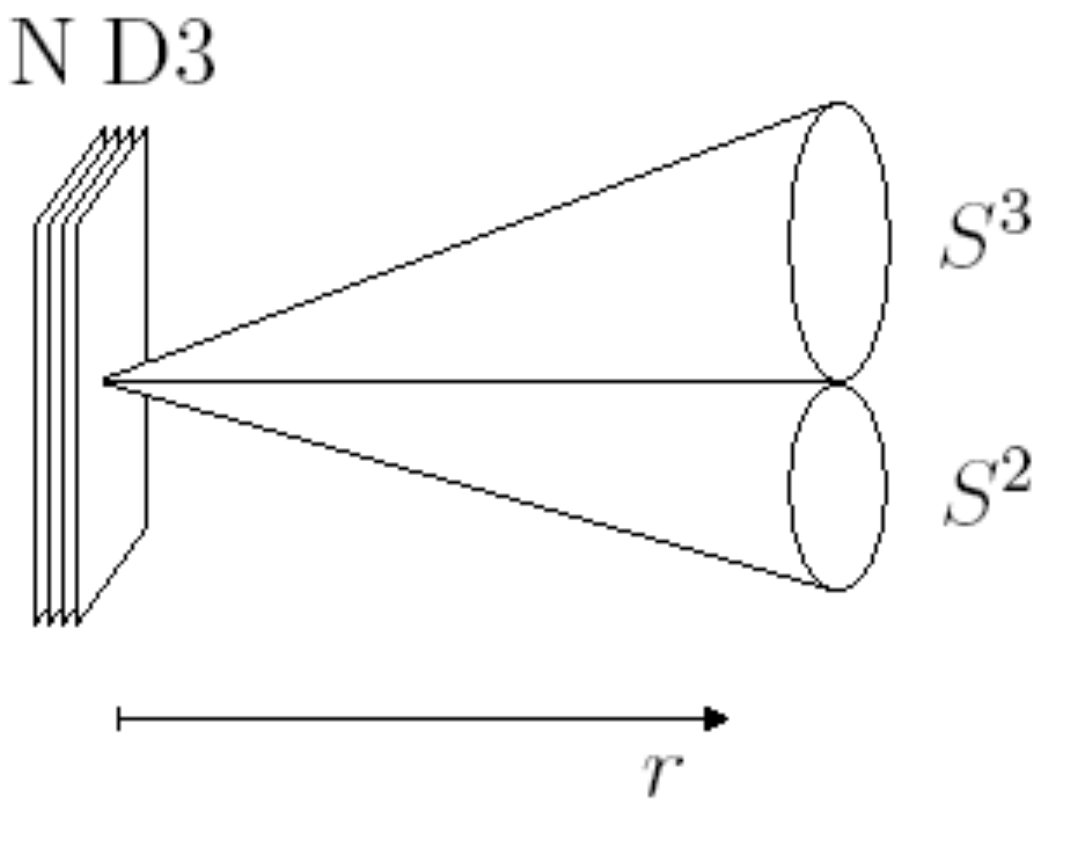}
 \caption{In the Klebanov-Witten solution, only a stack of $N$ parallel D3-branes is located at the tip of the  conifold.}
 \label{FigKW}
 \end{center}
\end{figure}

While Maldacena's ${\mathcal N}=4$ original conjecture was on $S^5$, which is the most symmetric internal manifold, other less symmetric manifolds, like the conifold $T^{1,1}$, also exhibit a clear holographic meaning with less supersymmetry. The Klebanov-Witten solution (KW) \cite{klebwit} is the first of a series of solutions on the conifold geometry. It also contains only D3-branes, but the conifold breaks the supersymmetry to  ${\mathcal N}=1$, allowing the gauge theory to have a superpotential. The ``magic" of AdS/CFT  tells us that the symmetries and the equations of motion for the fields in the superpotential are also the equations and symmetries that describe the position of the branes in the geometry. The fields describing the stack of $N$ D3-branes will then have the $SU(N)$ symmetry corresponding to the interchange of the $N$ position fields of each brane, as well as the symmetry of the angular rotations around its position. This brane configuration is illustrated in Figure \ref{FigKW}.

More precisely, from the gauge theory point of view, when the branes are located at the tip of the conifold, the resulting gauge theory is an $\mathcal{N}=1$ superconformal $SU(N)\times SU(N)$ gauge theory. It contains 4 chiral superfields: $A_1, A_2$, transforming under the $(N,\bar N)$ of the gauge group, and $B_1, B_2$ transforming in the conjugate representation. These superfields combine together in the gauge theory's superpotential $\mathcal{W}=\lambda\epsilon^{ij}\epsilon^{kl}TrA_iB_kA_jB_l$ which has the extra $SU(2)\times SU(2)\times U(1)$ global symmetry group of the conifold. This global symmetry actually contains 2 spurious $U(1)$ symmetries: one axial symmetry and an R-symmetry associated to chirality, which will remain unbroken here. 

From the dual gravity side, the D3-branes in type IIB supergravity are compactified on a warped conifold with a UV (or large radius) limit of $AdS_5 \times T^{1,1}$. The conifold is a 6-dimensional cone with base $T^{1,1}$ and $T^{1,1}$ is a $(SU(2)\times SU(2))/U(1)$ coset space with $S^2\times S^3$ topology and the same global symmetry as the gauge theory. 

Concretely, the warping of the metric means that the full 10-dimensional metric is of the form
\bg
ds_{10}^2=h^{-1/2}(r)\eta_{\mu\nu}dx^\mu dx^\nu + h^{1/2}(r)ds_6^2
\nd
where the warp factor $h(r)$ is given by
\bg
h(r)=1+\frac{L^4}{r^4}
\nd
with $L^4=4\pi g_sN\alpha'^2$ and $ds_6^2$ is given by the conifold metric:
\bg
ds_{conifold}^2=dr^2+r^2d_{T^{1,1}}^2
\nd
with
\bg\label{metriqueT11}
ds_{T^{1,1}}^2 = {1\over 9}(d\psi + {\cos}\theta_1 d\phi_1 +
\cos\theta_2 d\phi_2)^2 + {1\over 6}\sum_{i=1}^2(d\theta_i^2 + \sin^2\theta_i d\phi_i^2).
\nd

Connecting the gravitational and gauge theory perspectives, one can see \cite{klebwit} that the two couplings of the gauge theory are related to the integral of NS-NS $B_2$ and R-R $C_2$ 2-form moduli over the $S^2$ of $T^{1,1}$. In particular, their relation is given by
\bg\label{eqncouplings}
\frac{4\pi^2}{g_1^2} + \frac{4\pi^2}{g_2^2}&=&\pi e^{-\phi}\nonumber\\
\frac{4\pi^2}{g_1^2} - \frac{4\pi^2}{g_2^2}&=&e^{-\phi}\Big[\frac{1}{2\pi \alpha'}\Big(\int_{S^2} B_2\Big)-\pi\Big] \ \ \ (\text{mod} \ 2\pi).
\nd
In the present context, with only D3-branes, both $B_2$ and the dilaton $\phi$ happen to be constant, making the $g_1$ and $g_2$ couplings to be constant and the gauge theory to be conformal.

This regular D3-brane solution can however be made much more rich by adding fractional D3-branes to the regular ones. This solution is known as the Klebanov-Tseytlin (KT) solution \cite{kt} and will serve as the starting point in the UV of a series of Seiberg dualities, which Klebanov-Strassler (KS) \cite{ks} demonstrated to be an RG flow fully illustrating gauge/gravity duality.

%%%%%%%%%%%%%%%%%%%%%%%%%%%%%%%%%%%%%%%%%%%%%%%%%%%%%%%%%%%%%%%%%%%%%%%%%%%%%%%%%%%%%%
\section{The Klebanov-Tseytlin Solution}\label{KTsection}

The Klebanov-Tseytlin solution (KT) \cite{kt} considered the effect of adding $M$ fractional D3-branes, corresponding to D5-branes wrapping the 2-cycle of $T^{1,1}$ at the tip of the conifold, to the KW solution. These extra branes are additional possible locations for string endpoints and thus enhance the gauge group to $SU(N+M)\times SU(N)$. This affects the corresponding gauge theory on the branes by breaking conformality  while still preserving $\mathcal{N}=1$ supersymmetry. Note that on the gauge theory side, the $A_i$ and $B_i$ fields are now transforming under the $(N+M,\bar N)$ representation of the gauge group and its conjugate. This new configuration is illustrated below in Figure \ref{FigKT}.

\begin{figure}[h]
\begin{center}
\includegraphics[height=6cm,angle=0]{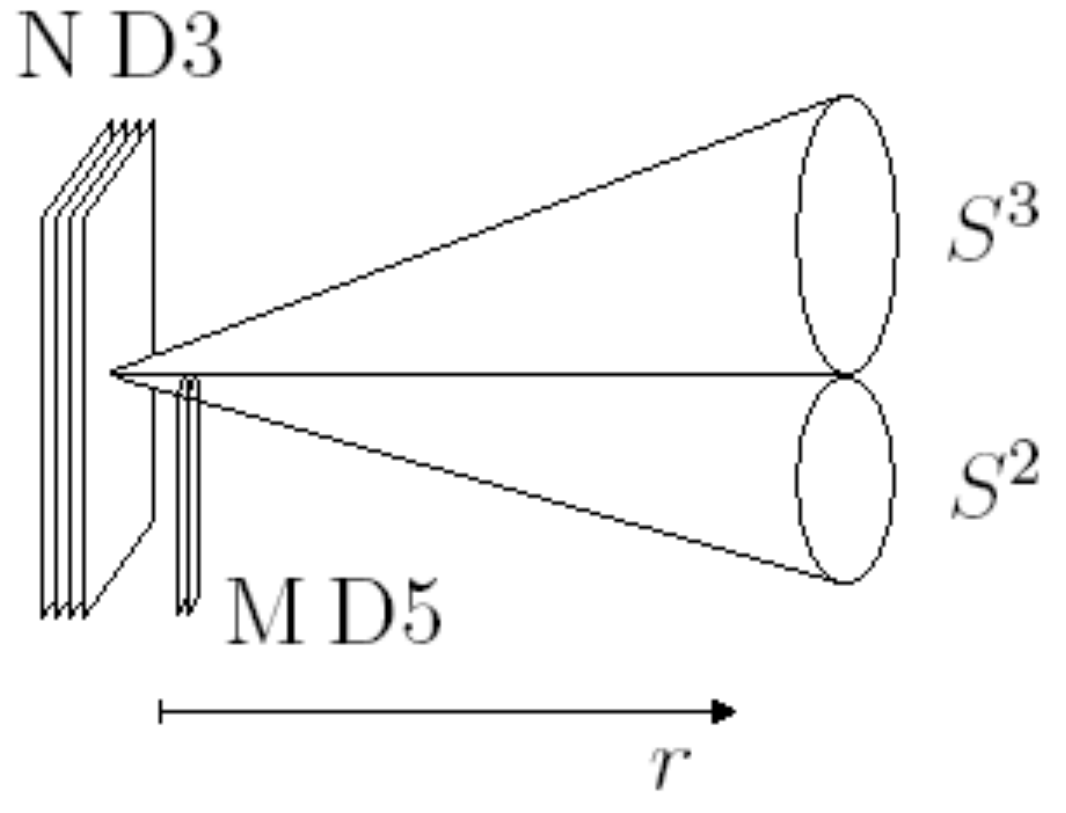}
\caption{In the Klebanov-Tseytlin solution, $M$ parallel D5-branes are placed on top of the $N$ D3-branes on the collapsed 2-cycle of $T^{1,1}$ at the tip of the conifold.}
\label{FigKT}
\end{center}
\end{figure}

An exact supergravity solution was found in KT for the 3- and 5-form fluxes, where the D5-branes source $M$ units of fluxes on $F_3$ as D5-branes are charged under the Hodge dual of $F_3$, i.e. under $F_7=\ast_{10}F_3 $. The 3-form flux solutions are then given by
\bg\label{eqnB2}
F_3 = Mw_3 &~&~~~~~~~B_2=3g_sMw_2\ln(r/r_0)\\
H_3&=&dB_2=3g_sM\frac{1}{r}dr\wedge w_2\nonumber
\nd
where $w_2$ and $w_3$ are closed and are also the basis of the $S^2$ and $S^3$ of $T^{1,1}$. Note also that these fluxes, combined into $G_3=F_3+ie^{-\phi}H_3$ are self-dual ($iG_3=\ast_6 G_3$), forcing the dilation to be constant.

A significant effect of these fluxes is their backreaction on the full 5-form $\tilde{F}_5=dC_4+B_2\wedge F_3$, which may now be written as:
\bg\label{cinqformdeKT}
\tilde{F}_5=\mathcal{F}_5+\ast_{10}\mathcal{F}_5, \ \ \ \ \mathcal{F}_5=\mathcal{K}(r)\text{vol}(T^{1,1}),\\
\mathcal{K}(r)=N+\frac{3}{2\pi}g_sM^2\ln(r/r_0)
\nd
This further contributes to the warp factor $h(r)$, which in the near-horizon takes  form
\bg
h(r)=\frac{L^4}{r^4}\Big(1+\frac{3g_s M^2}{2\pi N}\log r\Big)
\nd

The important feature to notice of this solution is the logarithmic dependence on the radial coordinate $r$, also called the RG scale, which can go infinitely negative at small $r$. More precisely, as $r$ decreases, i.e. towards the IR of the RG flow, both $\mathcal{F}_5$ and $h(r)$ decrease towards zero and eventually become negative. This singularity of the gravity solution is a clear hint that both the gravity and gauge theory description must be completed by new emerging elements as we move towards the IR. Furthermore, putting the $B_2$ flux (\ref{eqnB2}) into (\ref{eqncouplings}), we see that the fractional branes introduce a logarithmic running of the couplings, thereby explicitly breaking the conformality.

%%%%%%%%%%%%%%%%%%%%%%%%%%%%%%%%%%%%%%%%%%%%%%%%%%%%%%%%%%%%%%%%%%%%%%%%%%%%%%%%%%%%%%
\section{The Klebanov-Strassler Duality Cascade}\label{KSsection}

The apparent singularity of the KT solution was best understood later through the interpretation given by Klebanov and Strassler (KS) \cite{ks}. Their solution not only provides a mechanism to resolve and avoid the singularity, but also gives a geometrical realisation of confinement, as we would like to obtain from QCD-like theories. Let us explore how this occurs.

From the supergravity point of view, the amount of $\int_{S^2} B_2$ flux does not have to be periodic. As this flux decreases by one unit when $r$ decreases, so does the the 5-form flux by $M$ units: $\mathcal{K}(r)\rightarrow\mathcal{K}(r)-M$. Equivalently, the effective number of 3-branes also decreases: $N\rightarrow N-M$. This is interpreted as the gauge group changes from $SU(N+M)\times SU(N)$ to $SU(N-M)\times SU(N)$ as we go down one step in the RG scale. This process is called a Seiberg duality. Assuming $N$ is an integer multiple of $M$, this duality can be repeated until the IR limit where the gauge group simply becomes $SU(M)$ without quark flavor. This series of dualities is called a duality cascade. 

What happens at the end of this duality cascade is best illustrated from the gauge theory point of view. There, the $U(1)_R$ symmetry of the chiral fields is first broken to $Z_{2M}$ by the presence of the $M$ the fractional branes and further down $Z_2$ by the inclusion of the Affleck-Dine-Seiberg (ADS) \cite{ADS} contribution to the superpotential. This is exactly what we expect from confinement for pure $\mathcal{N}=1$ $SU(M)$ gauge theories and the reason we say there is confinement in the IR region of the duality cascade. 

\begin{figure}[h]
\begin{center}
\includegraphics[height=6cm,angle=0]{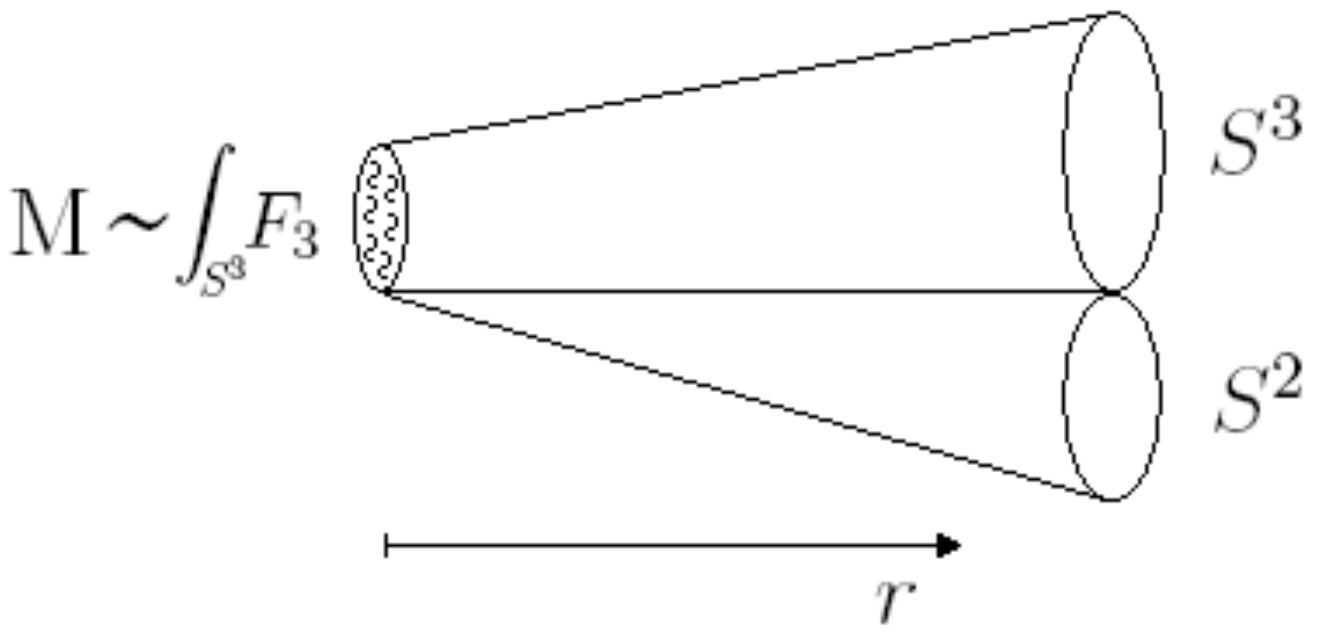}
\caption{At the end of the duality cascade, all the D3- and D5-branes have been dessolved into fluxes on the geometry, resulting in a finite size $S^3$ at the tip of the conifold.}
\label{FigKS}
\end{center}
\end{figure}

The ADS contribution to the superpotential leads to what is seen, from the supergravity perspective, as a resolution of the geometry. This changes the compactification geometry from that of the conifold to the deformed conifold with a finite size $S^3$ in the IR. More precisely, the geometry at the bottom of the cascade is of the form
\bg
d_{def}^2 &=& F_1(r)dr^2+F_2(r)(d\psi + {\cos}\theta_1 d\phi_1\nonumber
 + \cos\theta_2 d\phi_2)^2\\ &+& F_3(r)(d\theta_1^2 + \sin^2\theta_1 d\phi_1^2+d\theta_2^2 + \sin^2\theta_2 d\phi_2^2)\\
&+&F_4(r)[\cos\psi(d\theta_1d\theta_2+\sin\theta_1\sin\theta_2d\phi_1d\phi_2)-\cos\psi(\sin\theta_2d\phi_2d\theta_1
-\sin\theta_1d\phi_1d\theta_2)]\nonumber
\nd
Here, both 2-spheres have the same radius but the $U(1)_\psi$ shift is broken by the last line and allows for the blown up $S^3$ at $r=0$. In this limit, both $B_2$ and $F_5$ go to zero at $r=0$ while $F_3$ remains non-zero and is spread over the $S^3$. This is understood as there are no more regular D3-branes and the D5-branes wrapping the collapsed $S^2$ are smeared or dissolved over the $S^3$, leaving no branes remaining in the IR. This is illustrated in Figure \ref{FigKS}.

The full KS solution is an interpolation between the KT solution in the UV and the deformed solution in the IR. In the UV, branes wrap the conifold as a superstring solution and generate the gauge group on their world volume. As we move towards the IR, the branes dissolve into the geometry, deforming the conifold, reducing the gauge group and leaving us with a pure supergravity solution without branes as a source. This is the essence of the duality cascade.

Finally, if the UV completion of this setup could explicitly realize asymptotic freedom, String Theory would have provided a complete picture of the confinement/deconfinement phase transition of QCD-like theories. This however will be presented in the work of \cite{Mia:2009wj,jpsi1} as part of  chapter \ref{chapitreintrotemperature}.

%%%%%%%%%%%%%%%%%%%%%%%%%%%%%%%%%%%%%%%%%%%%%%%%%%%%%%%%%%
\section{The Pando Zayas-Tseytlin Solution}\label{ptbg}
%%%%%%%%%%%%%%%%%%%%%%%%%%%%%%%%%%%%%%%%%%%%%%%%%%%%%%%%%%

There are two natural ways to smooth the singularity of the conifold at $r=0$: make the $S^3$ of finite size (deformed conifold) or make the $S^2$ of finite size (resolved conifold) \cite{ankenon}. Similar to the Klebanov-Strassler model, Pando Zayas and Tseytlin (PT) \cite{pt} have shown that a warped geometry can be created by fluxes in the \emph{resolved} conifold background with fractional 3-branes. The type IIB supergravity solution was found to correspond to the KT and KS solution in the UV (at large $r$), but  differences arise in the IR (small $r$) where the resolution appears. The PT solution can then be seen as a new starting point for the duality cascade. Before we look to the details of the PT solution for fractional branes, let us review the resolved conifold geometry. Since this will be of much importance later, more details will be provided here than in previous sections. Notice too that the solutions presented so far correspond below to the case where we take the resolution to vanish. As we can see, taking $a=0$ in \eqref{resmetric} brings us back to the metric of KW \eqref{metriqueT11} and we recover the same geometry.

The resolved conifold is a manifold which looks asymptotically like the singular conifold, but is non--singular at the tip. Its geometry can be derived by starting with the singular version, a non--compact Calabi--Yau 3--fold, that can be embedded in $\mathbb{C}^4$ as \cite{candelas}
\begin{equation}
  \sum_{i=1}^4 z_i^2 \,=\, 0\,.
\end{equation}
This describes a cone over $S^2\times S^3$, which becomes singular at the origin.
By a change of coordinates this can also be written as
\begin{equation}
  yz-uv \,=\, 0\,,
\end{equation}
which is equivalent to non--trivial solutions to the equation 
\begin{equation}
\begin{pmatrix} z & u \\ v & y\end{pmatrix}
\begin{pmatrix} \xi_1 \\ \xi_2 \end{pmatrix}
\,=\, 0\,,
\end{equation}
in which $\xi_1, \xi_2$ are homogeneous coordinates on $\mathbb{CP}^1\sim S^2$. For $(u,v,y,z)\ne 0$ (away from the tip), they describe again a conifold. But at $(u,v,w,z)=0$ this is solved by any pair $(\xi_1,\xi_2)$. Due to the overall scaling freedom $(\xi_1,\xi_2)\sim (\lambda\xi_1,\lambda\xi_2)$ we can mod out by this equivalence class and $(\xi_1,\xi_2)$ actually describe a $\mathbb{CP}^1\sim S^2$ at the tip of the cone. 
The resolved conifold can be covered by two complex coordinate patches ($H_+$ and $H_-$), given by 
\begin{eqnarray}\label{coordpatch}
H_+ &=& \{ \xi_1 \neq 0 \} = \{ (u,y;\lambda) | u,y,\lambda \in \mathbb{C} \} \ , \ \lambda = \frac{\xi_2}{\xi_1}   \\
H_- &=& \{ \xi_2 \neq 0 \} = \{ (v,z;\mu) | v,z,\mu \in \mathbb{C} \} \ , \ \mu = \frac{\xi_1}{\xi_2} \, . 
\end{eqnarray}
On $H_+$ we have that 
\begin{equation}
z = -u \lambda \ , v = - y \lambda  \,,
\end{equation}
on $H_-$
\begin{equation}
y = -v \mu \ , u = - z \mu  \,,
\end{equation}
and on the intersection of these two patches, the coordinates are related by
\begin{equation}\label{overlap}\nonumber
(v,z;\mu) = (-y\lambda,-u\lambda ; 1/\lambda)  \,.
\end{equation}
The holomorphic coordinates are conveniently parametrized by
\begin{eqnarray}\label{holocoord}\nonumber
  z & =&  \left ( 9 a^2 \rho^4 + \rho ^6 \right ) ^{1/4} e^{i/2(\psi-\phi_1-\phi_2)}\,\sin(\theta_1/2)\sin(\theta_2/2) \\ \nonumber
  y & =&  \left ( 9 a^2 \rho^4 + \rho ^6 \right ) ^{1/4} e^{i/2(\psi+\phi_1+\phi_2)}\,\cos(\theta_1/2)\cos(\theta_2/2) \\
  u & =&  \left ( 9 a^2 \rho^4 + \rho ^6 \right ) ^{1/4} e^{i/2(\psi+\phi_1-\phi_2)}\,\cos(\theta_1/2)\sin(\theta_2/2) \\ \nonumber
  v & =&  \left ( 9 a^2 \rho^4 + \rho ^6 \right ) ^{1/4} e^{i/2(\psi-\phi_1+\phi_2)}\,\sin(\theta_1/2)\cos(\theta_2/2)\,.
\end{eqnarray}
Here, $\theta_i=0\ldots \pi$, $\phi_i=0\ldots 2\pi$ are the usual Euler angles on $S^2$, $\psi=0\ldots 4\pi$ describes a U(1) fibre over the two 2--spheres and $\rho=0\ldots\infty$ is the radial coordinate. Note that our radial coordinate $\rho$ is related to the commonly-used $r$ via $\rho^2=3/(2 r^2) F'(r^2)$, where $F(r^2)$ appears in the K\"ahler potential $K$ of the resolved conifold
\begin{equation}
  K(r) \,=\, F(r^2)+4a^2\log(1+|\lambda|^2)\,.
\end{equation}
Note that the K\"ahler potential is not a globally defined quantity, since $\lambda$ is only defined on the patch $H_+$ that excludes $\xi_1=0$. For completeness let us also quote \cite{candelas, pt}
\begin{eqnarray}\label{Fprime}
  F'(r^2) \,=\, \frac{\partial F(r^2)}{\partial r^2} &=& \frac{1}{r^2}\,\left(-2a^2 + 4a^2 N^{-1/3}(r) + N^{1/3}(r)\right) \quad\mbox{with}\\ \label{defN}
  N(r) &=& \frac{1}{2}\,\left(r^4-16 a^2+\sqrt{r^8-32a^6r^4}\right)\,.
\end{eqnarray}
The inverse relation between $\rho$ and $r$ is found to be
\begin{equation}\label{rhoandr}
  r \,=\, \left(\frac{2}{3}\right)^{3/4}\,(9a^2\rho^4+\rho^6)^{1/4}\,.
\end{equation}

\noindent In terms of these real coordinates the Ricci--flat K\"ahler metric on the resolved conifold reads
\begin{eqnarray}\label{resmetric}\nonumber
  ds^2_{\rm res} & = & \kappa(\rho)^{-1}\,d\rho^2 + \frac{\kappa(\rho)}{9}\,\rho^2\big(d\psi+\cos\theta_1\,d\phi_1
    +\cos\theta_2\,d\phi_2\big)^2 \\
  & &+ \frac{\rho^2}{6}\,\big(d\theta_1^2+\sin^2\theta_1\,d\phi_1^2\big) 
    +\frac{\rho^2+6a^2}{6}\,\big(d\theta_2^2+\sin^2\theta_2\,d\phi_2^2\big)\,,
\end{eqnarray}
with $\kappa(\rho)=(\rho^2+9a^2)/(\rho^2+6a^2)$.  In the limit $a\to 0$ one recovers the singular conifold metric. Note that as $\rho\to 0$, the $(\theta_2,\phi_2)$ sphere remains finite, whereas for the singular conifold both $(\theta_i,\phi_i)$ spheres scale with $\rho^2/6$. Therefore, $a$ is called ``resolution'' parameter and gives the radius of the blown--up 2--sphere at the tip. 

\noindent It will be useful later on to have a set of vielbeins that describes this metric, i.e.
\begin{equation}
  ds^2 \,=\, \sum_{i=1}^6 (e_i)^2\,.
\end{equation}
Following \cite{ankenon}, we choose
\begin{eqnarray}\label{resvielb}\nonumber
  e_1 &=& \kappa^{-1/2}\,d\rho\\ \nonumber
  e_2 &=& \frac{\rho\sqrt{\kappa}}{3}\,(d\psi+\cos\theta_1\,d\phi_1+\cos\theta_2\,d\phi_2) 
    \,=\, \frac{\rho\sqrt{\kappa}}{3}\,e_\psi\\ \nonumber
  e_3 &=&\frac{\rho}{\sqrt{6}}\,(\sin\psi/2\,\sin\theta_1\,d\phi_1+\cos\psi/2\,d\theta_1)\\ 
  e_4 &=&\frac{\rho}{\sqrt{6}}\,(-\cos\psi/2\,\sin\theta_1\,d\phi_1+\sin\psi/2\,d\theta_1)\\ \nonumber
  e_5 &=&\frac{\sqrt{\rho^2+6a^2}}{\sqrt{6}}\,(\sin\psi/2\,\sin\theta_2\,d\phi_2+\cos\psi/2\,d\theta_2)\\ \nonumber
  e_6 &=&\frac{\sqrt{\rho^2+6a^2}}{\sqrt{6}}\,(-\cos\psi/2\,\sin\theta_2\,d\phi_2+\sin\psi/2\,d\theta_2)
\end{eqnarray}
as they lead to a closed K\"ahler form $J$ as well as a closed holomorphic 3--form $\Omega$ with a simple complex structure induced by
\begin{equation}\label{J}
  J^{(1,1)} \,=\, e_1\wedge e_2+ e_3\wedge e_4+e_5\wedge e_6\,,\qquad
  \Omega^{(3,0)} \,=\, (e_1+ie_2)\wedge(e_3+ie_4)\wedge(e_5+ie_6)\,,
\end{equation}
in other words we define our complex vielbeins to be
\begin{equation}\label{cs}
  E_1 \,=\, e_1+i\,e_2\,,\qquad  E_2 \,=\, e_3+i\,e_4\,,\qquad  E_3 \,=\, e_5+i\,e_6\,. 
\end{equation}
This results in a coordinate expression for $J$ as
\begin{eqnarray}\label{Jres}\nonumber
  J &=& \frac{\rho}{3}\,d\rho\wedge (d\psi+\cos\theta_1\,d\phi_1+\cos\theta_2\,d\phi_2)\\
  & & + \frac{\rho^2}{6}\,\sin\theta_1\,d\phi_1\wedge d\theta_1+ \frac{\rho^2+6a^2}{6}\,\sin\theta_2\,d\phi_2\wedge 
    d\theta_2\,.
\end{eqnarray}

\begin{figure}[h]
\begin{center}
\includegraphics[height=6cm,angle=0]{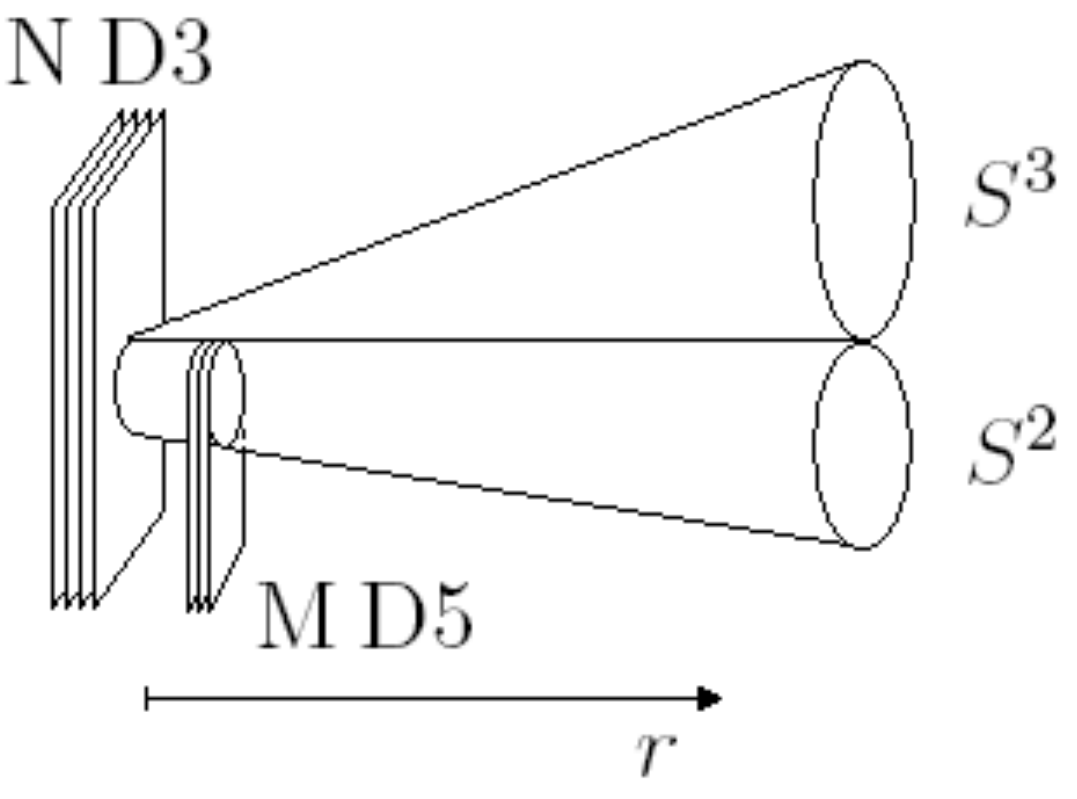}
\caption{The PT solution is a natural extension of the KT solution where the 2-cycle wrapped by the D5-branes is non-singular at the tip of the conifold.}
\label{FigPT}
\end{center}
\end{figure}

As mentioned above, the full supergravity solution for the resolved conifold was derived by Pando--Zayas and Tseytlin \cite{pt} (PT) and includes non--trivial RR and NS flux with constant dilaton. It can be understood as placing a stack of fractional D3--branes (i.e. D5--branes that wrap a 2--cycle) in this background. This construction is illustrated in Figure \ref{FigPT}. 

The ten--dimensional metric is found to be
\begin{equation}
\label{pzmet}
 ds_{10}^2=h^{-1/2}(\rho)\,\eta_{\mu\nu}dx^\mu dx^\nu +h^{1/2}(\rho)\,ds_6^2\,,
\end{equation}
where $ds_6^2$ refers to the resolved conifold metric found in (\ref{resmetric}). This asymmetry in the resolved geometry plays a crucial role here as it determines the asymmetry in the flux on the 2--cycles and is the source of supersymmetry breaking.
The 3--form fluxes in this background are\footnote{There is a typo in eq. (4.3) of \cite{pt}, concerning the sign of $F_3$.}
\begin{eqnarray}\label{fluxres}
  H_3 & = & d\rho\wedge[f'_1(\rho) \,d\theta_1 \wedge \sin\theta_1\,d\phi_1+f'_2(\rho)\,d\theta_2\wedge\sin\theta_2\,d\phi_2]  \\
  F_3 & = & P e_{\psi}\wedge (d\theta_1\wedge \sin\theta_1\,d\phi_1- d\theta_2\wedge\sin\theta_2\,d\phi_2 )
\end{eqnarray}
and the self--dual 5--form flux is given by
\begin{equation}
F_5 = {\cal F}+* {\cal F}\ , \quad  \ \ \ \ {\cal F} = K(\rho)\,e_{\psi}\wedge d\theta_1 \wedge\sin\theta_1\,d\phi_1 \wedge d\theta_2\wedge \sin\theta_2\,d\phi_2 \ ,
\end{equation}
where
\begin{eqnarray}\label{deffs}\nonumber
  f_1(\rho) & =& \frac{3}{2} g_s P \ln (\rho^2+9a^2)  \\
  f_2(\rho) & =&   \frac{1}{6} g_s P \bigg(  \frac{36a^2}{\rho^2} - \ln [\rho^{16} (\rho^2+9a^2)] \bigg)  \\ \nonumber
  K(\rho)   & =&  Q - \frac{1}{3}  g_s P^2  \bigg(  \frac{18a^2}{\rho^2}- \ln [\rho^{8} (\rho^2+9a^2)^5] \bigg) 
\end{eqnarray}
and where P is proportional to the number of fractional D3-branes $M$ and Q proportional to the number of regular D3-branes\footnote{Different notations have been used to represented the number of branes. It is understood that $Q\sim\alpha'^2N\sim L^4$ and $P\sim\alpha'M$ and it should be obvious from the context.}, and both are proportional to $\alpha^\prime$.

In the large $\rho$ limit ($\rho\gg3a^2$), we reproduce exactly the KT solution \cite{kt} with its characteristic logarithmic behavior. In the small distance limit ($\rho\ll3a^2$), the solution allows for singularities to appear for $K(\rho)$ and $h(\rho)$ at $\rho_K$ and $\rho_h$ respectively. Since $\rho_K>\rho_h$, one expects the geometry to stop at $\rho_K$. Expanding around $\rho_K$ in the IR, one gets a similar behavior as in the deformed conifold case \cite{ks} and should expect confinement.

Let us rewrite the above solution in a new notation that we defined earlier. Using \eqref{resvielb} we can rewrite the 3--form flux in terms of vielbeins
\begin{equation}
G_3 \,=\, -\frac{18 P}{\rho^3\sqrt{\kappa}}\,(e_2\wedge e_3\wedge e_4+i\, e_1\wedge e_5\wedge e_6)
 +\frac{18P\,(e_2\wedge e_5\wedge e_6+i\, e_1\wedge e_3\wedge e_4)}{\rho\sqrt{\rho^2+6a^2}\sqrt{\rho^2+9a^2}}\,.
\end{equation}
The vielbein notation is extremely convenient to see that this flux is indeed imaginary self-dual. The Hodge dual is simply found to be
\begin{equation}\nonumber
*_6 (e_{i_1}\wedge e_{i_2}\wedge\ldots\wedge e_{i_k}) = \epsilon_{i_1 i_2\ldots i_k}^{\phantom{i_1 i_2\ldots i_k}i_{k+1}\ldots i_6}\,e_{i_{k+1}}\wedge \ldots \wedge e_{i_6}
\end{equation}
and does not involve any factors of $\sqrt{g}$. We use the convention that $\epsilon_{123456}=\epsilon_{123}^{\phantom{123}456}=1$.
With the complex structure \eqref{cs} the PT flux becomes
\begin{eqnarray}\nonumber
  G_3 &=& \frac{-9 P}{\rho^3\sqrt{\rho^2+6a^2}\sqrt{\rho^2+9a^2}}\,\Big[(\rho^2+3a^2)\,(E_1\wedge E_2\wedge \overline E_2 
    - E_1\wedge E_3\wedge\overline E_3)\\
  & & \quad\phantom{\rho^3\sqrt{\rho^2+6a^2}\sqrt{\rho^2+9a^2}}+ 3a^2\,(E_2\wedge\overline E_1\wedge\overline E_2 + E_3\wedge\overline E_1\wedge\overline E_3)\Big]\,.
\end{eqnarray}
We make several observations: This flux is neither primitive\footnote{Since $J=\frac{\imath}{2}\,\sum_i(E_i\wedge\overline E_i)$ it
follows immediately that $J\wedge G_3$ has a non-vanishing $E_2\wedge E_3\wedge \overline{E_1}\wedge \overline{E_2}\wedge\overline{E_3}$ part that is proportional to $a^2$.} nor is it of type (2,1). It has a (1,2) \em
and \em a (2,1) part, which cannot be avoided by a different choice of complex structure. 

It was pointed out in \cite{cvetgiblupo} and confirmed in \cite{ankenon} that this solution breaks supersymmetry. The reason lies in the (1,2) part of the 3--form flux. Analysis of the explicit susy variations lead to require $J\wedge G_3=0$ (primitivity) and $G_3$ being purely (2,1)\cite{granapol, kst}. Therefore, the PT flux breaks susy ``in two ways'', which are actually equivalent statements for ISD fluxes. As we will shown in the next section, it is, nevertheless, a supergravity solution because the 3--form flux $G_3=F_3-iH_3$ obeys the imaginary self--duality condition $*_6 G_3=iG_3$.

We also observe that, in the limit $a\to 0$, the (1,2) part vanishes, the flux becomes primitive, and we recover the singular conifold solution. This indicates that the resolution forbids a supersymmetric supergravity solution, i.e. the blow--up of a nontrivial 2--cycle in a conifold geometry can lead to supersymmetry breaking. We will use this fact to our advantage. Before we do so, let us step back and elaborate on supersymmetric flux compactification on compact manifolds.

%%%%%%%%%%%%%%%%%%%%%%%%%%%%%%%%%%%%%%%%%%%%%%%%%%%%%%%%%%%%%%%%%%%%%%%%%%%%%
\section{GKP Flux Compactification}\label{SectionGKP}

Although the KS solution is a fascinating string solution for regular and fractional branes, it is nevertheless relies on the non-compact conifold. If this could not be solved, it would be a problem as non-compact manifolds are generically not suitable for reducing to four dimensions since they induce an infinite 4d Planck mass ($M^2_{p,4D}=M^2_{p,10D}V_6$). Fortunately, flux compactification allows for warped compactifications which are locally like KS, but globally compact. The work of Giddings, Kachru and Polchinski (GKP) \cite{GKP} provided a fully compact supergravity solution with fluxes which stabilized all the moduli, except one, and obtained a hierarchy from quantized fluxes. Since we will use the same procedure later to solve for the thermal supergravity equations in chapter \ref{chapitreintrotemperature}, we will review their compactification here.

We start with the $SL(2,\mathbb{Z})$-invariant type IIB action:
\bg \label{tubra}
S_{\rm IIB} &=& \frac{1}{2\kappa_{10}^2}\int d^{10}x\;
\sqrt{-g}\Bigg[R-\frac{\partial_a \tau \partial^a \bar\tau}{2|{\rm
Im}\tau|^2}-
\frac{G_3\cdot\bar{G}_3}{12{\rm Im}\tau}-\frac{\widetilde{F}_5^2}{4\cdot 5!}\Bigg]\nonumber\\
&+& \frac{1}{8i\kappa_{10}^2}\int \frac{C_4\wedge G_3 \wedge
\bar{G_3}}{{\rm Im}\tau } +S_{\rm loc}
\nd
where $S_{loc}$ is the action for all the localized sources, like D-branes.

We have defined the complex axion-dilaton field  as
\bg\label{eqntau}
\tau = C_0 + ie^{-\phi} ~~~ ,
\nd
the combined complex 3-form flux as
\bg
G_3=F_3-\tau H_3~~~,
\nd
and the full 5-form flux as
\bg
\widetilde{F}_5=F_5-\frac{1}{2}C_2\wedge H_3+\frac{1}{2}B_2\wedge F3~~.
\nd
The condition $\widetilde{F}_5=\ast_{10}\widetilde{F}_5$ must be imposed by hand on the equations of motion and $F_5=dC_4$. 

As always, we look for warped solutions of the form 
\bg\label{metricGKP}
ds_{10}^2=e^{2A}\eta_{\mu\nu}dx^\mu dx^\nu +e^{-2A}\tilde{g}_{mn}dy^mdy^n
\nd
where $x^\mu$ are the 4-dimensional spacetime coordinates maintaining the Poincar\'e symmetry and $y^m$ are on the compact 6-dimensional manifold $\mathcal{M}_6$. We will allow the warp factor $A$ and the fluxes to vary only on the compact directions. Poincar\'e invariance allows us to turn on a 5-form flux of the form
\bg\label{5form}
\widetilde{F}_5=(1+\ast_{10})(d\alpha\wedge dx^0\wedge dx^1\wedge dx^2\wedge dx^3)
\nd
where $\alpha$ is a function of the compact space. 

Turning our attention to the sources of $S_{loc}$, for a $Dp$-brane wrapping the spacetime directions and a $(p-3)$-cycle $\Sigma$ in ${\mathcal M}_6$ with vanishing fluxes on the brane, the leading order terms in $\alpha'$ in the action are
\bg
S_{loc}=-\int d^{p+1}\xi T_p\sqrt{-g}+\mu_p\int C_{p+1}~~.
\nd

Considering this action, let us now look for solutions to the equations of motion. Einstein's equations of motion, trace reversed, are
\bg
R_{MN}=\kappa_{10}(T_{MN}-\frac{1}{8}g_{MN}T)
\nd
where $T_{MN}=T_{MN}^{SUGRA}+T_{MN}^{loc}$ is the total stress tensor of the supergravity fields and the localized sources respectively. In particular, the contribution from the localized sources is defined as
\bg
T^{loc}_{MN}=-\frac{2}{\sqrt{-g}}\frac{\delta S_{loc}}{\delta g^{MN}} ~.
\nd
With the action \ref{tubra}, the Einstein equations of motion then reads
\bg \label{ricci_T}
R_{\mu\nu}&=&-g_{\mu\nu} \left[\frac{G_3 \cdot \bar{G_3}}{48\; {\rm
Im}\tau}+\frac{\widetilde{F}_5^2}{8\cdot 5!}\right]+\frac{\widetilde{F}_{\mu
abcd}\widetilde{F}_\nu^{\;abcd}}{4 \cdot 4!}
+\kappa_{10}^2 \left(T_{\mu\nu}^{\rm loc}-\frac{1}{8} g_{\mu\nu} T^{\rm loc}\right)\nonumber\\
R_{mn}&=&-g_{mn} \left[\frac{G_3 \cdot \bar{G_3}}{48 \;{\rm
Im}\tau}+\frac{\widetilde{F}_5^2}{8\cdot 5!}\right]+\frac{\widetilde{F}_{m
abcd}\widetilde{F}_n^{\;abcd}}{4 \cdot 4!}+\frac{G_m^{\;bc}\bar{G}_{nbc}}{4\;{\rm Im}\tau}
+\frac{\partial_m \tau \partial_n \bar\tau}{2\;|{\rm Im}\tau|^2}\nonumber\\
&+&\kappa_{10}^2 \left(T_{mn}^{\rm loc}-\frac{1}{8} g_{mn} T^{\rm
loc}\right) 
\nd
Using the fact that the five-form flux is self-dual and of the form (\ref{5form}), the spacetime components of the equations become
\bg
\label{Ricci_Min}
R_{\mu\nu}&=&-g_{\mu\nu}
\left[\frac{G_3 \cdot \bar{G_3}}{48\; {\rm Im}\tau}+\frac{e^{-8A}
\partial_m\alpha \partial^m\alpha}{4}\right]+\kappa_{10}^2
\left(T_{\mu\nu}^{\rm loc}-\frac{1}{8} g_{\mu\nu} T^{\rm loc}\right)
\nd
With our metric ansatz (\ref{metricGKP}), the Ricci tensor along spacetime reads
\bg\label{Ricciextr}
R_{\mu\nu}&=&-\eta_{\mu\nu}
e^{4A}\tilde{\triangledown}^2A=-\frac{1}{4}\eta_{\mu\nu}(\widetilde{\triangledown}^2e^{4A}-e^{-4A}\partial_{\tilde{m}}e^{4A}\partial^{\tilde{m}}e^{4A})
\nd
where the tildes refers to the compact metric $\tilde{g}^{mn}$ and where the Laplacian $\widetilde{\triangledown}^2$ is defined as:
\bg\label{laplabe}
\widetilde{\triangledown}^2\phi&=&\frac{1}{\sqrt{\widetilde{g}}}\partial_m\big(\sqrt{\widetilde{g}}\widetilde{g}^{mn}\partial_n\phi\big)\\
&=&\widetilde{g}^{mn}\partial_m\partial_n\phi
+\partial_m\widetilde{g}^{mn}\partial_n\phi+\frac{1}{2}\widetilde{g}^{mn}\widetilde{g}^{pq}\partial_n\widetilde{g}_{pq}
\partial_m\phi\nonumber
\nd
Using (\ref{Ricciextr}) and tracing (\ref{Ricci_Min}), we get
\bg\label{eqnein}
\widetilde{\triangledown}^2e^{4A}&=&
e^{2A}\frac{G_{mnp}\bar{G}^{mnp}}{12\textrm{Im}\tau} +e^{-6A}\big(\partial_m\alpha\partial^{m}\alpha+\partial_me^{4A}\partial^{m}e^{4A}\big)\\
&+&\frac{k^2_{10}}{2}e^{-2A}(T^m_m-T^{\mu}_{_\mu})^{loc}
\nd
This equation, without local sources $T^{loc}$, is the origin of a no-go theorem for warped compactifications \cite{deWitSmiwHarriDass, MaldNunez}. Integrating both sides of the equations on a compact manifold, the LHS vanishes because it is a total derivative, while the RHS is a sum of squares which each have to vanish individually, leaving no non-trivial solution. This theorem can fortunately be evaded in string theory because the source terms on the RHS can give negative contributions, allowing warped compactifications.

Turning away from the Einstein equations, let us look at the equations of motion (or Bianchi identity) for the 5-form flux. 
\bg\label{jhgjhgjhgjhg}
d\tilde{F}_5=H_3\wedge F_3+2\kappa^2_{10}T_3\rho_3^{loc} 
\nd
where $\rho_3$ is the 3-brane charge density from all localized sources, including {\it e.g.} possible 7-branes. Integrating (\ref{jhgjhgjhgjhg}) over ${\mathcal M}_6$, we obtain the type IIB tadpole-cancellation condition
\bg
\frac{1}{2\kappa_{10}^2T_3}\int_\mathcal{M} H_3\wedge F_3+Q_3=0~~,
\nd
where $Q_3$ is the total charge from $\rho_3$. This shows that if $Q_3\neq0$, 3-form flux must be turned on. Note also that if D3 charge is induced on D7-branes, it will have a negative contribution to $Q_3$. Lifting to F-theory on a Calabi-Yau four-fold X, we can show that $Q_3$ is given $Q_3=N_{D3}-\chi(X)$, where $\chi(X)$ is the Euler characteristic of the four-fold $X$ wrapped by the 7-brane in $\mathcal{M}$ and $N_{D3}$ is the D3-brane charge present.

Using our metric (\ref{metricGKP}) and 5-form flux ansatz \ref{5form}, the Bianchi identity (\ref{jhgjhgjhgjhg}) becomes
\bg
\widetilde{\triangledown}^2\alpha=ie^{2A}\frac{G_{mnp}(\ast_6\bar{G}^{mnp})}{12\textrm{Im}\tau}+2e^{-6A}\partial_m\alpha\partial^m4^{4A}+2\kappa_{10}^2e^{2A}T_3\rho_3^{loc}
\nd
where $\ast_6$ is the dual along the compact directions. Subtracting this from (\ref{eqnein}), one gets
\bg\label{eqnGKPextremal}
\widetilde{\triangledown}^2(e^{4A}-\alpha)&=& \frac{e^{2A}}{6\textrm{Im}\tau}|{i}
G_3-\ast_6G_3|^2+e^{-6A}|\partial (e^{4A}-\alpha)|^2\\
&+&2\kappa_{10}^2e^{2A}\big[\frac{1}{4}(T_m^m-T_\mu^\mu)^{loc}-T_3\rho_3^{loc}\big]\nonumber
\nd
Integrating over ${\mathcal M}_6$, the LHS integrates to zero while the RHS is non-negative, assuming the contribution of the local sources respects the BPS-like condition. Thus, taking the 3 terms on the RHS to be zero, we conclude that
\begin{itemize}
\item the 3-form flux is imaginary self-dual (ISD)
\bg
\ast_6G_3=iG_3 ~,
\nd
\item the warp factor and 4-form flux are related
\bg\label{anzalpha}
\alpha = e^{4A} ~,
\nd
\item and the BPS condition is saturated
\bg
\frac{1}{4}(T^m_m-T^\mu_\mu)^{loc}=T_3\rho_3^{loc} ~.
\nd
\end{itemize}
Note that D3-branes, D7-branes on 4-cycles and O3-planes all saturate the BPS condition.

The remaining equations to satisfy are the 3-form Bianchi identities
\bg
dF_3=dH_3=0~~~~,
\nd
the Einstein equations along the compact directions and the $\tau$-field equations
\bg \label{cond1}
\tilde{R}_{mn}&=&\frac{\partial_m \tau \partial_n \bar\tau}{2\;|{\rm Im}\tau|^2}
+\kappa_{10}^2 \left(T_{mn}^{D7}-\frac{1}{8} g_{mn} T^{D7}\right)\\ \label{cond2}
\widetilde{\triangledown}^2\tau &=&\frac{\widetilde{\triangledown}\tau \cdot \widetilde{\triangledown}\tau}{i{\rm Im}\tau}-\frac{4\kappa_{10}^2({\rm Im}\tau)^2}{\sqrt{-g}}\frac{\delta S_{D7}}{\delta\bar{\tau}}
\nd
To summarize, a necessary and sufficient condition to have a solution to all of the supergravity equations of motions  on a compact manifold is to require the manifold to satisfy (\ref{cond1}) and (\ref{cond2}), the $F_3$ and $H_3$ forms closed, an ISD $G_3$ flux and to satisfy the tadpole-cancellation and BPS conditions.

Let us now turn our attention to the superpotential formulation of the moduli stabilisation. The scalar potential of $\mathcal{N}=1$ 4d supergravity can be derived by direct dimensional reduction of the IIB supergravity action. It is induced by the flux kinetic term
\begin{equation}
  S_G \,=\, -\frac{1}{4\kappa_{10}^2}\int \frac{G_3\wedge *\overline{G}_3}{{\rm Im} \tau}\,,
\end{equation}
where the Hodge star is taken on the internal manifold, so this integral runs over the six internal dimensions.
This can be rewritten as a potential plus a topological term, if we split $G_3$ in its ISD and anti-ISD part
\begin{eqnarray}\nonumber
  && G_3 = G^{\rm ISD}+G^{\rm AISD}\,,\qquad\qquad G^{\rm (A)ISD} 
 \,\equiv \, \frac{1}{2}\big(G_3\pm i *G_3\big)\\
  && *G^{\rm ISD} =  iG^{\rm ISD}\,, \qquad\qquad\qquad  *G^{\rm AISD} \,=\,  -iG^{\rm AISD}\,.
\end{eqnarray}
Then this part of the action becomes
\begin{eqnarray}\nonumber
  S_G &=& -\frac{1}{2\kappa_{10}^2}\int \frac{G^{\rm AISD}\wedge *\overline{G}^{\rm AISD}}{{\rm Im}\tau}+\frac{i}{4\kappa_{10}^2}\int 
    \frac{G_3\wedge \overline{G}_3}{{\rm Im}\tau}\\[1ex]
  &=& -V -N_{\rm flux}\,.
\end{eqnarray}
The second term is topological and independent of the moduli. In a compact setup it will be cancelled by the localised charges, if we use the tadpole cancellation condition $\int H_3\wedge F_3=-2\kappa_{10}^2T_3 Q_3^{\rm loc}$. This condition is of course relaxed in a non--compact space, but we want to keep the point of view that we can consistently compactify our background in an F--theory framework. The potential for the moduli is given by the anti-ISD fluxes only\footnote{For a more precise treatment that also includes warping, the Einstein term and the $F_5$ flux term see \cite{dewolfe}. The qualitative result remains unchanged. It was actually shown that the GVW superpotential is not influenced by warping.}
\begin{equation}\label{aisdpot}
  V \,=\, \frac{1}{2\kappa_{10}^2}\int \frac{G^{\rm AISD}\wedge *\overline{G}^{\rm AISD}}{{\rm Im}\tau}\,.
\end{equation}
This means that the potential vanishes identically for ISD flux and the resulting condition $*G_3=iG_3$ fixes almost all moduli, namely complex structure moduli and dilaton.

If the basis of the complex structure moduli space is given by the holomorphic 3-form $\Omega$ (which is AISD) and $h^{2,1}$ primitive ISD (2,1) forms $\chi_i$, the flux $G_3$ is expanded in this basis. Upon this expansion, the scalar potential takes a form that only depends on the coefficients of the expansion of the anti--ISD part
\begin{equation}\label{expandG}
  G_3^{\rm AISD} \,=\, g_1\, \Omega + g_2^i\,\bar \chi_i
\end{equation}
and becomes
\begin{equation}\label{vexpand}
  V \,=\, \frac{i\int G_3\wedge\overline{\Omega}\int \overline{G}_3\wedge\Omega+\int G_3\wedge \chi_i\int\overline{G}_3\wedge \overline{\chi}^i}{2{\rm Im}\tau\,\kappa_{10}^2\int \Omega\wedge\overline{\Omega}}\,.
\end{equation}
This is identical to the standard scalar potential of $\mathcal{N}=1$ 4d supergravity in terms on the superpotential $W$ and the K\"ahler potential $\mathcal{K}$
\begin{equation}\label{sugrapot}
  V \,=\, e^{\mathcal{K}}\left(\sum_\alpha\vert D_\alpha W\vert^2-3\vert W\vert^2\right)\,,
\end{equation}
if the superpotential is the usual Gukov--Vafa--Witten \cite{GVW} potential
\begin{equation}\label{gvwsuppot}
  W \,=\, \int G_3\wedge \Omega~~.
\end{equation}
The K\"ahler potential would be given by 
\bg
\mathcal{K}=-\log(-i\int \Omega\wedge\bar\Omega)-\log[-i(\tau-\bar\tau)]-3\log[-i(\sigma-\bar\sigma)]~~,
\nd
where $\sigma$ is the K\"ahler modulus associated with the overall volume of the Calabi--Yau. 
The (2,1) forms $\chi_i$ enter through the derivative of $\Omega$, because the derivative of $\Omega$ with respect to a complex structure parameter $z_j$ has a (3,0) and a (2,1) part (see {\it e.g.} \cite{ossa})
\begin{equation}\label{derivomega}
  \frac{\partial \Omega}{\partial z_j} \,=\, k_j(z,\bar z)\Omega^{(3,0)}+\chi_j^{(2,1)}\,.
\end{equation}
In \eqref{sugrapot} the index $\alpha$ runs over all K\"ahler moduli $k_a$, complex structure moduli $z_i$ and the dilaton $\Phi$. The K\"ahler covariant derivate is $D_\alpha W=\partial_\alpha W+W\,\partial_\alpha \mathcal{K}$. For no--scale models one finds a cancellation between the covariant derivatives with respect to the K\"ahler moduli against the last term, i.e.
\bg\nonumber
\sum_{k_a}\vert D_{k_a} W\vert^2 = 3 \vert W \vert^2
\nd
so that
\begin{equation}
   V \,=\, e^{\mathcal{K}}\sum_i\vert D_i W\vert^2\,,
\end{equation}
where now $i$ only runs over the complex structure moduli and $\Phi$ only. It is therefore easy to see that even with zero cosmological constant, $V=0$, and satisfing the supergravity equations of motion, we can still have broken supersymmetry, as $D_{k_a} W$ can be nonvanishing. Finally, by solving all $D_i W=0$ equations, we effectively stabilize all these moduli, leaving only the K\"ahler moduli unstabilized.

~\newline\\
\begin{flushleft}
{\it Summary}
\end{flushleft}

Throughout this chapter we developed gravitational constructions which are dual to gauge theories. Each improvement brings us a step closer to accurately describing the observed QCD of the Standard Model. The KW model gaves us the colours, the KT model broke conformality and The KS model realised confinement. The PT and GKP solutions further refined the gravitational details by respectively resolving the singularity of the conifold, and addressing the compactness and supersymmetry realization. We will now turn our attention towards including D7-branes on the the resolved conifold. This will correspond to a non-singular geometry which is dual to a KS-type gauge theory with flavor.

%%%%%%%%%%%%%%%%%%%%%%%%%%%%%%%%%%%%%%%%%%%%%%%%%%%%%%%%%%%%%%%%%%%%%%%%%%%%%%%%%%%%%%%%%%%%%%%%%%%%%%%%%%%%%%%%%%
\chapter{D7-branes on the Resolved Conifold}\label{DtermPaper}
%%%%%%%%%%%%%%%%%%%%%%%%%%%%%%%%%%%%%%%%%%%%%%%%%%%%%%%%%%%%%%%%%%%%%%%%%%%%%%%%%%%%%%%%%%%%%%%%%%%%%%%%%%%%%%%%%%

Our motivation in studying the warped resolved conifold with soft supersymmetry breaking is to come a step closer to a consistent string theory background that can be used to study inflation later, embeddings of the Standard model. 
Current D--brane inflation models ({\it e.g.} \cite{KKLMMT,bdkm, axel, BCDF}) are usually embedded in a particular type IIB string theory setup that has become known as the ``warped throat''. It is a background on which fluxes create a strongly warped Calabi--Yau geometry via their backreaction on the metric. The Calabi--Yau in question is taken to be the conifold or its cousin the deformed conifold, in which the tip of the throat is non--singular. Placing an anti--D-brane at the bottom of the throat and a D-brane at some distance from it, breaks supersymmetry. Consequently, the D-brane is attracted towards the bottom of the throat with the inter--brane distance serving as the inflaton. As has been pointed out in a variety of papers \cite{KKLMMT, bdkm}, it is very hard to achieve slow roll in these models. 

As an alternative one can break supersymmetry spontaneously by turning on appropriate fluxes, {\it e.g.} instead of lifting the potential with an anti--D-brane, one can turn on D--terms. (This idea was put forward in \cite{quevedo}, but needed some corrections \cite{achucarro, fernando}. In short, one can only generate D-terms in a non-susy theory, i.e. if there are also F-terms present \cite{nilles}.)

There has been much interest in D--terms coming from string theory \cite{uranga, lust, hans, haacklust, berglund} both for particle phenomenology and cosmological applications. D--terms can generically be created by non--primitive flux on D--brane worldvolumes. It turns out, however, that in the case of only D3--branes, the D--terms will vanish in the vacuum \cite{uranga}. Even with D7--branes and D3/D7 setups, the cycles wrapped by the branes need to fulfill non--trivial topological conditions to achieve a D-term uplifting \cite{hans}. Although D-brane inflation mostly considers D3--branes, D7--branes have been established as a key ingredient for moduli stabilisation. Non--perturbative effects (gaugino condensation) on their worldvolume allow the stabilization of the overall radial modulus.

In light of this knowledge, we propose a background that breaks supersymmetry, but still solves the supergravity equations of motion. It contains D7--branes, which allow for the creation of D--terms. With cosmological applications in mind, this background is a ``relative'' of the warped throat, i.e. it looks asymptotically like a conifold, but has a different behaviour near the tip. The key ingredient is the blow--up of a 2--cycle (in contrast to the 3--cycle of the deformed conifold), which will introduce non--primitive flux into the theory. This flux still solves the equation of motion as it is imaginary self--dual (ISD). Generically, such a flux cannot exist on a compact Calabi--Yau. We therefore have to generalise our manifold to some non--CY compactification, or keep the whole setup non--compact. For simplicity, we will follow the latter approach, giving some speculations about what a consistent non--CY compactification might induce.

\section{The Type IIB Picture}\label{iib}
In this section, we show how fluxes can be added without violating the supergravity equations of motion and how D7-branes contribute to this flux.

%%%%%%%%%%%%%%%%%%%%%%%%%%%%%%%%%%%%%%%%%%%%%%%%%%%%%%%%%
\subsection{The scalar potential and supersymmetry with (1,2)-flux}\label{cc}
%%%%%%%%%%%%%%%%%%%%%%%%%%%%%%%%%%%%%%%%%%%%%%%%%%%%%%%%%

We have just argued in section \ref{ptbg} that the non-primitive (1,2) flux breaks supersymmetry. One might therefore wonder if it can be used to uplift our potential to a positive vacuum. The answer is no because, as is obvious from \eqref{aisdpot}, the scalar potential always remains zero when the flux is ISD, regardless of whether or not the vacuum breaks supersymmetry. But how can we understand this from the point of view of the SuGra potential as expressed in \eqref{sugrapot}? 
Clearly, there is no F--term associated to derivatives w.r.t. the K\"ahler parameter or the dilaton, as the superpotential \eqref{gvwsuppot} does not depend on them. But what about an F--term $D_{z_j}W$?
Let us for a moment assume we are still talking about a CY, although (1,2) ISD flux cannot exist on a compact CY. So we still assume our moduli space to be parameterised by $\Omega$ and $\chi_i$. Let us furthermore assume the superpotential is still given by \eqref{gvwsuppot}. Then it is easy to see that there could be a non--vanishing derivative of $W$ w.r.t. a complex structure parameter. Using \eqref{derivomega} one finds
\begin{equation}
  \partial_{z_i} W \,=\, k_i(z,\bar z)\,W+\int G_3\wedge \chi_i^{(2,1)}\,,
\end{equation}
which could be nonvanishing for $G_3$ of type (1,2). But (1,2) flux can only be ISD if it is proportional to the K\"ahler form, $G^{(1,2)}=J^{(1,1)}\wedge \bar m^{(0,1)}$, so this becomes
\begin{equation}
  \partial_{z_i} W \,=\,\int J^{(1,1)}\wedge \bar m^{(0,1)}\wedge \chi_i^{(2,1)} \,=\, 0
\end{equation}
when we use the fact that $\chi_i$ is primitive, $J^{(1,1)}\wedge\chi_i^{(2,1)} \,=\, 0$. If there is no (0,3) part present, $W$ vanishes identically and 
\begin{equation}
  D_{z_i} W \,=\,  \partial_{z_i} W+W\,\partial_{z_i}\mathcal{K} \,=\,0\,,
\end{equation}
so all F--terms vanish in our setup. Note that in the non--compact scenario the term $-3\vert W\vert^2$ is absent (we neglected $M_P$ in above formulae). However, our argument does not depend on the no--scale structure of the model. $W$ is identically zero, because we don't have any (0,3) flux turned on, and all F-terms vanish individually.

This discussion has two weak points: First of all, we can no longer assume our moduli space is only parameterised by $\Omega$ and $\chi_i$ if we allow for a (1,2) flux. Once we compactify, there has to be a basis for the one--form $m^{(1,0)}$ as well (for simplicity of the argument let us assume there is only one such 1--form in the following). This would modify the derivative of $\Omega$, the natural guess respecting the  (3,0)+(2,1) structure\footnote{In the case of a complex manifold, the original derivation \cite{ossa} holds and \eqref{newdomega} would not acquire an extra term.} being 
\begin{equation}\label{newdomega}
  \frac{\partial \Omega}{\partial_{z_j}} \,=\, k_j(z,\bar z)\Omega^{(3,0)}+\chi_j^{(2,1)}+\nu_j\, J^{(1,1)}\wedge m^{(1,0)}\,.
\end{equation}
If we keep using the GVW superpotential, we get an additional term
\begin{equation}
  \partial_{z_j} W \,=\,  \int G_3\wedge (\nu_j\, J^{(1,1)}\wedge m^{(1,0)}) \,=\, \int J^{(1,1)}\wedge \bar 
    m^{(0,1)}\wedge \nu_j\, J^{(1,1)}\wedge m^{(1,0)}\,,
\end{equation}
which will in general be non--zero for the type of $G_3$ flux we have turned on. However, the superpotential will also change since we have to expand $G_3$ in this new basis as well. Equation \eqref{expandG} changes to
\begin{equation}
  G_3^{\rm AISD} \,=\, g_1\, \Omega + g_2^i\,\bar \chi_i + g_3\, J\wedge \bar m\,.
\end{equation}
Plugging this into the scalar potential \eqref{aisdpot} does not give \eqref{vexpand}, but additional terms due to $\bar m$. To bring this into the form of the standard SuGra F--term potential we would need to know the metric on the new moduli space, which does not correspond to a CY anymore. Finding the relevant moduli space would allow one to see how $W$ changes. It is likely that it will contain terms with $J$, and thus will introduce a dependence on K\"ahler structure moduli. This breaks the no--scale structure and we have to re--examine the cancellation between $D_{k_a} W$ and $W$. Regardless, we know that the combination $\sum_\alpha |D_\alpha W|^2-3|W|^2$ has to vanish, as \eqref{aisdpot} remains valid. ISD flux cannot give a non--zero potential.

In addition, it is worth noting that we may have to modify the superpotential as to include a term enforcing primitivity. In the compact CY setting this is already taken care of, because an ISD (2,1) form is always primitive. The ISD (1,2) form, on the other hand, is not. If we allow for this type of flux, we should introduce a term that reproduces the primitivity condition as a susy condition $DW=0$. This was already considered in an M/F--theory context \cite{GVW}, where it was conjectured that
\begin{equation}\label{tildew}
  \widetilde{W} \,=\, \int J\wedge J\wedge G_4\,.
\end{equation}
Then $D_J\widetilde{W}=0$ leads to the primitivity condition $J\wedge G_4=0$ for the 4-form flux on the 8--manifold. It is not obvious how this term reduces to type IIB. It will not give rise to a superpotential, but rather to a D--term, as it depends on the K\"ahler moduli and not the complex structure moduli.
For a  $K3\times K3$ orientifold, the dimensional reduction of $\widetilde{W}$ has been carried out \cite{lust} and the result agrees with that obtained in type IIB from a D7--worldvolume analysis \cite{hans}. Also in the F--theory setup, only the non--primitive fluxes on the D7--branes create a D--term in the effective four--dimensional theory. We can therefore safely conclude that the supersymmetry breaking due to the (1,2) flux will not be visible in the scalar potential that appears from the reduction of the IIB bulk action. 

There is also an enlightening discussion in \cite{haack} where it was illustrated that, from an F--theory point of view, a flux of type (0,4), (4,0) or proportional to $J\wedge J$ can break supersymmetry without generating a cosmological constant. It is the latter case that corresponds to non--primitive ISD flux in IIB. We do not have an explicit map between these two types of fluxes, but we give some arguments in section \ref{lift}. It should be clear that ISD flux lifts to self--dual flux in F-theory and that the non-primitivity property is preserved in this lift.

To summarise, the supersymmetry breaking associated to non--primitive (1,2) fluxes will not give rise to an F--term uplift, as the scalar potential generated by the flux in the IIB bulk action remains zero, so does the superpotential if we rely on the CY property of the resolved conifold. We can, however, in the spirit of KKLMMT \cite{KKLMMT} allow a non--vanishing $W_0$ that is created by fluxes in the compact bulk that is glued to the throat. It does not appear in the scalar potential because of the no--scale structure of these models (but it will, once the no--scale structure is broken by non--perturbative effects or because the superpotential is not simply the one from GVW \cite{GVW} anymore). The (1,2) flux gives rise to an ``auxiliary D--term'' \cite{kst}, which is absent in the 4d scalar potential but can be understood as an FI--term from an anomalous $U(1)$ on the D7 worldvolume (the pullback of the B-field on the D7 worldvolume enters into the DBI action). Let us therefore turn to the question how to embed a D7 in the resolved conifold background; we will then turn to the computation of the D--terms in section \ref{dterms}.

%%%%%%%%%%%%%%%%%%%%%%%%%%%%%%%%%%%%%%%%%%%%%%%%%%%%%%%%%%%%%%%%%%%%%%%
\subsection{Ouyang embedding of D7--branes on the resolved conifold}\label{ouyangbed}
%%%%%%%%%%%%%%%%%%%%%%%%%%%%%%%%%%%%%%%%%%%%%%%%%%%%%%%%%%%%%%%%%%%%%%%%

\begin{figure}[h]
\begin{center}
\includegraphics[height=6cm,angle=0]{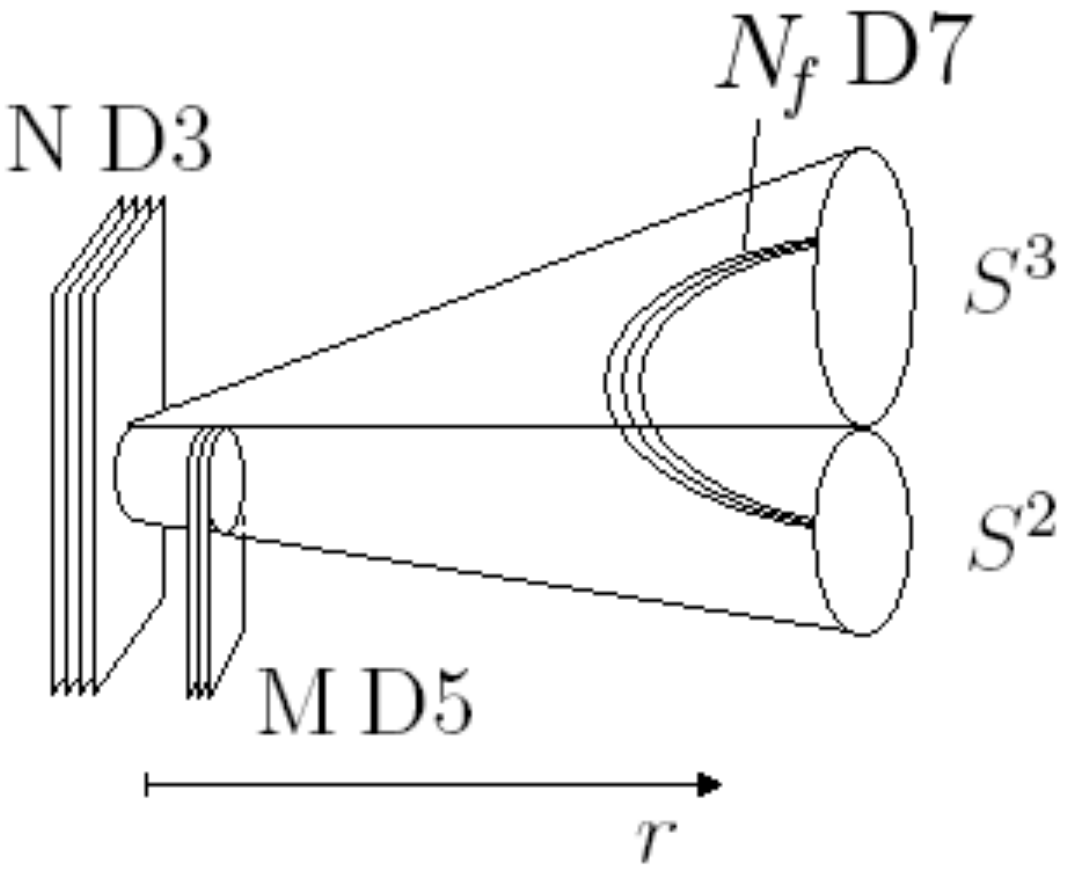}
\caption{Our construction is the extenson of the KT solution in 2 ways: 1) A resolution of the 2-cycle is introduced, making the conifold non-singular in the IR. 2) We add D7-branes in the UV through the Ouyang embedding.}
\label{FigDterm}
\end{center}
\end{figure}

We consider now that addition of D7--branes to the PT background. In \cite{ouyang}, a holomorphic embedding of D7--branes into the singular conifold background was presented. Such an embedding is necessary to preserve supersymmetry on the submanifold, although not alone sufficient (complete BPS conditions are found in \cite{gomis, Marino:1999af}). The particular holomorphic embedding chosen in \cite{ouyang} is described by
\begin{equation}
  z \,=\, \mu^2\,,
\end{equation}
where $z$ is one of the holomorphic coordinates defined in \eqref{holocoord}. Although we already know that the PT background breaks supersymmetry, we will use precisely the same embedding (we consider only $\mu=0$ for simplicity). The configuration and ingredients used here are illustrated in Figure \ref{FigDterm}.

It is worth emphasising that this embedding, first considered on the singular conifold, remains holomorphic on the resolved conifold (details are found in Appendix \ref{embed}). As a consistency check we should always be able to recover the original singular solution in the limit $a\to 0$. This singular solution from \cite{ouyang} is actually not supersymmetric, though one might have expected otherwise. The embedding is holomorphic, but supersymmetry requires in addition that the pullback of the flux is (1,1) and primitive on the cycle wrapped by the D7. The latter condition is not met by the singular Ouyang embedding in \cite{ouyang}. However, as we will demonstrate in section \ref{dterms}, this susy breaking in \cite{ouyang} does not manifest itself in a D--term.

The D7--brane induces a non--trivial axion--dilaton
\begin{equation}\label{dilbehavior}
  \tau \,=\,  \frac{i}{g_s}+\frac{N}{2\pi i} \log z\,,
\end{equation} 
where $N$ is the number of embedded D7-branes. Since $\tau$ is given by \eqref{eqntau}, we can extract the individual running of the axion and dilaton fields associated to the Ouyang embedding:
\bg\label{7brembDterm}
C_0&=&\frac{N}{4\pi}(\psi-\phi_1-\phi_2)\\
e^{-\phi}&=&\frac{1}{g_s}-\frac{N}{8\pi}\log(r^6+9a^2r^4)-\frac{N}{2\pi}\log\Big(\sin\frac{\theta_1}{2}\sin\frac{\theta_2}{2}\Big)
\nd
As pointed out in \cite{BCDF}, this shows that there is an additional running of the dilaton when the two--cycle in the ``resolved warped deformed conifold'' is blown up. However, as we focus on the limit where the geometry looks like the resolved conifold, we recover the PT supergravity solution, which has a constant dilaton. We will therefore concentrate on the running of the dilaton \eqref{dilbehavior} as generated by the D7--brane embedding. This running dilaton was not taken into account by \cite{bdkm}, where the D7 is embedded in the singular conifold and a D3--brane is attracted towards an anti--D3 at the bottom of the throat. The given reasoning is that the dilaton contribution should be exactly cancelled by a change in geometry when approaching the supersymmetric limit (if the D7--brane embedding is supersymmetric and the D3--brane preserves the same supersymmetry, the scenario has to be stable when the susy--breaking anti--D3 is removed). Our setup, on the other hand, is non--supersymmetric from the start and therefore we are not led to conclude that the running of the dilaton should vanish from a similar line of argument. It will, however, be suppressed by the susy breaking scale. For a viable inflationary scenario one should rather use the resolved warped deformed conifold; its running dilaton will be the primary reason for a D3 to move towards the tip\footnote{Such a scenario has been studied in \cite{BCDF}, where the running dilaton due to a blown--up 2--cycle was parameterized by $\delta N(a)\,\log z$, where $a$ is a small resolution. This analysis was based on the original Ouyang embedding \cite{ouyang}, which we will now reconsider for the resolved conifold.}. In this section we simply want to study the backreaction of the dilaton onto the background.

We determine the change the dilaton induces in the other fluxes and the warp factor at linear order $g_sN$, see appendix \ref{embed} for details of the calculation. We neglect any backreaction on the geometry beyond a change in the warp factor, i.e. we will assume the manifold remains a conformal resolved conifold. A distortion of the conifold with Ouyang embedding has been studied in {\it e.g.} \cite{nunez}, where the D7--branes are smeared over the angular directions, such that the dilaton does not exhibit the behaviour \eqref{dilbehavior}, but runs as $\log\rho$ only. Instead of choosing this approximation we will rather attempt to make some statement about the expected manifold from an F--theory perspective. We first embed D7--branes in the non--susy PT setup, neglecting any back--reaction on the internal manifold and then lift the resulting warped resolved conifold with non--trivial axion--dilaton to F--theory. The resulting four--fold is in general not a fibration over a Calabi--Yau three--fold, even in the orientifold limit (see section \ref{ftheory} for this discussion). Solving the full equations of motion would require us to determine the Ricci tensor of the internal manifold from
\begin{equation}\label{zokepe}
  R_{mn} \,=\, \frac{\partial_m\tau\partial_n\bar\tau+\partial_n\tau\partial_m\bar\tau}{4(\rm{Im}\,\tau)^2}
    +\left(T_{mn}^{\rm D7}-\frac{1}{8}\,g_{mn}T^{\rm D7}\right)\,,
\end{equation}
where $T_{mn}^{\rm D7}$ is the energy momentum tensor of the D7 evaluated in our non--trivial background. 
However, we can rely on the fact that in a consistent F-theory compactification this equation is automatically satisfied \cite{GKP} when several stacks of D7-branes and O7-planes are taken into account. An actual computation of the RHS of 
\eqref{zokepe} is generically difficult. This is because to compute $T_{mn}$ of the D7 branes we would first need to
evaluate the non-abelian Born-Infeld action for $N$ D7 branes, and secondly extend the action to curved space because the 
D7 branes wrap non-trivial surfaces in the internal space. We have not been able to perform this direct computation 
(because of the absence of adequate technology), but we give an indirect confirmation of our background from F-theory in
the next section.

Consider first the Bianchi identity, which in leading order becomes ($H_3$ indicates the unmodified NS flux from \eqref{fluxres}, whereas the hat indicates the corrected flux at leading order)
\begin{eqnarray}
  d\hat G_3 & =& d\hat F_3 - d \tau \wedge\hat H_3 - \tau \wedge d\hat H_3 = -d \tau \wedge H_3 +\mathcal{O}((g_s N)^2)  \\ \nonumber
  & = & -\bigg( \frac{N}{2 \pi i } \frac{dz}{z} \bigg) \wedge \big( df_1(\rho) \wedge  d\theta_1 \wedge \sin\theta_1\,d\phi_1 
    + df_2(\rho) \wedge  d\theta_2 \wedge \sin\theta_2\,d\phi_2 \big) +\mathcal{O}((g_s N)^2)\,.
\end{eqnarray}
In order to find a 3--form flux that obeys this Bianchi identity, we make an ansatz
\begin{equation}
  \hat G_3 \,=\, \sum \alpha_i\,\eta_i
\end{equation}
where $\{\eta_i\}$ is a basis of imaginary self--dual (ISD) 3--forms on the resolved conifold. In accordance with the observations about the cohomology of $G_3$, we do not restrict ourselves to (2,1) forms, but allow for $\eta_i$ of (1,2) cohomology as well. With the convention \eqref{cs} we define
\begin{eqnarray}\label{eta}\nonumber
  \eta_1 &=& E_1\wedge E_2\wedge \overline{E}_2 - E_1\wedge E_3\wedge \overline{E}_3\\ \nonumber
  \eta_2 &=& E_1\wedge E_2\wedge \overline{E}_3 - E_1\wedge E_3\wedge \overline{E}_2\\ \nonumber
  \eta_3 &=& E_1\wedge E_2\wedge \overline{E}_1 + E_2\wedge E_3\wedge \overline{E}_3\\ \nonumber
  \eta_4 &=& E_1\wedge E_3\wedge \overline{E}_1 - E_2\wedge E_3\wedge \overline{E}_2\\ 
  \eta_5 &=& E_2\wedge E_3\wedge \overline{E}_1
\end{eqnarray}
\begin{eqnarray} \nonumber
  \eta_6 &=& E_1\wedge \overline{E}_1\wedge \overline{E}_3 + E_2\wedge \overline{E}_2\wedge \overline{E}_3\\ \nonumber
  \eta_7 &=& E_1\wedge \overline{E}_1\wedge \overline{E}_2 - E_3\wedge \overline{E}_2\wedge \overline{E}_3\\ \nonumber
  \eta_8 &=& E_2\wedge \overline{E}_1\wedge \overline{E}_2 + E_3\wedge \overline{E}_1\wedge \overline{E}_3\\ \nonumber
\end{eqnarray}
Note that there are five (2,1) ISD forms, but only three (1,2) ISD forms. This is due to the fact that a form of type (1,2) can only be ISD if it is proportional to $J$.

Not surprisingly, there is no solution to the Bianchi identity involving only the (2,1) forms.
We find a particular solution in terms of only four of above eight 3--forms
\begin{eqnarray}\label{particular}
  P_3 &=& \alpha_1(\rho)\,\eta_1 + e^{-i\psi/2}\alpha_3(\rho,\theta_1)\,\eta_3 + e^{-i\psi/2}\alpha_4(\rho,\theta_2)\,\eta_4 
    +\alpha_8(\rho)\,\eta_8\,,
\end{eqnarray}
with 
\begin{eqnarray}\nonumber\label{alphas}
  \alpha_1 &=& \frac{3g_sNP}{8\pi\rho^3}\frac{ \left[18 a^2 - 36(\rho^2+3a^2)\log\left(\frac{\rho}{a}\right) 
    + (10\rho^2+72a^2)\log\left(\frac{\rho^2}{\rho^2+9a^2}\right)\right]} {\sqrt{\rho^2+6a^2}\sqrt{\rho^2+9a^2}}\\[1ex]
% \end{eqnarray}
% \begin{eqnarray}
  \nonumber
  \alpha_3 &=& -3\sqrt{6} g_sNP\,\frac{72 a^4-3\rho^4+a^2\rho^2(\log(\rho^2+9 a^2)-56\log \rho)}
    {8\pi\rho^3(\rho^2+6 a^2)^2}\,\cot\frac{\theta_1}{2}\\[1ex]
  \alpha_4 &=& -9\sqrt{6} g_sNP\,\frac{\rho^2-9a^2\log(\rho^2+9 a^2)}{8\pi\rho^4\sqrt{\rho^2+6 a^2}}\,
    \cot\frac{\theta_2}{2}\\[1ex] \nonumber
  \alpha_8 &=& \frac{3a^2}{\rho^2+3a^2}\left[3g_sNP\frac{-9(\rho^2+4a^2)+28\rho^2\log\rho+(81a^2+13\rho^2)\log(\rho^2+9a^2)}
    {8\pi\rho^3\sqrt{\rho^2+6a^2}\sqrt{\rho^2+9a^2}}+\alpha_1\right]\,.
\end{eqnarray}
Note that $a_8$ is implicitly given by $\alpha_1$.
Furthermore, we find a homogeneous solution
\begin{eqnarray}
  G_3^{hom} &=& \beta_1(z,\rho)\,\eta_1 + e^{-i\psi/2}\beta_3(\rho,\theta_1)\,\eta_3 
    + e^{-i\psi/2}\beta_4(\rho,\theta_2)\,\eta_4 \\ \nonumber 
  & &  + e^{-i\psi}\beta_5(\rho,\theta_1,\theta_2)\,\eta_5 +\beta_8(z,\rho)\,\eta_8\,,
\end{eqnarray}
with $\beta_i$ given in \eqref{betas}. This solution has the right singularity structure at $z=0$ and $\rho=0$, but it does not transform correctly under $SL(2,\mathbb{Z})$. When $\psi\to\psi+4\pi$, the axion--dilaton transforms as $\tau\to\tau+N$. This would imply that $G_3$ has to be invariant under this shift, which is true for the particular solution, but not the homogeneous one. We therefore conclude that the correction to the 3--form flux, which is in general a linear combination of $P_3$ and $G_3^{hom}$, is given by \eqref{particular} only
\begin{equation}
  \hat{G}_3 \,=\, G_3+P_3\,.
\end{equation}
Note that in terms of $\eta_i$ the original 3--form flux was given by
\begin{equation}
  G_3 \,=\, -9 P\,\frac{(\rho^2+3a^2)\,\eta_1+ 3a^2\,\eta_8}{\rho^3\sqrt{\rho^2+6a^2}\sqrt{\rho^2+9a^2}}\,.
\end{equation}
We can now determine the change in the remaining fluxes and the warp factor, at least to linear order in $(g_sN)$. 
We find the corrected RR and NS flux from the real and imaginary part of $\hat G_3$, respectively
\begin{equation}
  \hat H_3 \,=\, \frac{ \overline{\hat{G}}_3 - \hat{G}_3 }{\tau - \bar{\tau}}\qquad\mbox{and}\qquad 
    \widetilde{F}_3 \,=\, \frac{\hat{G}_3 + \overline{\hat{G}}_3}{2}\,.
\end{equation}
This results in the closed NS-NS 3--form
\begin{eqnarray}\label{H3formDterm}\nonumber
  \hat H_3 &=& d\rho\wedge e_\psi\wedge(c_1\,d\theta_1+c_2\,d\theta_2) + d\rho\wedge(c_3\sin\theta_1\,d\theta_1\wedge d\phi_1-c_4\sin\theta_2\,d\theta_2\wedge d\phi_2)\\
  & & +\left(\frac{\rho^2+6a^2}{2\rho}\,c_1\sin\theta_1\,d\phi_1 
    -\frac{\rho}{2}\,c_2\sin\theta_2\,d\phi_2\right)\wedge d\theta_1\wedge d\theta_2
\end{eqnarray}
and the non--closed RR 3--form (note that $\widetilde{F}_3=\hat F_3-C_0 \hat H_3$, where $\hat F_3$ is closed)
\begin{eqnarray}\label{F3formDterm}\nonumber
  \widetilde{F}_3 &=& -\frac{1}{g_s}\,d\rho\wedge e_\psi\wedge(c_1\sin\theta_1\,d\phi_1+c_2\sin\theta_2\,d\phi_2)\\ \nonumber
  & &  +\frac{1}{g_s}\,e_\psi\wedge(c_5\sin\theta_1\,d\theta_1\wedge d\phi_1-c_6\sin\theta_2\,d\theta_2\wedge d\phi_2)\\ 
  & & -\frac{1}{g_s}\,\sin\theta_1\sin\theta_2\left(\frac{\rho}{2}\,c_2 \,d\theta_1-\frac{\rho^2+6a^2}{2\rho}\,c_1\,d\theta_2 \right)
    \wedge d\phi_1\wedge d\phi_2\,,
\end{eqnarray}
see \eqref{defc} for the coefficients $c_i$.
This allows us to write the NS 2--form potential ($dB_2=\hat H_3$)
\begin{eqnarray}\label{bfield}
  B_2 &=& \left(b_1(\rho)\cot\frac{\theta_1}{2}\,d\theta_1+b_2(r)\cot\frac{\theta_2}{2}\,d\theta_2\right)\wedge e_\psi\\ \nonumber
  & + &\left[\frac{3g_s^2NP}{4\pi}\,\left(1+\log(\rho^2+9a^2)\right)\log\left(\sin\frac{\theta_1}{2}\sin\frac{\theta_2}{2}\right)
    +b_3(\rho)\right]\sin\theta_1\,d\theta_1\wedge d\phi_1\\ \nonumber 
  & - & \left[\frac{g_s^2NP}{12\pi\rho^2}\left(-36a^2+9\rho^2+16\rho^2\log\rho+\rho^2\log(\rho^2+9a^2)\right)
    \log\left(\sin\frac{\theta_1}{2}\sin\frac{\theta_2}{2}\right)+b_4(\rho)\right]\\ \nonumber
  & & \qquad\qquad \times \sin\theta_2\,d\theta_2\wedge d\phi_2\,,
\end{eqnarray}
where the coefficients are given in \eqref{defb}. This mirrors closely the result for the singular conifold \cite{ouyang} and we can indeed show that we produce this result in the $a\to 0$ limit. Away from the singular limit, we find an asymmetry between the $(\theta_1,\phi_1)$ and $(\theta_2,\phi_2)$ spheres, which was to be expected since our manifold (the resolved conifold or its more complicated cousin, the resolved warped deformed conifold) does not have the $\mathbb{Z}_2$ symmetry that exchanges the two 2--spheres in the singular conifold geometry. The lesser degree of symmetry is naturally also expressed in the fluxes.

The five--form flux is as usual given by ($\tilde{*}_{10}$ indicates the Hodge star on the full 10--dimensional {\em warped} space)
\begin{equation}
  \hat{F}_5 \,=\, (1+\tilde{*}_{10})(d\hat h^{-1}\wedge d^4x)\,,
\end{equation}
which requires knowledge of the warp factor $\hat h(\rho)$ that is consistent with these new fluxes. In order to solve the supergravity equations of motion one requires
\begin{equation}\label{wfactor}
  \hat h^2\,\Delta \hat h^{-1}-2 \hat h^3\,\partial_m \hat h^{-1}\,\partial_n \hat h^{-1} g^{mn}\,=\,-\Delta\hat{h}
    \,=\,*_6 \left(\frac{\hat G_3\wedge \overline{\hat G}_3}{6(\overline{\tau}-\tau)}\right)\,=\,\frac{1}{6}\,*_6 d\hat{F}_5\,,
\end{equation}
where $\Delta$ is the Laplacian on the unwarped resolved conifold and all indices are raised and lowered with the unwarped metric.
After some simplifications the Laplacian on the resolved conifold takes the form
\begin{equation*}
  \Delta\hat{h} \,=\,\kappa\,\partial_\rho^2 \hat{h} + \frac{5\rho^2+27a^2}{\rho(\rho^2+6a^2)}\,\partial_\rho\hat{h} 
    + \frac{6}{\rho^2}\,\left(\partial_{\theta_1}^2 \hat{h}+\cot\theta_1\,\partial_{\theta_1}\hat{h}\right) 
    + \frac{6}{\rho^2+6a^2}\,\left(\partial_{\theta_2}^2 \hat{h}+\cot\theta_2\,\partial_{\theta_2} \hat{h}
    \right)\,.
\end{equation*}
This should be evaluated in linear order in N, since we solved the SuGra eom for the fluxes also in linear order. As the the right hand side of
\begin{eqnarray}\nonumber
  \frac{1}{6}\,*_6 d\hat{F}_5 &=& 
    \frac{54g_sP}{\pi\rho^6(\rho^2+6a^2)(\rho^2+9a^2)}\Bigg\{12\pi\rho^4+9a^2\rho^2(8\pi-g_sN)+54a^4(4\pi+g_sN)\\ \nonumber
  & & +g_sN\bigg[(25\rho^4+66a^2\rho^2-54a^2)\log\rho+(10\rho^4+102a^2\rho^2+189a^4)\log(\rho^2+9a^2)\\ 
  & & \qquad\quad +6(\rho^4+6a^2\rho^2+18a^4)\log\left(\sin\frac{\theta_1}{2}\sin\frac{\theta_2}{2}\right)\bigg]\Bigg\} 
\end{eqnarray}
appears sufficiently complicated, we need to employ some simplification. The obvious choice is to consider $\rho\gg a$, i.e. we only trust our solution sufficiently far from the tip. As in the limit $a\to 0$ we recover the singular conifold setup, we know our solution takes the form \cite{ouyang}
\begin{eqnarray}\label{warpfacteurOuyang}
  \hat h(\rho,\theta_1,\theta_2) &=& 1+\frac{L^4}{r^4}\,\left\{1+\frac{24g_sP^2}{\pi\alpha'Q}\,\log\rho\left[1+\frac{3g_sN}{2\pi\alpha'}
    \left(\log\rho+\frac{1}{2}\right)\right.\right.\\ \nonumber
  && \left.\left. \qquad\qquad\qquad\qquad +\frac{g_sN}{2\pi\alpha'}\,\log\left(\sin\frac{\theta_1}{2}\sin\frac{\theta_2}{2}\right)\right]\right\} 
    +\mathcal{O}\left(\frac{a^2}{\rho^2}\right)
\end{eqnarray}
with $L^4=27\pi g_s\alpha'Q/4$. Apart from the $a^2/\rho^2$--correction, this is the same result as for the singular conifold \cite{ouyang}. We have not been able to find an analytic solution at higher order, but considering that most models work with even cruder approximations of the warp factor (i.e. $h(r)\sim\log r/r^4$), we believe this should suffice.

%%%%%%%%%%%%%%%%%%%%%%%%%%%%%%%%%%%%%%%%%%%%%%%%%%%%%%%%%%%%%%%%%%%%%%%
\subsection{D-terms from non--primitive background flux on D7--branes}\label{dterms}
%%%%%%%%%%%%%%%%%%%%%%%%%%%%%%%%%%%%%%%%%%%%%%%%%%%%%%%%%%%%%%%%%%%%%%%

Soft supersymmetry breaking via D--terms on D7--branes has been considered in \cite{uranga}, and was later applied to more realistic type IIB orientifolds \cite{hans, haacklust} or their F--theory lift \cite{lust, berglund} (see also \cite{zwirner} for a IIA scenario); the most general study for generalised CYs has appeared in \cite{luca1}. The established consensus is that non--primitive flux on the D7--worldvolume gives rise to D-terms in the effective 4--dimensional theory, which can only under certain conditions remain non--zero in the vacuum. One way to phrase the necessary condition is to require that the 4--cycle wrapped by the D7--branes admits non--trivial 2--forms that become trivial in the ambient Calabi--Yau, i.e. the $H^2$--cohomology on the four--cycle is bigger than just the pullback of $H^2(CY)$. (Equivalently \cite{hans} states that the 4--cycle needs to intersect its orientifold image over a 2--cycle that supports non--trivial flux. The same is true in the case of two stacks \cite{haacklust} intersecting over a 2--cycle.) This condition can be satisfied for the Ouyang embedding in the $\mu\ne 0$ case: The resolved conifold admits only one non--trivial 2--cycle, the sphere that remains finite at the tip. The 4--cycle that the D7  wraps, on the other hand, can also have a non-trivial cycle spanned by $(\theta_1,\phi_1)$, if the D7 in the Ouyang embedding do not reach all the way to the bottom of the throat. On the D7, this cycle will never shrink completely. Nevertheless, we are mostly concerned with the case $\mu=0$ here. In contrast to \cite{hans, haacklust}
we consider the pullback of a background field with non--vanishing fieldstrength, not the zero mode fluctuations, i.e. we do not expand the worldvolume flux in a basis of $H^2$. This gives rise to a D-term that depends on the overall volume of the manifold and the resolution parameter $a$.  
Though an orientifold will be necessary to consistently compactify our background, we will not specify any orientifold action here, as we do not know a specific compactification our background.

Following the derivation in \cite{haacklust, luca1}, we extract the D-terms from the DBI action. Suppose our stack of $N$ D7--branes wraps a 4-cycle $\Sigma$ as specified by the Ouyang embedding in section \ref{ouyangbed}. The full DBI action for the 8--dimensional worldvolume (in string frame) reads
\begin{equation}\label{d7dbi}
  S_{D7} \,=\, -\mu_7\int_{\Sigma\times\mathcal{M}_4} d^8\xi\,e^{-\Phi}\sqrt{|\check g+\check B-2\pi\alpha'F|}
\end{equation}
where the symbol $\,\check{\phantom{}}\,$ indicates the pullback of the metric and the NS field onto the D7, $F$ is the worldvolume gauge flux. With this product ansatz for the spacetime this expression becomes
\begin{equation}
  S_{D7} \,=\, -\mu_7\int d^4x\,e^{-\Phi}\sqrt{|\check g_4|}\,\sqrt{\left|1+2\pi\alpha'\check g_4^{-1}F_4\right|}\,\Gamma\,,
\end{equation}
where $g_4$ and $F_4$ indicate the 4--dimensional part of the metric and gauge flux and one defines
\begin{equation}\label{defGamma}
  \Gamma \,=\, \int_\Sigma d^4\xi\,\sqrt{|\check g_\Sigma+\mathcal{F}|}\,,
\end{equation}
where we have introduced $\mathcal{F}=\check B-2\pi\alpha'F$. In the following, the pullback is always understood as onto the 4--cycle $\Sigma$. We do not consider any gauge fields along the external space $\mathcal{M}_4$.
The quantity \eqref{defGamma} is the main parameter for the D--terms. Expanding the full action \eqref{d7dbi} at low energies yields the potential contribution
\begin{equation}\label{d7pot}
  V_{D7} \,=\, \mu_7e^{3\Phi}\mathcal{V}^{-2}\Gamma\,,
\end{equation}
where the volume $\mathcal{V}$ of the resolved conifold is defined as
\begin{equation}
  \mathcal{V} \,=\, \frac{1}{6}\int_Y J\wedge J\wedge J\,=\,\frac{(4\pi)^2}{108}\,\int_0^R \rho^3(\rho^2+6a^2)\,d\rho
    \,=\, \frac{8\pi^3}{81}\,R^4(R^2+9 a^2)\,.
\end{equation}
This integral has to be regularised by an explicit cut--off, as we study the non--compact case. Simply cutting off the radial direction does probably destroy the holomorphicity condition, but we will ignore this subtlety here.

One can write \cite{haacklust} $\Gamma=\tilde\Gamma e^{-i\zeta}=|\tilde\Gamma|e^{i(\tilde\zeta-\zeta)}$, where $\zeta$ is determined from the BPS calibration condition and
\begin{equation}\label{tildegamma}
  \tilde\Gamma \,=\, \frac{1}{2}\int_\Sigma \left(\check J\wedge \check J-\mathcal{F}\wedge \mathcal{F}\right)
     +i \int_\Sigma \check J\wedge \mathcal{F}\,.
\end{equation}
Then the condition for the D7 to preserve the same supersymmetry as the O7 corresponds to $\zeta=\tilde\zeta=0$, or equivalently ${\rm Im} \tilde\Gamma=0$. Allowing for a small supersymmetry breaking one expands the D7--potential \eqref{d7pot} in ${\rm Im}\tilde\Gamma\ll{\rm Re}\tilde\Gamma$ and finds
\begin{eqnarray}\nonumber
  V_{D7} &=& \mu_7e^{3\Phi}\mathcal{V}^{-2}\Gamma \,=\, \mu_7e^{3\Phi}\mathcal{V}^{-2}\,
    \sqrt{({\rm Re}\tilde\Gamma)^2+({\rm Im}\tilde\Gamma)^2}\\
  &=& \mu_7e^{3\Phi}\mathcal{V}^{-2}\,{\rm Re}\tilde\Gamma + \frac{1}{2}\,\mu_7e^{3\Phi}\mathcal{V}^{-2}\,
    \frac{({\rm Im}\tilde\Gamma)^2}{{\rm Re}\tilde\Gamma}\,.
\end{eqnarray}
The first term in this expansion will be cancelled by the tadpole cancellation condition in a consistent compactification. The second term is interperted as the susy--breaking D--term.
The real and imaginary part of $\tilde\Gamma$ are easily read off from \eqref{defGamma} (the integrals are real) and can be calculated for our explicit case at hand. All we need to know is the pullback of the K\"ahler form onto the 4--cycle and the worldvolume flux $\mathcal{F}$.

We would like to consider the simple case such that
\begin{equation}
  \check B \,\ne\,0\,,\qquad\qquad F\,=\,0\,,
\end{equation}
as we have an explicit solution of this form. There could be gauge flux on the D7--brane to could restore supersymmetry in the $a\to 0$ limit. It is noted again that to preserve supersymmetry, holomorphicity is not enough. One also needs the worldvolume flux to be of pure (1,1) type and primitive \cite{gomis}. The reason that it is so difficult to achieve non--trivial D--terms with closed $\check B$ is that $F$ could always cancel the non--primitive part of $\check B$ \cite{hans}, unless some non--trivial topological conditions are met.

In calculating the D--terms, we must treat the D7 as a probe. Thus the B--field that is pulled back is not the one we calculated in \eqref{bfield}, but the original PT solution
\begin{equation}
  B \,=\, f_1(\rho) \sin\theta_1\,d\theta_1\wedge d\phi_1+ f_2(\rho) \sin\theta_2\,d\theta_2\wedge d\phi_2\,,
\end{equation}
where $f_1$ and $f_2$ were defined in \eqref{deffs}. The embedding $z=0$ we use has actually 2 branches, since  
\begin{equation}
   z \,=\, 0 \, =\,  \left ( 9 a^2 \rho^4 + \rho ^6 \right ) ^{1/4} \,\sin\frac{\theta_1}{2}\,\sin\frac{\theta_2}{2}
\end{equation}
can be satisfied by either $\theta_1=0$ or $\theta_2=0$. This implies that also $\phi_1=$fixed or $\phi_2=$fixed, as $\theta_i$ being zero refers to the pole of one of the 2--spheres where the circle described by $\phi_i$ collapses. The full holomorphic cycle is then a sum over these 2 branches.

Consider the 2 four--cycles $\Sigma_1=(\rho,\,\psi,\,\phi_1,\,\theta_1)$ and $\Sigma_2=(\rho,\,\psi,\,\phi_2,\,\theta_2)$ that correspond to the branches $\theta_2=0$ and $\theta_1=0$, respectively. The complex structure induced on them is actually a trivial pullback of the complex structure on the resolved conifold. Using the complex vielbeins \eqref{cs}, we see that 
\begin{equation}
  \Sigma_1 \,=\, (E_1\vert_{\theta_2=0}, E_2)\,,\qquad\qquad \Sigma_2 \,=\, (E_1\vert_{\theta_1=0}, E_3)\,,
\end{equation}
where in $E_1\vert_{\theta_2=0}$ and $E_1\vert_{\theta_1=0}$ the imaginary part is truncated to
\begin{equation}\nonumber
  {\rm Im} E_1\vert_{\theta_2=0} \,=\, \frac{\rho\sqrt{\kappa}}{3}\,(d\psi+\cos\theta_1\,d\phi_1)\qquad\mbox{and}\qquad
  {\rm Im} E_1\vert_{\theta_1=0} \,=\, \frac{\rho\sqrt{\kappa}}{3}\,(d\psi+\cos\theta_2\,d\phi_2)\,,
\end{equation}
respectively.
It is easy to show that the induced complex structure on the four--cycle still allows for a closed K\"ahler form.
With this observation we find the pullback of $B$ onto both branches
\begin{equation}\label{Bpullback}
  \check B\vert_{\Sigma_1} \,=\, \frac{-3i}{\rho^2}\,f_1\,E_2\wedge \bar E_2\,,\quad\qquad
  \check B\vert_{\Sigma_2} \,=\, \frac{-3i}{\rho^2+6 a^2}\,f_2\,E_3\wedge \bar E_3\,,
\end{equation}
which turn out to be of type (1,1). But that does not mean they are primitive. In fact, as we will see shortly, the pullback of B is not primitive on each individual branch, but in the limit $a\to 0$ the D-term generated by them vanishes when summing over both branches. So it appears that the Ouyang embedding in the singular conifold \cite{ouyang} breaks supersymmetry due to this non--primitivity, but generates neither an F-term nor a D-term. 
Supersymmetry could possibly be restored by choosing appropriate gauge flux,
but we solved the equations of motion only for the case $F=0$, so we will keep working with this assumption. In general, $F$ would mix with the metric in the e.o.m., changing our original setup.

If we consider the B--field \eqref{bfield} that reflects the D7--backreaction, we find its pullback onto $\Sigma_1$ (the case of $\Sigma_2$ is completely analogous)
\begin{eqnarray}\label{fullBpull}
   \check B_2\vert_{\Sigma_1} &=& b_1(\rho)\cot\frac{\theta_1}{2}\,d\theta_1\wedge (d\psi+\cos\theta_1\,d\phi_1)\\ \nonumber
  & + &\left[\frac{3g_s^2NP}{4\pi}\,\left(1+\log(\rho^2+9a^2)\right)\log\left(\sin\frac{\theta_1}{2}\cdot 0\right)
    +b_3(\rho)\right]\sin\theta_1\,d\theta_1\wedge d\phi_1\,.
\end{eqnarray}
We encounter the usual problem that B contains terms with $\log z$, so naturally we find a log--divergent term if we pull back onto a cycle that is described by $z=0$. However, this is not our concern here. We just want to point out, that this B-field is not of pure (1,1) type anymore, but rather contains (2,0) and (0,2) terms as well:
\begin{eqnarray}\nonumber
  \check B_2\vert_{\Sigma_1} &=& \frac{3\sqrt{3}i\,b_1(\rho)}{2\rho^2\sqrt{2\kappa(\rho)}}\,\cot\frac{\theta_1}{2}\left[e^{i\psi/2}
    (E_1\wedge\bar E_2-\bar E_1\wedge \bar E_2)+e^{-i\psi/2}(E_1\wedge E_2+E_2\wedge \bar E_1)\right]\\ 
  &-&\frac{3i}{\rho^2}\left[\frac{3g_s^2NP}{4\pi}\,\left(1+\log(\rho^2+9a^2)\right)\log\left(\sin\frac{\theta_1}{2}\cdot 0\right)
    +b_3(\rho)\right]\,E_2\wedge \bar E_2\,.
\end{eqnarray}
For our considerations the probe approximation shall suffice. We could not obtain any sensible result with the B--field \eqref{fullBpull} anyway, as we would have to integrate over the divergent points $\theta_i=0$. Naturally, this is some kind of self--interaction and divergent. 

Let us now turn to the calculation of the D-terms for the embedding $\mu=0$. The crucial integral for the D-term coming from \eqref{tildegamma} is given by the pullbacks of $J$ and $B$. We still need to give the pullback of $J$ onto both branches:
\begin{eqnarray}\nonumber
  \check J\vert_{\Sigma_1} &=& \frac{\rho}{3}\,d\rho\wedge(d\psi+\cos\theta_1\,d\phi_1)+\frac{\rho^2}{6}\,\sin\theta_1\,d\phi_1\wedge 
    d\theta_1\\
  \check J\vert_{\Sigma_2} &=& \frac{\rho}{3}\,d\rho\wedge(d\psi+\cos\theta_2\,d\phi_2)+\frac{\rho^2+6a^2}{6}\,\sin\theta_2\,d\phi_2\wedge 
    d\theta_2\,.
\end{eqnarray}
And we repeat the pull--back of $B$ in terms of real coordinates:
\begin{equation}\label{Bpullreal}
  \check B\vert_{\Sigma_1} \,=\, f_1(\rho)\,\sin\theta_1\,d\theta_1\wedge d\phi_1\,,\quad\qquad
  \check B\vert_{\Sigma_2} \,=\, f_2(\rho)\,\sin\theta_2\,d\theta_2\wedge d\phi_2\,.
\end{equation}
The D-term is now obtained from ${\rm Im}\tilde\Gamma$ in \eqref{tildegamma}
\begin{eqnarray}\nonumber\label{dterm}
  D &=& \int_{\Sigma_1} \check J\vert_{\Sigma_1}\wedge \check B\vert_{\Sigma_1}+ \int_{\Sigma_2} \check J\vert_{\Sigma_2}\wedge \check 
    B\vert_{\Sigma_2}\\
  &=& \int_{\Sigma_1} \frac{\rho}{3}\,f_1\sin\theta_1\,d\rho\wedge d\psi\wedge d\theta_1\wedge \phi_1
    + \int_{\Sigma_2} \frac{\rho}{3}\,f_2\sin\theta_2\,d\rho\wedge d\psi\wedge d\theta_2\wedge \phi_2\,.
\end{eqnarray}
We see immediately that for the case $f_1=-f_2$, i.e. the singular $a\to 0$ limit of the KT solution, the D-term vanishes after summing both cycles, even though the pullback of $B$ is non-primitive in this case.
% A gauge flux that would make $\check B\vert_{\Sigma_1}$ primitive would be
% \begin{equation}
%   F \,=\, -\frac{3g_sP}{\rho}\,\log(\rho^2+9a^2)\,d\rho\wedge d\psi\,.
% \end{equation}
For the case $a\ne 0$ we can perform the integrals, again introducing a cut--off $R$ for the radial direction. We find
\begin{equation}\label{dresult}
  D \,=\, \frac{32\pi^2g_sP}{9}\,\big[9a^2\log (9+a^2)-(9a^2-2R^2)\log R-(9a^2+R^2)\log(9a^2+R^2)\big]\,.
\end{equation}

To obtain the full D-term potential, we also need ${\rm Re}\tilde\Gamma$ from \eqref{tildegamma}. Looking at the pullbacks of the B--fields \eqref{Bpullback} we see that $\check B\wedge \check B$ vanishes for both branches, so
\begin{eqnarray}\nonumber
  {\rm Re}\tilde\Gamma &=& \frac{1}{2}\int_{\Sigma_1} \check J\vert_{\Sigma_1}\wedge  \check J\vert_{\Sigma_1} + \frac{1}{2}\int_{\Sigma_2} 
    \check J\vert_{\Sigma_2} \wedge  \check J\vert_{\Sigma_2}\\ 
  &=& \frac{4\pi^2}{9}\,R^2(R^2+6a^2)\,.
\end{eqnarray}
The total D-term potential then reads
\begin{eqnarray}\nonumber
  V_{D7} &=& \frac{1}{2}\,\mu_7e^{3\Phi}\mathcal{V}^{-2}\,\frac{({\rm Im}\tilde\Gamma)^2}{{\rm Re}\tilde\Gamma}\\
  &=& \frac{59049\,\mu_7\,e^{3\Phi}}{512 \pi^8}\,\frac{D^2}{R^{10}(R^2+6a^2)(R^2+9a^2)^2}
\end{eqnarray}
with the D-term $D$ from \eqref{dresult}. In the probe approximation, $\Phi$ is just the constant background dilaton and can be set to zero. This is one of the main results of our paper. We find a non--zero D--term created by non-primitive (1,2) flux when pulled back to non-primitive flux on D7--branes. Their magnitude is highly suppressed in a large volume compactification. It would be most desireable to find a consistent compactification for our setup, in which we do not have to introduce a cut--off by hand that spoils holomorphicity. Let us stress again that these (1,2) fluxes did not lead to the creation of a bulk cosmological constant, because they are ISD. We would expect, however, a modification of the superpotential, i.e. in general D-terms on D7--branes also create F-terms \cite{lust, hans, haacklust}. 

We have so far neglected  any zero modes. Once we study D3/D7 inflation, there will also be degrees of freedom that become light when the two branes approach each other. The D-- and F--terms in this case have to be re-evaluated. As already outlined in the beginning of this section, we believe that the conditions to have non--zero D--terms in the vacuum (i.e. intersection over a two--cycle with non--trivial flux or a cohomology $H^2(\Sigma)$ of the 4--cycle that is greater than the pullback of the CY cohomology $H^2(CY)$) can be met when $\mu\ne 0$. For $\mu=0$ it appears rather the opposite:
There is only one non--trivial 2--cycle in the resolved conifold, the blown--up $(\phi_2,\,\theta_2)$--sphere. With $\mu=0$, the cycle $\Sigma_1$ is topologically trivial (it contains the shrinking 2--sphere), the cycle $\Sigma_2$ is not. However, once we compactify we will introduce another cycle on which the (0,1) form is supported. This should be in $(\rho,\,\psi)$ direction, as $G^{(1,2)}\sim J\wedge \bar E_1$, and $E_1$ extends along $\rho$ and $\psi$. However, from \eqref{Bpullreal} we see that this 2--cycle does not support any flux.

We believe this puzzle might be clarified once the original Ouyang embedding in the singular conifold background is made supersymmetric with appropriate gauge fluxes. Note however, that there is an essential difference between the singular KT and the resolved PT backgrounds: the B--field in the bulk is primitive, i.e. $J\wedge J\wedge B=0$, for the first case but not for the latter.

The next step would be to consider the embedding $\mu\ne 0$. The integrals becomes much more complicated and cannot be solved analytically. Only for the case $a=0$ have we been able to show by numerical integration that $D=0$. For $a\ne 0$ the integrand's strong oscillatory behaviour has prevented us from finding a solution so far. Note that the pullback of $J$ and $B$ is much more involved. We have to use the embedding equations
\begin{equation}\label{rhopsi}
  (\rho^6+9a^2\rho^4) \,=\, \left(\frac{|\mu^2|}{\sin\frac{\theta_1}{2}\,\sin\frac{\theta_2}{2}}\right)^4\,,\quad
  \psi \,=\, \phi_1+\phi_2+\,\mbox{const}\,.
\end{equation}
It is then tedious but straightforward to calculate the pullback
\begin{equation}
  \check J_{\alpha\beta} \,=\, \partial_\alpha y^m \partial_\beta y^n\,J_{mn}\,, 
\end{equation}
where $m,n=\rho,\psi,\theta_1,\theta_2,\phi_1,\phi_2$ run over the whole 3--fold, whereas $\alpha,\beta=\theta_1,\theta_2,\phi_1,\phi_2$ parameterise the 4--cycle. A similar formula gives the pullback of the NS field $\check B$. Note, however, that the pullback will contain terms with $(\sin\theta_i)^{-1}$, which diverge at the integration boundaries $\theta_i=0$. For the case $a=0$ this seems to be under control, for the resolved case we cannot make any definite statement.

%%%%%%%%%%%%%%%%%%%%%%%%%%%%%%%%%%%%%%%%%%%%%%%%%%%%%%%
\section{A view from F--theory}\label{ftheory}
%%%%%%%%%%%%%%%%%%%%%%%%%%%%%%%%%%%%%%%%%%%%%%%%%%%%%%%

Now that we have more or less the complete type IIB picture, 
we should deviate to address the F-theory \cite{Vafa}
lift of our background. Studying F-theory lift has
many advantages:

\vskip.1in

\noindent $\bullet$ It can give us a precise way to study the compact version of our background. 
Recall that the background 
that we constructed is non-compact. The compact form of our background can be formulated if we can find a compact 
four-fold associated with the resolved conifold background.

\noindent $\bullet$  It is directly related to M-theory by a $S^1$ reduction \cite{Vafa}. 
In M-theory the structure of the four-fold
remains the same, but there are a few advantages. 
We can determine the precise warped form of the metric \cite{BeckerM, DRS}, the precise
superpotential \cite{GVW} and the complete perturbative \cite{DRRS}
and non-perturbative terms on the IIB seven branes. 

\vskip.1in

\subsection{Construction of the fourfold}

\vskip.1in

With the above advantages in mind, we aim to determine the fourfold in F-theory and study the subsequent properties associated with the 
fourfold in M-theory. The generic structure of the fourfold can be of the following form:
\begin{equation}
\label{mfourfold}
ds^2 = e^{2A} \eta_{\mu\nu} dx^\mu dx^\nu + e^{2B} g_{mn} dy^m dy^n + e^{2C} \vert dz\vert^2
\end{equation} 
where $A, B, C$ are the warp factors that could be in general functions of time as well as the internal 
coordinates ($y^m, z$) and ($\mu, \nu$) = (0, 1, 2). 
The fourfold is a $T^2$ fibration over a base. We denote the complex coordinate of the 
$T^2$ by $z$ and the base has a metric $g_{mn}$. The corresponding type IIB metric is expected to be of the 
form (see also \cite{chen}):
\begin{equation}
ds^2 ~= ~e^{2A+C} ~(-dx_0^2 + dx_1^2 + dx_2^2)~ +~ 
\frac{e^{-3C}}{\vert\tau\vert^2}~dx_3^2 ~+~
e^{2B+C} ~g_{mn} dy^m dy^n
\end{equation}
which tells us that in principle the $3+1$ dimensional Lorentz could be broken by 
choosing a generic warp factor of the fibre torus in M-theory. The fibre torus, in M-theory, is parametrised by 
a complex structure $\tau$ which is proportional to the axio-dilaton in type IIB:
\begin{equation}
dz ~=~ dx^{11} + \tau dx^{3}, ~~~~~~~ {\tau} ~ = ~ C_0 ~ + ~ \frac{i}{g_s}
\end{equation}
Clearly if the torus was non-trivially fibred over the threefold base (with metric $g_{mn}$) we would expect non-zero 
cross terms in the type IIB metric. For our case we simply choose a trivial $T^2$ fibration of the 
fourfold, so the cross-terms are absent. For a compact manifold we would require the axion charge to vanish. This 
would mean that the contribution to $C_0$ from a single D7 brane is very small. This would change our metric 
to 
\begin{equation}
ds^2 ~= ~e^{2A+C} ~(-dx_0^2 + dx_1^2 + dx_2^2)~ +~ 
\frac{e^{-3C}}{({\rm Im}~\tau)^2}~dx_3^2 ~+~
e^{2B+C} ~g_{mn} dy^m dy^n
\end{equation} 
Furthermore, restoring full $3+1$ dimensional Lorentz 
invariance will tell us that the type IIB 
metric has the following form:
\begin{equation} 
ds^2 ~ = ~ \frac{e^{3A/2}}{\sqrt{{\rm Im}~\tau}} ~ \eta_{\mu\nu}dx^\mu dx^\mu ~ + ~ 
\frac{e^{2B - A/2}}{\sqrt{{\rm Im}~\tau}} ~g_{mn} dy^m dy^n
\end{equation}
Comparing the above form of the metric with the metric that we have \eqref{pzmet}, it is easy to work out the 
corresponding M-theory warp factors in terms of $h$ and the axio-dilaton $\tau$ as:
\begin{equation}
\label{warpy}
e^A ~ = ~ \Bigg[\frac{{\rm Im}~\tau}{h}\Bigg]^{\frac{1}{3}}, ~~~~~ e^B ~ = ~ 
\Big[h ({\rm Im}~\tau)^2\Big]^{\frac{1}{6}}, ~~~~~ e^C ~ = ~  
\Bigg[\frac{\sqrt{h}}{({\rm Im}~\tau)^2}\Bigg]^{\frac{1}{3}}
\end{equation}
Now combining \eqref{warpy} with \eqref{mfourfold} we can easily see that the fourfold is a given by the following 
metric:
\begin{eqnarray}\label{fourfold}\nonumber
  ds^2_{\rm 4-fold} & = & \frac{h^{1/3}}{({\rm Im}~\tau)^{4/3}}~ 
\big\vert dx^{11} + \tau~dx^{3}\big\vert^2 + 
{h^{1/3}({\rm Im}~\tau)^{2/3}}\Big[ 
\frac{d\rho^2}{\kappa} ~ + \frac{\rho^2}{6}\,\big(d\theta_1^2+\sin^2\theta_1\,d\phi_1^2\big) ~ + \\
& & + ~ \frac{\kappa}{9}\,\rho^2\big(d\psi+\cos\theta_1\,d\phi_1
 +\cos\theta_2\,d\phi_2\big)^2 
    +\frac{\rho^2+6a^2}{6}\,\big(d\theta_2^2+\sin^2\theta_2\,d\phi_2^2\big)\Big]\,,
\end{eqnarray}
where the other variables have already been defined above. The type IIB NS and RR three-form fluxes would converge 
to give us $G$-fluxes $G_{mnpq}$ on the fourfold. The equations of motion of $G$-fluxes are determined from the 
gravitational quantum corrections in M-theory as well as $M2$ brane sources. To analyse this on the fourfold 
background \eqref{fourfold} becomes too cumbersome, so let us simply illustrate the case of a metric \eqref{mfourfold} with a warp factor 
of the fibre torus $e^{2B}$ i.e $C = B$. In this case the $G$-fluxes satisfy the following two equations:
\begin{eqnarray}\label{gfluxe}\nonumber
&&(1)~ D_q\Big[e^{3A}\big(G^{mnpq} ~ - ~ (\ast G)^{mnpq}\big)\Big] ~ = ~ \frac{2k^2 T_2}{8!}\epsilon^{mnpa_1....a_8}
(X_8)_{a_1....a_8} \\
&& (2)~ \square ~e^{6B} ~ = ~ -\frac{1}{2\cdot 4!} G_{mnpq}(\ast G)^{mnpq} ~ - ~ 
\frac{2k^2 T_2}{8!}\cdot \frac{X_8}{\sqrt{-g}} + ...
\end{eqnarray}
where $k^2 T_2$ are constants appearing in the M-theory Lagrangian, and we have made 
all fields and the Hodge star operations 
w.r.t. the unwarped metric, except for the $X_8$ term.  
The $X_8$ term in the above two equations is the eight form expressed entirely in terms of 
the curvature tensor of the warped metric. This is the quantum correction that we can put to zero when the background 
is non-compact. A simple observation of \eqref{gfluxe} will tell us that for a compact manifold, a vanishing $X_8$ term 
will lead to contradiction. 

We have also left some dotted terms in the second equation of \eqref{gfluxe}. These unwritten terms account for sources, like $M2$
branes, in the theory. These $M2$ branes are precisely the D3 branes that we will need to eventually put in to study inflation in
our model. 

Observe now that when we make $X_8$ negligibly small (or in other words, when we ignore quantum corrections), 
the equations of motion of the $G$-fluxes \eqref{gfluxe}, tell us that the covariant derivatives of $G$-fluxes have to vanish. This condition can be satisfied by two different varieties of $G$-flux:
\begin{equation}\label{susycond}
G_{mnpq} ~ = ~ (\ast G)_{mnpq}, ~~~~~{\rm or} ~~~~~ G_{mnpq} ~ - ~ (\ast G)_{mnpq} ~ = ~ e^{-3A}\gamma_{mnpq}
\end{equation}
where $\gamma_{mnpq}$ is a covariantly constant tensor. 
The first condition means that the $G$-fluxes have to be self-dual. If it is also 
primitive then this is the condition to 
preserve susy \cite{BeckerM}\footnote{Recall that primitivity implies self-duality but not vice-versa on a 4-fold, in contrast to primitivity and \emph{imaginary} self--duality on a 3--fold.}. The second condition concerns us here. Generically, this implies that
the $G$-fluxes are not primitive and therefore 
susy is spontaneously broken in our model. However, if we can rewrite $\gamma_{mnpq}$ as 
\begin{equation}\label{covgama}
\gamma_{mnpq} ~\equiv~ e^{3a}\gamma^{(1)}_{mnpq} ~ - ~ e^{3a}\left[\ast\gamma^{(1)}\right]_{mnpq}
\end{equation}
with $e^{3a}$ being a function that we will specify below, 
then self-duality is restored in the presence of a new $G$-flux that is of the form 
\begin{equation}\label{newgg}
{\cal G}_{mnpq} \equiv G_{mnpq} - e^{-3(A-a)}\gamma^{(1)}_{mnpq}
\end{equation}
although this may not be primitive.
Indeed, if we demand $\gamma^{(1)}$ to be of the form
\begin{equation}\label{jwedj}
\gamma^{(1)} ~\equiv ~{\cal J} \wedge {\cal J}
\end{equation}
with ${\cal J}$ being the fundamental 2--form in M/F-theory and $e^{-3(A - a)}$ 
is a closed zero form then susy can be broken with a 
non-primitive self-dual (2, 2) form \cite{beckersusy}\footnote{A non-self-dual flux of the form 
$G_{mnpq} = \frac{e^{-3A}}{2}\left(\gamma - \ast \gamma\right)_{mnpq}$ can also break susy and satisfy the 
second condition in \eqref{susycond}. However, such a choice of flux does not satisfy the equation of motion.}. 
A similar condition can be derived on the fourfold with three warp factors, as in \eqref{mfourfold} and \eqref{fourfold}. With three warp factors the analysis remains the same. One 
can easily verify this from the $G$-fluxes constructed out of type IIB three-forms. In the following we will try to 
justify the existence of this (2, 2) non-primitive form. 

\vskip.1in

%%%%%%%%%%%%%%%%%%%%%%%%%%%%%%%%%%%%%%%%%%%%%%%%%%%%%%%%%%%%%%%%%%%%%%%%%%%%%%%%%%%%
\subsection{Normalisable harmonic forms and seven branes}
%%%%%%%%%%%%%%%%%%%%%%%%%%%%%%%%%%%%%%%%%%%%%%%%%%%%%%%%%%%%%%%%%%%%%%%%%%%%%%%%%%%%
\vskip.1in

So far, our study in M-theory has followed in parallel to that in type IIB. To see some novelty 
from the M-theory picture, let us look for the remnants of the seven branes in M-theory. Since M-theory 
does not support any branes other than two and five-branes, the information of type IIB seven branes can only come from
the gravity solution. In type IIB theory, recall that the seven branes were embedded via the Ouyang embedding \cite{ouyang}.
This means the embedding equation is:
\begin{equation}\label{oembed}
(\rho^6 + 9a^2 \rho^4)^{1/4} ~{\rm exp} \bigg[\frac{i(\psi - \phi_1 - \phi_2)}{2}\bigg]~{\rm sin}~\frac{\theta_1}{2} 
~  {\rm sin}~ \frac{\theta_2}{2} = \mu^2
\end{equation}
In the limit $\mu \to 0$ the seven branes should be embedded via the two branches:
\begin{eqnarray}\label{sevencoor}\nonumber
&&{\rm Branch ~1}:~~~\theta_1 ~ = ~ 0, ~~~~ \phi_1 ~ = ~ 0 \\ 
&&{\rm Branch ~2}:~~~\theta_2 ~ = ~ 0, ~~~~ \phi_2 ~ = ~ 0 
\end{eqnarray}
and both run along the radial direction\footnote{It is easy to see why. 
A generic configuration of seven branes would be able to lower their actions by going to 
smaller $\rho$. Therefore, they cannot be fixed at a specific $\rho \equiv \rho_0$.}. 
The full geometrical analysis of the embedding is difficult, but we can see that for branch 1
 the seven branes wrap a four-cycle along directions ($\theta_2, \phi_2$) and ($\psi, \rho$) inside
the resolved conifold background and are stretched along the spacetime directions $x^{0, 1, 2, 3}$. One can easily see
that the axionic charges of the seven branes could all globally cancel by allowing a trivial F-theory monodromy
so that there is no contradiction with
Gauss' law. Subtleties come when we want to study compact manifolds in the presence of seven-branes {\it and} 
non-primitive fluxes. In the absence of non-primitive fluxes one can compactify the manifold with a sufficient number 
of seven branes and orientifold planes. The more subtle situation with non-primitive fluxes will 
be discussed later.

For the present case let us look at the metric along directions orthogonal to the type IIB seven branes. The M-theory
metric given above \eqref{fourfold} will immediately tell us the orthogonal space to be:
\begin{eqnarray}\label{orthospace}\nonumber
ds^2 ~ & = &~ \frac{h^{1/3}}{({\rm Im}~\tau)^{4/3}} ~\big\vert dx^{11} + \tau dx^{3}\big\vert^2 ~ + ~ 
{h^{1/3} ({\rm Im}~\tau)^{2/3}} \Big[\frac{\rho^2}{6}\,
d\theta_1^2 + \frac{\rho^2}{6} {\rm sin}^2 \theta_1~d\phi_1^2 \Big] \\ 
&= & \frac{h^{1/3}}{({\rm Im}~\tau)^{4/3}} ~(dx^{11} + {\rm Re}~\tau ~dx^{3})^2 + 
{h^{1/3} ({\rm Im}~\tau)^{2/3}} \Big[\frac{\rho^2}{6}d\theta_1^2 + \frac{\rho^2}{6} {\rm sin}^2 \theta_1 ~d\phi_1^2\Big]
 \\ \nonumber
&& ~+ ~\frac{h^{1/3}}{({\rm Im}~\tau)^{4/3}} ~ ({\rm Im}~\tau)^2 (dx^{3})^2
\end{eqnarray} 
where Re $\tau$ and Im $\tau$ are related to the axion and dilaton respectively in the following way:
\begin{eqnarray}\label{axiodil}\nonumber
&&{\rm Re}~\tau ~ \equiv ~ C_0 ~ = ~ \frac{N}{2\pi} (\psi - \phi_1 - \phi_2) \\
&&{\rm Im}~\tau ~ \equiv ~ e^{-\Phi} ~ = ~ \frac{1}{g_s} - \frac{N}{2\pi}~ 
{\rm log}\Bigg[(\rho^6 + 9a^2 \rho^4)^{\frac{1}{4}} ~{\rm sin}~
\frac{\theta_1}{2}~{\rm sin}~\frac{\theta_2}{2}\Bigg]
\end{eqnarray}
and $N$ is the number of the seven branes, as discussed in \cite{ouyang}. The above choice of 
axion-dilaton is {\it not} the full story. For the time being, however, we 
will continue using this result because the corrections to axion-dilaton are subleading. Some aspects 
of these corrections have been discussed in \cite{BCDF} using results of \cite{DKS}.

To study the geometry further, let us analyse the background close to the point ($\phi_1=0,$ $\theta_1=0$). 
The resulting metric in the local
neighbourhood of the point ($\phi_1, \theta_1$) has the following form:
\begin{equation}\label{locmeti}
ds^2 ~=  {h^{1/3} ({\rm Im}~\tau)^{2/3}} \Bigg[\frac{\rho^2}{6} d\theta_1^2 + 
\frac{\rho^2}{6} {\rm sin}^2\theta_1~d\phi_1^2 + (dx^{3})^2\Bigg] + 
\frac{h^{1/3}}{({\rm Im}~\tau)^{4/3}} ~\Bigg(dx^{11} + 
\frac{N}{2\pi} (\psi - \phi_1 - \phi_2) dx^{3}\Bigg)^2
\end{equation}
which can be compared to a Taub-NUT metric:
\begin{equation}\label{Tn}
ds^2_{\rm Taub-NUT} = V({\tilde r}) \Big(dx^{11} + A_{3} dx^{3}\Big)^2 + V({\tilde r})^{-1} 
\Big[ d{\tilde r}^2 + {\tilde r}^2 d\theta^2 + 
{\tilde r}^2 ~{\rm sin}^2~\theta (dx^{3})^2\Big]
\end{equation}
with $V({\tilde r})$ being the typical harmonic function. We see that \eqref{locmeti} does have a strong resemblance to 
\eqref{Tn}, with the $A_{3}$ charge of the Taub-NUT being given by the axionic charge of $N$ type IIB seven-branes,
as expected. However, the local metric is more complicated than the standard TN space because of the non-trivial 
back-reaction of the G-flux. In particular, the warp factors and some of the coordinates 
appearing in \eqref{locmeti} are 
not quite of the form in \eqref{Tn}. Nevertheless, \eqref{locmeti} does capture some of the key features of a 
Taub-NUT space, namely, the $U(1)$ fibration structure and the gauge charge. In 
\eqref{Tn} the gauge charge has a proportionality $A_{3} \propto {\rm cos}~\theta$. Such a choice of Taub-NUT charge helps us to determine
an anti-self-dual harmonic form in this space \cite{gibbsone, senone, DRRS}. Comparing this to \eqref{locmeti}, 
we see that the charge is given by $C_0 \equiv \frac{N}{2\pi}\left(\psi - \phi_1 - \phi_2\right)$. 
A small change in this charge can be related to a small change in $\phi_1$, keeping other variables constant (recall 
that we are measuring the charge away from the D7 brane). 

We now define the vielbeins in the following way:
\begin{eqnarray}\label{vieldef} \nonumber 
&& e^y \equiv \frac{h^{1/6}}{({\rm Im}~\tau)^{2/3}}\Bigg(dx^{11} + 
\frac{N}{2\pi} (\psi - \phi_1 - \phi_2) dx^{3}\Bigg), ~~~~~ e^3 \equiv h^{1/6} ({\rm Im}~\tau)^{1/3}~dx^3 \\
&& e^{\theta_1}\equiv \frac{h^{1/6} ({\rm Im}~\tau)^{1/3} \rho}{\sqrt{6}} ~d\theta_1, ~~~~~~~~~~  
e^{\phi_1}\equiv \frac{h^{1/6}({\rm Im}~\tau)^{1/3} \rho ~{\rm sin}~\theta_1}{\sqrt{6}} ~d\phi_1  \, .
\end{eqnarray}
Using these vielbeins we are now ready to construct our harmonic forms.
These harmonic forms have to be 
self-dual (or anti self-dual)
as well as normalisable. Let us make the following ans\"atze for the one-form:
\begin{equation}\label{hform}
\omega ~ = ~ l({\theta_1}) ~\Bigg(dx^{11} + \frac{N}{2\pi} (\psi - \phi_1 - \phi_2) dx^{3}\Bigg) \, .
\end{equation}
The harmonic two-form will then be given by $d\omega$ and is therefore exact as well as harmonic. To require this to 
be anti self-dual, we want $\ast d\omega = - d\omega$ in this space with the Hodge star 
being given by the warped metric \eqref{locmeti}. This gives us:
\begin{equation}\label{ltheta}
l(\theta_1) ~ = ~ {\rm exp}~\Bigg[\mp \frac{N}{2\pi} \int^{\theta_1} \frac{d\theta_1}{{\rm sin}~\theta_1 ~
{\rm Im}~\tau}\Bigg] \, .
\end{equation}
This implies that the one-form is:
\begin{equation}\label{oneform}
\omega ~ = ~ {\rm exp} ~\Bigg[\mp\int^{\theta_1} \frac{d\theta_1}{{\rm sin}~\theta_1 
\left({\rm log~ sin}~\frac{\theta_1}{2} + ...\right)}\Bigg]
\Bigg(dx^{11} + \frac{N}{2\pi} 
(\psi - \phi_1 - \phi_2) dx^{3}\Bigg) \, ,
\end{equation}
which clearly means that an anti self-dual two-form is normalisable, whereas a 
self-dual two-form in not. 
Existence of such normalisable forms guarantees many things:
firstly it confirms the existence of seven branes in this background. Once the harmonic forms are defined over a 
compact two-sphere then the resulting background can be compactified so that an effective four-dimensional theory 
could be defined. In the presence of a non-compact background, the harmonic forms are very useful to determine 
the world volume theory on the seven branes \cite{imamura, sentwo, DRRS}. Secondly, existence of harmonic forms
guarantees the {\it non-commutative} deformations on the seven-branes \cite{DRRS}. Recall that the world-volume
theory on the type IIB seven-branes is non-commutative because of the presence of non-primitive fluxes. This is 
perfectly consistent with the original D3/D7 inflationary model \cite{DHHK} that was also non-commutative 
due to the presence of a non-primitive background. The key difference between our present background and the original
D3/D7 system is that (apart from being the fact that the original D3/D7 system was defined on $K3 \times T^2/Z_2$)   
in the original D3/D7 system the non-primitivity was treated as a tunable 
parameter (although it might violate the equations of motion) and could be switched off to regain supersymmetry. In our present scenario we see no way to switch off the non-primitivity. In other words, our present background 
is inherently non-supersymmetric.  

At this point we wish to make several comments: Firstly,
the above analysis is only for one of the embedding branches. It is not difficult to see that a similar 
analysis could be performed for the other branch. The total normalisable anti-selfdual harmonic form is 
presumably a linear combination of these two forms.  
Secondly, $-$ and this is important $-$ 
the above analysis relies heavily on the particular embedding that we took, namely the embedding \eqref{oembed}. This
embedding is the trivial embedding that should be modified when $\mu \ne 0$ in \eqref{oembed}. An immediate 
modification of the embedding equation \eqref{sevencoor}, which was for $\mu = 0$, will be the following set 
of equations:
\begin{equation}\label{corembed}
(\rho^6 + 9a^2 \rho^4)^{1/4}~{\rm sin}~\frac{\theta_1}{2}~{\rm sin}~\frac{\theta_2}{2} ~ = ~ \vert \mu\vert^2, ~~~~~~~ 
\psi - \phi_1 - \phi_2~ = ~ \tilde\theta
\end{equation}
where 
$\tilde\theta \equiv -i~{\rm log} ~\frac{\mu}{\vert\mu\vert} - 2n\pi$ is a phase factor 
fixed by the orientation of the seven branes in the angular directions. As soon as 
$\mu \ne 0$, the embedding equations are no longer the simplified equation \eqref{sevencoor}, but rather the surface 
\eqref{corembed}. Thus we see in a resolved conifold that the seven branes wrap along a nontrivial curved four-cycle
in the internal space\footnote{This is clearly a four-cycle because there are six unknowns and two equations in 
\eqref{corembed}.}. 

For this case one can also work out the normalisable harmonic form. The analysis is more complicated but can be worked 
out as before. We will not attempt this here, but end this part of the discussion by noting that these normalisable 
harmonic forms would give rise to second cohomologies (i.e the second Betti numbers) once we compactify the 
non-compact resolved conifold background. 

\vskip.1in

%%%%%%%%%%%%%%%%%%%%%%%%%%%%%%%%%%%%%%%%%%%%%%%%%%%%%%%%%%%%%%%%%%%%%%%%%%%%%%%%%%%%
\subsection{One forms and M-theory uplift of fluxes}\label{lift}
%%%%%%%%%%%%%%%%%%%%%%%%%%%%%%%%%%%%%%%%%%%%%%%%%%%%%%%%%%%%%%%%%%%%%%%%%%%%%%%%%%%%
\vskip.1in

At this point we should come back to the issue that we briefly 
alluded to earlier: compactifying our manifold in type IIB theory. From the F/M-theory point of view, this is equivalent to finding a consistent compact base. This problem has already been solved earlier in \cite{gtpaper1, gtpaper2} and 
\cite{dot1, dot2, dot3}. The compact base $-$ which we call $B$ henceforth $-$ should have at least one smooth 
curve ${\bf P}^1$ with normal bundle ${\cal O}(-1)\oplus {\cal O}(-1)$. The Weierstrass model for the fourfold 
can be obtained as a Calabi-Yau hypersurface with the equation:
\begin{equation}\label{weierfor}
y^2 ~ = ~ 4 x^3 -g_2 x - g_3
\end{equation}
where $y$ is the coordinate on the bundle ${\cal O}_B(3K_B)$, $x$ is the coordinate on the bundle ${\cal O}_B(2K_B)$  
and $g_k$ is a section of ${\cal O}_B(-2kK_B)$ for $k = 2, 3$. 

The elliptic fibre is generically smooth, but is a cuspidal cubic over points where $y^2 = 4 x^3$ and nodal cubic over 
points where $g_2^3 = 27 g_3^2$ with $g_k$ not zero. These latter are, of course, the points where the discriminant of the 
Weierstrass equation vanishes. The zero locus of the discriminant is a complex surface $S$ containing the curve 
$D$ defined by $y^2 = 4x^3$. Once we know $S$ and $D$, the Euler characteristics of the fourfold can be completely 
written in terms of the Euler characteristics of these submanifolds, i.e
\begin{equation}\label{euler}
\chi ~ = ~ \chi(S) + \chi(D) ~ = ~ 19728 ~ = ~ 24 \times 822
\end{equation}
which would tell us that the total number of branes and fluxes should add up to 822 for this manifold\footnote{
Incidentally, if we make a conifold transition to the base to go to a fourfold that is a $T^2$ fibration over a
deformed conifold base, the Euler number remains unchanged. See \cite{gtpaper2, dot3} for more details.}. 

Observe, however, that the fourfold that we choose with a K\"ahler base is not the most generic answer. In general, the 
base could be a non-K\"ahler manifold. What we need from our present analysis is the existence of one-forms in our 
manifold which could be used to express the (1,2) fluxes in the type IIB set-up. Presently, in the type IIB 
set-up, we can think of the following three choices of one-forms in our manifold:

The first of the three one forms can be written in terms of the holomorphic coordinates ($z, y, u, v$) given in \eqref{holocoord}, in the following way \cite{ionel}:
\begin{equation}\label{oneforma}
\omega_1 \equiv r^{-2}\left(N^{1/3} + 4a^4 N^{-1/3} - 2a^2\right)~{\rm Im}~\left(\bar z dz + \bar y dy  + 
\bar u du  + \bar v dv \right)
\end{equation}
where $N = N(r) = \frac{1}{2}\left(r^4 - 16a^6 + \sqrt{r^8 - 32 a^6 r^4}\right)$. See \eqref{rhoandr} for the relation between $r$ and our radial coordinate $\rho$. The above one form 
contributes an exact part to the K\"ahler form on the resolved conifold. This one form is invariant under the 
underlying $SO(4, \mathbb{R})$. 

Another one form can be constructed using the homogeneous coordinates $\zeta_+ = \frac{\xi_2}{\xi_1}$ and 
$\zeta_- = \frac{\xi_1}{\xi_2}$ that respectively define the two patches $H_+$ where $\xi_1 \ne 0$ and 
$H_-$ where $\xi_2 \ne 0$ on the $S^2$ of the resolved conifold. (See section \ref{ptbg} for more details on the geometry.) 

We construct one forms on the two patches $H_\pm$ in the following way:
\begin{equation}\label{oneformb}
\omega_\pm ~ = ~ \frac{1}{2}~ {\rm Im}~ \frac{\zeta_\pm d\bar\zeta_\pm}{1 + \vert\zeta_\pm\vert^2}
\end{equation}
One can also show that these forms are also invariant under $SO(4)$ just like $\omega_1$ above. 

Finally, the third category of one forms in our background are of the form:
\begin{equation}\label{oneformc}
\omega_3^i ~ = ~ g_i(\rho) E_i, ~~~~~ \bar\omega_3^j ~ = ~ h_j(\rho) \bar E_j
\end{equation}
with no sum over $i, j$ (although one can combine these one forms to write another one form). The $E_i$ are 
the complex vielbeins described in section \ref{ptbg}. These one forms can only exist on the compactified base if they have 
a compact support. In the following we will discuss the asymptotic behaviours of $\omega_3$ and $\bar\omega_3$. 

To study the asymptotic behaviour it is important to divide our type IIB fluxes into ($2,1$) and ($1, 2$) parts. Let 
us also scale the radial coordinate $\rho$ as $\rho \to \lambda \rho$ so that large $\lambda$ means that we are 
exploring UV geometries. In this limit clearly 
\begin{equation}\label{vielasym}
E_i ~ \to ~ \lambda E_i, ~~~~~~~~~ \eta_i ~ \to ~ \lambda^3 \eta_i
\end{equation}
where the ISD forms $\eta_i$ were defined in \eqref{eta}. The ($2, 1$) part of 
${\hat G}_3$ is then\footnote{Recall that we are using hatted quantities to indicate the background flux with 
backreaction from the embedded seven branes.}: 
\begin{equation}\label{21part}
{\hat G}_3^{(2, 1)} = \Bigg[\alpha_1(\rho) ~- 
~\frac{9P(\rho^2+3a^2)}{\rho^3\sqrt{\rho^2+6a^2}\sqrt{\rho^2+9a^2}}\Bigg]\eta_1
+ e^{-i\psi/2}\alpha_3(\rho,\theta_1)\,\eta_3 + e^{-i\psi/2}\alpha_4(\rho,\theta_2)\,\eta_4 
\end{equation} 
with the functional forms of $\alpha_1, \alpha_3$ and $\alpha_4$ derived in Appendix \ref{embed}, see \eqref{alphas}. For large $\rho$ or
large $\lambda$, the behaviour of ${\hat G}_3^{(2,1)}$ is of the form:
\begin{equation}\label{large}
{\hat G}_3^{(2, 1)} ~ \to ~ {\rm constant} ~ + ~ {\rm log}~ \lambda
\end{equation}
and therefore ${\hat G}_3^{(2,1)}$ diverges logarithmically. This divergence is not problematic because eventually we are compactifying our 
manifold to a non-CY threefold. One should also observe that the ($2,1$) part of the fluxes in the original PT solution \cite{pt} asymptotically
goes to a constant. 

On the other hand, the asymptotic behaviour of the ($1,2$) part of the fluxes is more interesting. The explicit 
form of the ($1, 2$) part is given by:
\begin{equation}\label{12part}
{\hat G}_3^{(1, 2)} ~ = ~ \Bigg[\alpha_8 -\frac{27 P a^2}{\rho^3\sqrt{\rho^2+6a^2}\sqrt{\rho^2+9a^2}}\Bigg] \eta_8\,,
\end{equation}
with $\alpha_8$ given in \eqref{alphas}. Asymptotically ${\hat G}_3^{(1, 2)}$ now behaves in the following way:
\begin{equation}\label{small}
{\hat G}_3^{(1, 2)} ~ \to ~ \frac{1}{\lambda^2}
\end{equation}
and therefore goes to zero very fast. In fact the ($1, 2$) part of the fluxes in \cite{pt} also has the same behaviour 
asymptotically. 

Such an asymptotic behaviour of ${\hat G}_3^{(1, 2)}$ is good for us. 
This means that, since the fluxes vanish at the boundary, they should naturally exist once we compactify the resolved conifold to a compact threefold.
Furthermore we see that the ($1, 2$) part of the three form flux can be expressed alternatively as:
\begin{equation}\label{12form}
{\hat G}_3^{(1, 2)} ~ = ~ J \wedge \bar m
\end{equation}
with $\bar m$ being a ($0, 1$) form as one would have indeed expected. From our above consideration the ($0, 1$) 
form and $J$ are given in terms of the three one-forms in the following way:
\begin{equation}\label{jwbar}
\bar m ~ \equiv ~ h_1(\rho) \bar E_1 ~ = ~ 
\Bigg[\alpha_8 -\frac{27 P a^2}{\rho^3\sqrt{\rho^2+6a^2}\sqrt{\rho^2+9a^2}}\Bigg] \bar E_1, ~~~~~~~~~~
J ~ = ~ d \omega_1 ~ + ~ 4a^2 d \omega_\pm
\end{equation}
on the two patches $H_\pm$. The latter definition of $J$ is identical to the definition of $J$ in terms of the 
complex vielbeins $E_i$ given in \eqref{cs}\footnote{Note that the volume form is unique despite the existence of  multiple one-forms. The volume form is given by: $V = du \wedge dy \wedge d\zeta_+ = dv \wedge dz \wedge d\zeta_-$.}. 
It is also clear that the ($2, 1$) form cannot be expressed as 
\eqref{12form} using a one form because the ($2,1$) form is primitive. Observe, however, that 
the existence of a normalisable ($0, 1$) form doesn't always imply the existence of a non-trivial one-cycle in the 
manifold\footnote{Although, in the language of the fourfold the 
threefold base does have a non-vanishing first Chern class.}. 

Once we have the explicit ($1, 2$) forms, we still must see how this is uplifted in the M-theory picture. This is where 
things become somewhat subtle. The generic uplift of type IIB three-forms was given in \cite{DRS, GKP} in the following form:
\begin{equation}\label{gfourone}
G_4 ~ = ~ -\frac{{\hat G}_3 \wedge d\bar z}{\tau - \bar\tau} ~ 
+ ~ \frac{\bar {\hat G}_3 \wedge dz}{\tau - \bar\tau}
~ = ~ {\hat F}_3 \wedge dx^3 ~ + ~ {\hat H}_3 \wedge dx^{11}
\end{equation}
where we have used the usual definitions of $G_3$ and $dz$, namely: 
${\hat G}_3 = {\hat F}_3 - \tau {\hat H}_3$ and 
$dz = dx^{11} + \tau dx^3$ (although $d\tau \ne 0$). Thus $\widetilde{F}_3 = {\hat F}_3 - C_0 {\hat H}_3$ and 
${\hat F}_3 = d\hat C_2$ to comply with the notation used in section \ref{iib}.
With these definitions, the T-duality from IIB to M-theory
works in an expected way.

However, because of the presence of $d\bar z$ and $dz$ in \eqref{gfourone}, the uplift of a ($2, 1$) form is indeed 
a ($2, 2$) form, but the naive uplift of a ($1, 2$) form becomes a (1, 3) or a (3, 1) form, none of which are 
suitable for our case because these forms are ASD in M-theory. In the literature such subtlety was never observed 
because the ISD fluxes were never taken to have (1, 2) components. For our case, as we saw above, such forms are allowed because of their localised 
and normalisable nature. 

Indeed, such localisation of fluxes will eventually help us to show that the (1, 2) forms would also lift to F-theory 
as (2, 2) forms. To see this, observe that F-theory allows the following two important topological couplings:
\begin{equation}\label{topcop}
{\cal L}_1 ~ \equiv ~ \int_{{\cal M}_{12}} C_4 \wedge G_4 \wedge G_4, ~~~~~~~ 
{\cal L}_2 ~ \equiv ~ \int_{{\cal M}_{12}} G_4 \wedge G_4 \wedge G_4  \, ,
\end{equation}
where $C_4$ is the self-dual four-form in type IIB theory and ${\cal M}_{12}$ is the twelve dimensional space
(see \cite{feramin} and references therein for more details on these couplings).

The coupling ${\cal L}_1$ is well known. 
This leads to the standard Chern-Simmons term on D7 branes when we decompose the 
four-form as $G_4 = F \wedge d\omega$, where $d\omega$ is the normalisable two-form derived earlier and $F$ is the 
gauge flux on a D7 brane. The second coupling, ${\cal L}_2$, concerns us here. In type IIB there are no fundamental 
massless four-forms other than $C_4$ discussed above. How do we interpret $G_4$? The coupling that we 
are concerned with here is 
\begin{equation}\label{newcoupla}
\int_{{\cal M}_8} G_4 \wedge F \wedge F  \, ,
\end{equation}
where ${\cal M}_8$ is an eight dimensional surface. The only eight dimensional surface that we have in type IIB 
is the surface of the D7 brane. Therefore, we should expect the coupling \eqref{newcoupla} to show up 
on the surface of the D7 brane as some kind of {\it compact} four-form coupling to it. 

Existence of such compact four-forms can arise from the Chern-Simons terms on the D7 branes. One can easily see
that there is a coupling of the form:
\begin{equation}\label{cscoup}
\int_{{\cal M}_8} \widetilde{F}_3 \wedge {\cal A} \wedge F \wedge F \equiv 
\int_{{\cal M}_8} \left({\hat F}_3 - C_0 ~{\hat H}_3\right) \wedge {\cal A} \wedge F \wedge F
\end{equation}
when we choose the orientation of the D7 branes such that the arbitrary phase factor $\tilde\theta$ in 
\eqref{corembed} is a constant and our gauge invariant field on any D7 brane is ${\cal F} =\check B - F$ where 
$\check B$ is the pullback of the NS 2--form\footnote{We take $2\pi \alpha' = 1$ henceforth.}.

The above form of the coupling \eqref{cscoup} is of the type \eqref{newcoupla} provided the one-form ${\cal A}$ 
also becomes localised. Observe that both the three-forms appearing in \eqref{cscoup} are the localised (1,2) 
forms. Let us then assume that the one-form is ${\cal A } = l_1(\theta_1) dx^3$, 
where $l_1(\theta_1)$ is some localised
function that we will specify soon. We have also made a gauge choice to orient ${\cal A}$ along $x^3$ direction. With 
this we see that one choice of localised four-form flux is:
\begin{equation}\label{locfor} 
G_4^{(1)} ~ \equiv ~ l_1(\theta_1) \widetilde{F}_3 \wedge dx^3 ~ 
= ~ l_1(\theta_1) \left({\hat F}_3 \wedge dx^3 - C_0 ~{\hat H}_3 \wedge dx^3\right) \, .
\end{equation}
There is another choice of localised four-form flux that we can have in addition to \eqref{locfor}. This choice
can be motivated from the Born-Infeld terms of the D7 branes, and is given by:
\begin{equation}\label{locfortwo}
G_4^{(2)} ~ = ~ {\hat H}_3 \wedge \omega \, ,
\end{equation} 
where $\omega$ is the one-form derived in \eqref{oneform}. Once we compactify the internal space, the total 
axionic charge has to vanish. In that case both $G_4^{(1)}$ and $G_4^{(2)}$ simplify. In the presence of axion 
field, the total localised four-form flux is given by:
\begin{equation}\label{gfour}
G_4 ~ \equiv ~ G_4^{(1)} + G_4^{(2)} ~ = ~ {\hat H}_3 \wedge \omega +  
l_1(\theta_1) \left(-{\hat F}_3 \wedge dx^3 + 
C_0 ~{\hat H}_3 \wedge dx^3\right) \, ,
\end{equation}
which can be put in a very suggestive form if we define $l_1(\theta_1) = l(\theta_1)$ 
with $l(\theta_1)$ being the  
function of $\theta_1$ given in \eqref{ltheta} and \eqref{oneform}:
\begin{equation}\label{gfupp}
G_4~ = ~ l(\theta_1)\Big({\hat H}_3 \wedge dx^{11} -{\hat F}_3 \wedge dx^3 + 2C_0~{\hat H}_3 \wedge dx^3\Big)~ = ~ 
- l(\theta_1)~ \frac{{\hat G}^{(1, 2)}_3 \wedge dz}{\tau - \bar\tau} ~ + ~ {\rm c.c}
\end{equation}
with $dz = dx^{11} + \tau dx^3$ and ${\hat G}^{(1, 2)}_3$ being the (1, 2) form. 
The above four-form is clearly a (2, 2) form
as one would have expected from the earlier discussions \cite{beckersusy, haack, Dinerohm}. Notice however that the 
four-form flux is {\it not} closed. 

It is also interesting to note that since ${\hat G}_3^{(1, 2)}$ 
is of the form $J \wedge \bar m$ (see \eqref{12form}), the
localised (2, 2) form in M-theory becomes:
\begin{equation}\label{locg}
G_4~ \equiv ~ \frac{1}{2}~{\rm Re}~\Big(i e^\phi ~l(\theta_1) ~J \wedge \bar m \wedge dz\Big)
\end{equation}
At this point we may want to connect the four-form with the results given in \cite{beckersusy, haack}. The 
four-form should be related to ${\cal J} \wedge {\cal J}$ in M-theory where ${\cal J}$ is the fundamental
(1, 1) form for the fourfold. Defining ${\cal J} = J + dz \wedge d\bar z$, we have
\begin{equation}\label{jj}
{\cal J} \wedge {\cal J} ~ = ~ J \wedge J ~ + ~ 2 J \wedge dz \wedge d\bar z \, .
\end{equation}
It is easy to follow these fluxes to see how they appear in type IIB side. The second component in 
\eqref{jj} i.e $J \wedge dz \wedge d\bar z$ becomes a three-form field strength in T-dual type IIA theory: 
\begin{equation}\label{curF}
(\tau - \bar\tau)~J \wedge dx^3
\end{equation}
whose origin will be discussed in the next section. 
Similarly, the first component 
in \eqref{jj} ($J \wedge J$) becomes a five-form in type IIB side which has one component 
along $x^3$ direction and other components inside the threefold. This takes the form:
\begin{eqnarray}\label{five}\nonumber 
G_5 & = & \frac{\rho^3}{9} ~{\rm sin}~\theta_1~d\rho \wedge e_\psi \wedge d\phi_1 \wedge d\theta_1 \wedge dx^3 +  
\frac{\rho(\rho^2 + 6a^2)}{9}~{\rm sin}~\theta_2~d\rho \wedge e_\psi \wedge d\phi_2 \wedge d\theta_2 \wedge dx^3 \\
&& ~ + ~ \frac{\rho^2(\rho^2 + 6a^2)}{18} ~{\rm sin}~\theta_1~{\rm sin}~\theta_2~d\phi_1 \wedge d\theta_1
 \wedge d\phi_2 \wedge d\theta_2 \wedge dx^3 \, .
\end{eqnarray}   
This five-form (or the equivalent four-form) is 
strongly reminiscent of the four-form that we called $G_4^{(1)}$ in \eqref{locfor}, which does have one component
along $x^3$ direction. Indeed, the five-form\footnote{This is clearly non-vanishing because the underlying four-form
is not closed as we saw above.}:
\begin{equation}\label{fnow}
\frac{dl}{d\theta_1}~d\theta_1 \wedge {\hat F}_3 \wedge dx^3 ~ 
+ ~ \frac{l}{2} ~ \left(d{\hat G}_3 + d{\bar{\hat G}_3}\right)
\wedge dx^3
\end{equation} 
that we get from our background flux does match with \eqref{five}, but \eqref{fnow} has more terms than \eqref{five}.
This difference appears because, once we compactify our manifold, the fundamental form $J$ would change which, would
change the five-form \eqref{five}. 

The connection we have established here gives a stronger justification for why the cosmological constant should vanish in the bulk. It 
may be interesting to see if the arguments of \cite{Dinerohm} could be applied to our scenario also.  

%popo

\vskip.1in

%%%%%%%%%%%%%%%%%%%%%%%%%%%%%%%%%%%%%%%%%%%%%%%%%%%%%%%%%%%%%%%%%%%%%%%%%%%%%%
\section{Applications}\label{cosmo}
%%%%%%%%%%%%%%%%%%%%%%%%%%%%%%%%%%%%%%%%%%%%%%%%%%%%%%%%%%%%%%%%%%%%%%%%%%%%%%%%

\subsection{Compactification and non-K\"ahlerity}

There remain issues that were given only partial attention in our earlier sections. The first such issue is the nature of a possible compactification of our background, which will certainly not be a Calabi--Yau, nor even K\"ahler. In the F-theory section we discussed that the six--dimensional base 
cannot be a Calabi--Yau manifold as it has a non-vanishing first Chern class. By reducing to IIA we can argue that the T--dual IIB background will indeed be non--K\"ahler. This construction follows the ones laid out in \cite{andrei, gtpaper1}.

The three form flux \eqref{curF}
that we get in type IIA will dissolve in the metric once we T-dualise to type IIB theory, making the background 
non-K\"ahler.\footnote{In M-theory once $d{\cal J} \ne 0$ the four-form flux ${\cal J} \wedge {\cal J}$ is not 
closed. This is of course consistent with our choice of four-form flux \eqref{gfour}.} Once the background is non-K\"ahler there would be {\it extra} sources of fluxes in addition to the fluxes that 
we mentioned in \eqref{gfour}, namely ``geometric flux''. 
One can replace the type IIB three form NS flux by 
\begin{equation}\label{comthree}
{\tilde H}_3 ~ \equiv~ {\hat H}_3 + i d(e^{-\phi}J) \, .
\end{equation}
This complexification of the three form flux is not new and has been observed earlier in heterotic compactifications 
\cite{het1, het2, het3, het4}, which in turn gave rise to a new superpotential in the heterotic theory \cite{het5, 
curiolust1}. An interesting observation here is that the type IIB background itself becomes non-K\"ahler now as 
compared to the heterotic background where the type IIB background was conformally K\"ahler. 

We also remarked on possible generalisations of the IIB superpotential in section \ref{cc}. It seems clear that the GVW superpotential will get corrected if the moduli space is enlarged by non--trivial one--forms. For the case of a background that is mirror to a Calabi--Yau with NS flux (so it acquires a non--trivial $T^3$ fibration when the mirror symmetry is interpreted as three T--dualities --- the NS B--field becomes part of the metric in the mirror manifold \cite{andrei}), a superpotential has been proposed \cite{berglund}. Whether or not this is suggestive for our case requires further study. Thus far, we have no reason to believe that our IIB manifold (globally) admits an SU(3) structure. The space of generalised Calabi--Yau manifolds is much larger, though some work on superpotentials in this case appeared in \cite{louisgrana1, iman, louisgrana, luca}. If we could infer that our IIB background admits an SU(3) structure, then it would be guaranteed to be complex \cite{andrew, grana, dallagata} if it preserved supersymmetry. However, in the presence of susy--breaking flux we cannot infer the structure of the manifold. A complex manifold would have the advantage to give us control over the complex structure deformations.

\subsection{Inflationary dynamics} 

The major motivation for constructing the background in 
this paper was to study a model of inflation that may give slow roll dynamics with less fine tunings than the usual $D3-\overline{D3}$ scenarios \cite{KKLMMT, bdkm}. Let us therefore sketch a 
possible model of inflation using the resolved conifold background with D7 branes and additional D3 branes. 

Recall that D3/D7 inflation has primarily been studied in toroidal manifolds (see \cite{DHHK, HaackprKesh} and citations therein) of the 
form $T^n/{\Gamma}$ of which $K3 \times T^2/\mathbb{Z}_2$ is a 
special case. The F-term and D-term potentials appearing from the 
gaugino condensate and susy breaking fluxes, respectively, conspired to 
give a consistent resolution of the anomalies associated with the 
FI terms. 

We outline a possible scenario to achieve slow roll inflation when we combine the ideas of $D3-\overline{D3}$ in the ``warped throat'' (KKLMMT \cite{KKLMMT}) with D3/D7 models  \cite{DHHK,HaackprKesh})\footnote{Similar idea has been proposed independently by Cliff Burgess.}. 
%Recall also that in the KKLMMT scenario, inflationary dynamics is achieved (via heavy fine-tuning of course) by keeping an anti--D3 brane at the tip of the throat (i.e the far IR of the dual gauge theory) and allowing a D3 brane to move towards it. The motion in general is {\it not} slow-roll and therefore inflation is difficult to achieve here. \marginpar{already in the intro! everybody knows this by now}
We want to balance a D3 that is attracted towards the D7 (because of the non-primitive flux on the D7 worldvolume) with another force that drives the D3 toward the tip. This can be achieved by placing an anti-D3 there or by using a background in which the addition of a D3 explicitly breaks supersymmetry, such as the resolved warped deformed conifold \cite{DKS}. The motion of D3--branes towards the tip in the latter background is a consequence of the 
running dilaton. However, this potential alone is still too steep for slow--roll inflation. 

Combining both forces, however, we might hope to slow down the motion of the D3 in either the one or the other direction. There are two possible scenarios, depending on which force dominates:
%Let us now consider the following modification to the standard KKLMMT scenario: we wrap D7--branes on a four--cycle of the non-compact warped Calabi-Yau, and allow ${\cal F}_{mn}$ fluxes on it, exactly as for the D3/D7 system \cite{DHHK} with D7 branes wrapped on the K3. We see that there are two kinds of forces now:
\begin{itemize}
\item The D--term potential created by the non--primitive flux dominates and attracts the D3--brane towards the 
  wrapped D7 brane. Inflation ends when the D3 dissolves into the D7 as non-commutative instantons and supersymmetry is restored.
\item The attraction towards the anti--D3 brane at the bottom of the throat (or possibly a running dilaton in a more general background)   dominates. Inflation ends as all or some D3 branes getting annihilated by the 
anti--D3 brane(s) at the tip of the throat.
\end{itemize}
Naively one might hope that the motion would 
be slow because the D3 branes
are pulled in both directions. However, it may also turn out that the initial position of the D3 has to be heavily fine--tuned in this setup.

The F-term potential associated with the motion of the D3--branes towards the 
tip of the throat has recently been computed with the inclusion of holomorphically embedded D7--branes \cite{bdkm, axel, BCDF} using the 
analysis of \cite{BDKMMM}. 
% The D-terms forces can also be computed with some
% effort by knowing the local warped geometry near the throat. From this it
% seems possible that the system will show slow-roll without much fine tuning. 
If we want to combine the D-term and F-term potentials we are faced with an issue pointed out by \cite{nilles}: for a supersymmetric background it is impossible to have a $D$-term potential 
that could be used to pull the D3 brane towards the D7 branes. Thus if we want to switch on non--primitive
fluxes on the wrapped D7 branes we have to embed the D7 branes in a non-supersymmetric 
background. Our 
problem becomes threefold:
\begin{itemize}
\item Construct a supergravity background with embedded D7 branes that breaks supersymmetry 
  spontaneously.
\item Allow for a possible D-term uplifting by avoiding the no-go theorem of 
  \cite{nilles}, as pointed out by \cite{fernando}. Note that the D7 worldvolume theory will not only contribute the D but also possible F-terms, such that the issue of \cite{nilles} might be resolved.
\item Balance the D3 brane using the two forces: one from the D-term potential and the other 
  from the attractive force at the tip of the deformed conifold in the KKLMMT setup.
\end{itemize}
In this paper we have addressed the first two problems by constructing a non-supersymmetric background with 
D-terms on the D7 branes given by the pullback of a non--primitive flux. To analyse the last problem, we might have to go to a more generic 
background with both the two and the three cycles non vanishing.  
% In our last paper \cite{BCDF} we took the limit \eqref{oldpap}.  
% In this limit the background becomes a (non-compact) singular conifold, and one can add D7 branes using the technique 
% discussed in \cite{ouyang}. This is the simplest choice and works well in the situation when we are far from the 
% tip of the throat and the resolution parameter $\epsilon$ is very small, where 
% \begin{equation}\label{resolution}
%   F_1 ~ - ~ F_2 ~ \equiv ~ \epsilon f
% \end{equation}
% with $f$ being a another function of $\rho$. \marginpar{already in intro!}
Most of the literature deals with the limit where the manifold looks like a singular conifold.
This isn't the most generic situation so we have to go away from the usual conifold background. However, taking a resolved warped deformed conifold creates non-trivial dilaton profile from two sources now:
\begin{itemize}
\item From the D7 branes, and
\item  From the unequal sizes of the two-cycles.
\end{itemize}
The running of dilaton from the first case can already be seen at a supersymmetric level in the Ouyang background
\cite{ouyang}, which was originally analysed for a non--compact singular conifold background. Once we blow up resolution cycles of the conifold and switch on fluxes, the second case mentioned above kicks in, and 
we must discuss the combined effects to get the full background geometry. This makes the problem much harder to solve. 

The warped resolved conifold however may still be a good model of inflation with D-term uplifting. We would have to extend our analysis beyond the case $\mu=0$ (in this case the D7 extend all the way down the throat, which would not allow us to place a D3 \emph{between} the D7 and the tip) and to other embeddings, such as the Kuperstein embedding \cite{kuperstein}. Our preliminary analysis indicates that the value of the D-terms should depend on the choice of embedding.

\subsection{Supersymmetry restoration}

When the D3--brane falls into the D7--branes at the end of inflation
we expect supersymmetry to be restored. Such a susy restoration was first described 
in \cite{DHHK}. For our case, the situation is more involved. From the F-theory point of view, the {\it total} G-flux at the 
end of the inflation can be succinctly presented as: 
\begin{equation}\label{totg}
G_{\rm total} ~\equiv ~ G^{(2,2)}_{P} ~ + ~ c_1 ~{\cal J} \wedge {\cal J} ~ + ~ c_2 ~ F\wedge d\omega ~ + ~ 
c_3~ {\hat H}_3 \wedge \omega \, ,
\end{equation}
where $c_i$ are some defined functions of the coordinates ($\theta_1, \phi_1$) or ($\theta_2, \phi_2$) depending on 
which branch \eqref{sevencoor} we are on, 
$G^{(2,2)}_{P}$ is the primitive
part of the $G$-flux that come from the uplift of the type IIB (2, 1) forms,   
and $F$ is the gauge flux 
induced by dissolving the D3 brane inside the D7 branes. 
The 1-form $\omega$ was defined in \eqref{oneform}. The last term coming from the ${\hat H}_3$ coupling is 
non-primitive, and because of that
in the absence of $F$ flux, the $G$ flux was 
(2, 2) but non-primitive. Observe that in the presence of $F$ flux we can in fact demand:
\begin{equation}\label{primgfll}
{\cal J} \wedge G_{\rm total} ~ = ~ 0
\end{equation}
and therefore restore supersymmetry with (2,2) fluxes.

The $F$ flux used to restore supersymmetry in the above paragraph could be interpreted in two ways: switching 
on second Chern class or switching on first Chern class. The former, which leads to instantons, is the end point of the 
D3 brane dissolving on the D7 branes. 

The latter, however, gives rise to a bound state of a D5 brane with the D7 branes. 
Such a technique of restoring supersymmetry has already been discussed in \cite{Dgwyn1, Dgwyn2} and could 
probably be used to restore supersymmetry in the limit where the resolution parameter $a$ goes to zero. This 
would then be one simple way of restoring supersymmetry in the original Ouyang construction \cite{ouyang}.

\chapter{Towards Thermal Flux Compactification with Flavor}\label{chapitreintrotemperature}

As we mentioned in the introduction and chapter \ref{chapitreintrobase}, flux compactification solutions of the type presented in chapter \ref{DtermPaper} also are of particular interest towards understanding the relations between gauge and gravity theories through holography. In the rest of this manuscript, we will use these results and extend them towards even more realistic gravity duals of the Standard Model. But first, in this chapter, we will review previous work in this direction.

\section{Introducing Temperature on the Gravity Side}
An obviously interesting way to extend the type of string solutions we developed would be to include non-zero temperatures. While thermal field theories usually include temperature via periodicity in the Euclidean time direction, finite temperature in the dual supergravity side is best understood through non-extremal geometries \cite{GubsterKlebPeet, Witten:1998zw, Peet:2000hn}. Periodicity in the time direction is usually read from the metric of a disk, $ds^2=dr^2+r^2d\theta^2$, and the temperature is defined as the inverse of the period in $\theta$, $T=\frac{1}{2\pi}$. In supergravity, we also have a Wick rotated form of the cone's metric ($t\rightarrow it_E$) allowing us to read off the temperature from the prefactor of $dt^2$:
\bg
ds^2=-4\pi^2 T^2r^2dt^2+dr^2 ~~.
\nd

Solutions of this form appear naturally in supergravity as non-extremal branes solutions. An example of prime interest to us is the standard regular non-extremal D3-branes solution compactified on $X^5$, sourced only by the 5-form flux. In this case, the metric takes the form
\bg\label{metriquedeBH}
ds_{10}^2=h^{-1/2}(r)\big(-g(r)dt^2+dx^i dx^i\big) + h^{1/2}(r)\Big[\frac{dr^2}{g(r)}+r^2(dM_5)^2\Big]
\nd
with
\bg
h(r)=1+\frac{\tilde{L}^4}{r^4}\\
g(r)=1-\frac{r_h^4}{r^4}\label{nonextrfact}
\nd
where $i=1,2,3$ and $\tilde{L}^4=\sqrt{L^8+r_h^8/4}-r_h^4/2$. The new element here is the non-extremality or black hole factor: $g(r)$. On the boundary at $r=\infty$, this non-extremal factor has the effect of turning on a temperature of
\bg
T=\frac{g'(r_h)}{4\pi\sqrt{h(r_h)}}
\nd 
from the supergravity perspective. Since $g(r)$ introduces a singularity in the metric at the horizon $r_h$, we have to restrict our space to $r>r_h$. The $r<r_h$ region corresponds to the black hole volume. %This is illustrated in Figure \ref{FigBlackD3}.

%\begin{figure}[htb]
%\begin{center}
%\includegraphics[height=6cm,angle=0]{BlackD3.pdf}
%\caption{Thermal D3-branes solutions introduce temperature through a non-extremal black hole which cloaks the singularity of the conifold, ending the the radial direction at the horizon $r_h$. This is a finite temperature version of Figure \ref{FigKW}. Note that although the branes have been illustrated in the following Figures, they only contribute to the geometry through the fluxes and do not explicitely enter  into the action.}
%\label{FigBlackD3}
%\end{center}
%\end{figure}

In general, we will consider that having a metric of the form \eqref{metriquedeBH} with $g(r)\neq 1$ corresponds to introducing a temperature on the gravity side. Note that in order to have a regular horizon, we need $g(r)$ and $h(r)$ not to vanish at the same value of $r$.

In the same manner as when we went from the KW to the KT solutions of sections \ref{SectionKW} and \ref{KTsection}, it would be very interesting to generalize the regular and fractional D3-branes KT solution to include non-extremality. Such a solution would provide a gravity dual of $SU(N+M)\times SU(N)$ thermal gauge theories and potentially be the starting point of a thermal version of the KS cascade of section \ref{KSsection}.  

A. Buchel first addressed the question in \cite{BuchelSeul} with the interesting proposition that since non-extremality cuts the the RG flow at $r_h$, the KT singularity might be cloaked behind the horizon, thus preventing the necessity of the duality cascade of KS. Realizing this would provide a supergravity manifestation of chiral symmetry restoration at finite temperature. Unfortunately, the particular supergravity construction of Buchel was not regular as both the warp factor $h(r)$ and the non-extremal factor $g(r)$ vanished at the same $r_h$. Furthermore, this first attempt at a non-extremal KT solution failed to reduce to the standard non-extremal D3-brane solution of \eqref{metriquedeBH} at $M=0$.

Fortunately, the idea was developed by Klebanov and collaborators \cite{KT-non-ex1,KT-non-ex2} who successfully provided a supergravity dual of $SU(N+M)\times SU(N)$ gauge theories that is asymptotically KT and has a well defined horizon. The metric, consistent with the $U(1)_R$ symmetry of the $\psi$-rotation and with the interchange of the two $S^2$'s, was taken to be of the form of \eqref{metriquedeBH}, with the compact metric taken to include a squashing factor $e^{-10w}$ for $\psi$:
\bg
(dM_5)^2=e^{-10w}e_\psi^2 + e_{\phi_1}^2 + e_{\theta_1}^2+ e_{\phi_2}^2 + e_{\theta_2}^2
\nd
Their ansatz was the same as for the KT case and they also satisfied the Bianchi identity for the 4-form. %This configuration is illustrated in Figure \ref{BlackKTFig}. 

%\begin{figure}[h]
%\begin{center}
%\includegraphics[height=6cm,angle=0]{BlackKT.pdf}
%\caption{The solution of \cite{KT-non-ex} is a non-extremal generalization of the Klebanov-Tseytlin solution to finite temperature. Here, depending on its size, the black hole might protect us from the singularities in the fluxes and geometry and avoid the need of a duality cascade {\it a la} Klebanov-Strassler. The squashing of the 1-cycle has not been illustrated here.}
%\label{BlackKTFig}
%\end{center}
%\end{figure}

With these ansatz, Klebanov and collaborators \cite{KT-non-ex1,KT-non-ex2} could reduce to an effective one-dimensional action in $r$ and find a non-extremal fractional D3-brane solution. However, their solution was only valid at high temperature due to the perturbative expansion that their solution required. Their expansion parameter is $\lambda=P^2/\mathcal{K}_*$ where $\mathcal{K}_*$ is the 5-form flux \eqref{cinqformdeKT} at the horizon. Since $\mathcal{K}_*$ gets bigger as we push $r_h$ further towards the UV, it is clear that this is a large temperature expansion.

Three key observations about thermal solutions must be made from this. First, the squashing of the $U(1)$ fiber indicated a necessity to modify the $T^{1,1}$ metric. Second, the dilaton is no longer constant. Third, the 3-form fluxes should no longer be self-dual. These results will help us later to define our starting point when we will generalize this solution. 

Before we move on, let us have a closer look at what happens when we go towards the IR to see how the cloaking of the singularity suggested by Buchel actually occurs. There are three possibilities: i) $\mathcal{K}(r)$, the 5-form flux \eqref{cinqformdeKT}, vanishes before $r_h$, ii) $\mathcal{K}(r)$ vanishes after $r_h$, or iii) $\mathcal{K}(r)$ vanishes at $r_h$. In the situation iii), the horizon coincides with the singularity, as in \cite{BuchelSeul}, and is not a regular solution. Let us then define the radius at which $\mathcal{K}(r)$ vanishes by its temperature $T_c$. Then, the situation ii) corresponds to $T>T_c$, in which case the flux singularity is hidden and cascading is not necessary. For i) however, corresponding to $T<T_c$, $\mathcal{K}(r)$ reaches zero before the non-extremality affects the solution significantly. The construction then needs to cascade to remove the singularity and confinement occurs. Chiral symmetry restoration in the dual field theory interpretation is then a part of the phase transition that can occur at $T_c$.

\section{The Dasgupta--Mia Set up}\label{sectionMia}

As we just saw, previous work was successful only perturbatively at large temperature and without flavors. It would be much more interesting however if we could solve these issues as the IR limit of large $N$ thermal QCD is closer to the parameter space currently being explored by the RHIC experiment. 

Previous work by K. Dasgupta, M. Mia and collaborators addressed this issue extensively and provided a construction that is currently the closest set up to a UV complete gravity dual of large $N$ thermal QCD \cite{Mia:2009wj,jpsi1,Mia:2010zu,Mia:2011iv}. Key elements of this work can be seen as a non-extremal version of the set up we constructed in chapter \ref{DtermPaper} and generalized to take into account observations made in \cite{KT-non-ex1,KT-non-ex2}. More precisely, this setup presents what we expect for an embedding of regular and fractional D3-branes for colors with running couplings, along with D7-branes for flavor and non-extremality for temperature. The construction is compactified on a non-extremal squashed warped resolved conifold. %These ingredients are illustrated in Figure \ref{PremiereImageMia}. %garder nom d'image parce que reutiliser plus bas

%\begin{figure}[h]
%\begin{center}
%\includegraphics[height=6cm,angle=0]{BlackD3D5D7.pdf}
%\caption{In this construction, all the ingredient we would like to have have been included. D3- and D5-branes accound for the colors, D7-branes for the flavor, non-extremality for temperature and a squashing of the 2-cycle to allow to solve the supergravity equations. This is a finite temperature version of Figure \ref{FigDterm} with squashing of the $S^2$ 2-cycle. }
%\label{PremiereImageMia}
%\end{center}
%\end{figure}

Like in chapter \ref{DtermPaper}, the flavor D7-branes are included via the Ouyang embedding, but in this case, they will backreact differently on the geometry. Furthermore, the squashing of the $U(1)$ found by Klebanov et al in \cite{KT-non-ex1,KT-non-ex2} will be replaced by a squashing between the two 2-spheres, which will now take the backreactions from the non-extremal geometry and from the D7-branes into account. This squashing of the geometry defines new cycles on which the fluxes will be warped and, as we will see in the next chapter, will directly be related to the 3-from fluxes to no longer be ISD. This, of course, also makes our compactification manifold no longer Calabi-Yau, but this is not a concern for us as we do not try to preserve supersymmetry (which is broken by thermal effects), but only to find solutions to the equations of motion of String Theory. Note that this work is different from \cite{KT-non-ex1,KT-non-ex2} by the fact that it includes flavor, defines a precise squashing of the 2-spheres in the internal directions and address the question by the whole set of supergravity equations of motions, instead of a one-dimensional reduced action.

To make our geometric ansatz more precise, the metric which takes all these elements into account has the following form:
\bg\label{bhmet1}
ds^2 &=& {1\over \sqrt{h}}
\Big[-g(r)dt^2+dx^2+dy^2+dz^2\Big] +\sqrt{h}\Big[g(r)^{-1}dr^2+ d{\cal M}_5^2\Big]\nonumber\\
& = & e^{2A}(-e^{2B} dt^2 + \delta_{ij} dx^idx^j) + e^{-2A-2B} {\widetilde g}_{mn} dx^m dx^n
\nd
where $g(r)$ is the non-extremal factor introducing temperature and $d{\cal M}_5^2$ is the metric of the internal angular directions. The warp factors $A, B$ are related to the typical notation by
\bg\label{abdef}
h=e^{-4A},~~~~~ g=e^{2B}.
\nd
The metric along the internal directions is taken to be:
\bg\label{inmate}
&&{\widetilde g}_{mn} dx^m dx^n ~ = ~dr^2 + r^2 e^{2B} \Big[\frac{1}{9}(d\psi + {\rm cos}~\theta_1 d\phi_1 +{\rm cos}~\theta_2 d\phi_2)^2  +{\cal O}(g_sM^2/N,r_h^4/r^4)\nonumber\\
&&~~~~~~~~~~~+ \frac{1}{6}(d\theta_1^2 + {\rm sin}^2\theta_1 d\phi_1^2) +
{1\over 6}(1 + F(r))\left(d\theta_2^2 + {\rm sin}^2\theta_2 d\phi_2^2\right)\Big] ~~.
\nd
If we take the squashing factor $F(r)=\frac{6a^2}{r^2}$ with $a$ constant, this would correspond almost exactly to the resolved conifold metric \eqref{resmetric}. Here however, $F(r)$ is kept as a free parameter to include the backreactions of the non-extremality and D7-branes.

The above setup looks a lot like what was done in chapter \ref{DtermPaper}, except for the squashing (running resolution of the 2-cycle) and non-extremality. As before, we will need to require the same supergravity limit
\bg
&&(N,M,N_f, g_sN, g_sM)\sim{\rm Large}\\
&&(g_s, g_sM^2/N, g_sN_f, M/N)\sim{\rm Small}\label{limitesdeSUGRA}
\nd
but we will add that the deviation from the extremal case computed previously will be small. This will become more explicit in \eqref{rrange}. This will allow us to take the previous solution of chapter \ref{DtermPaper} and \cite{DtermPaper} as the zeroth order solution and to solve the supergravity equations for the first order corrections in the order of the small parameters above. More precisely, this allows us to understand that $F(r)$ will be the resolved conifold to zeroth order and the flux to be given by \eqref{F3formDterm} and \eqref{H3formDterm}, plus corrections that will vanish if we were to take $r_h=0$ or zero temperature.

\begin{figure}[h]
\begin{center}
\includegraphics[height=7cm,angle=0]{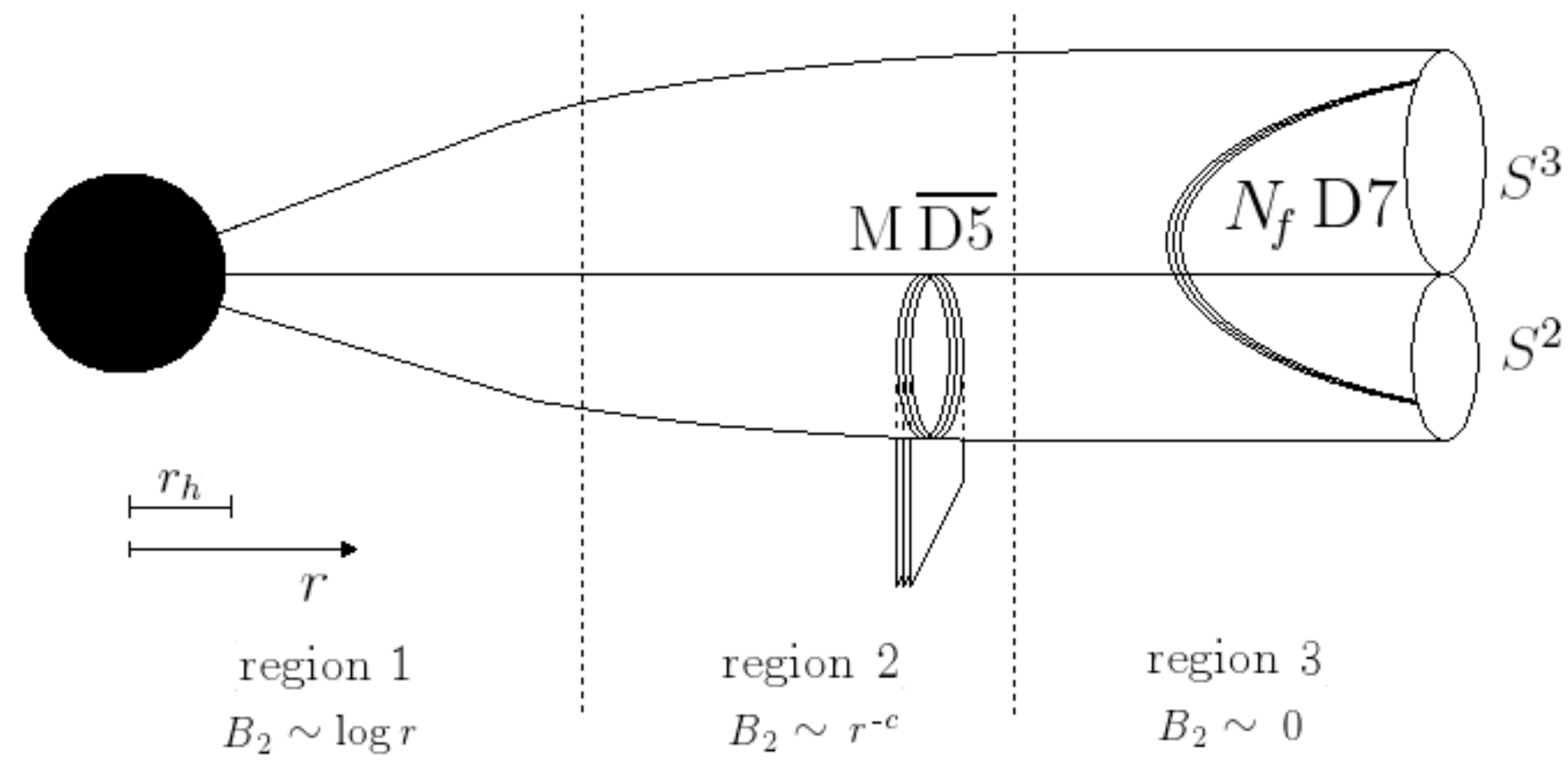}
\caption{In \cite{Mia:2009wj,Mia:2010zu}, it was conjectured that UV completion of KS-type solutions (region 1) could be accomplished by cancelling the 5-brane charge in the UV. This would allow for a decay of the 3-form flux in the intermediate (region 2) and a well defined AdS completion in the UV, without logarithmic running (region 3). }
\label{ZonesUVCompl}
\end{center}
\end{figure}

The work of \cite{Mia:2009wj,Mia:2010zu} further addressed the important question of the UV completion of this solution. As the KS duality cascade and QCD have similar IR behavior ($SU(M)$ color with fundamental matter), their UV behaviors are somehow different. While the couplings of QCD appear to be asymptotically free in the UV, an inverse KS cascade towards the UV immediately implies that one of the two gauge theory coupling will go to infinity (referred to as a Landau Pole). In order to avoid this problem, a modification of the construction is necessary towards the UV. Seen from the supergravity perspective, this Landau pole appears because the $B_2$ flux in the KT set up is logarithmic in the RG scale (the radial direction $r$) and thus goes to infinity. It was conjectured in \cite{Mia:2009wj,Mia:2010zu} that this could be cancelled by adding (p,q)-branes, at a certain distance far away in the UV, which would make the total 3-form flux now go to zero as we move towards the UV. The gravity picture is illustrated in Figure \ref{ZonesUVCompl} while the brane picture is illustrated in Figure 4.2 \cite{FangMia}. 

\begin{figure}[htb]\label{ImageSansChiffre}
        \begin{center}
		\vspace{- 0.5 cm}
		\centering
		\subfloat[N D3 branes placed at the tip of the conifold]{\includegraphics[width=0.33\textwidth]{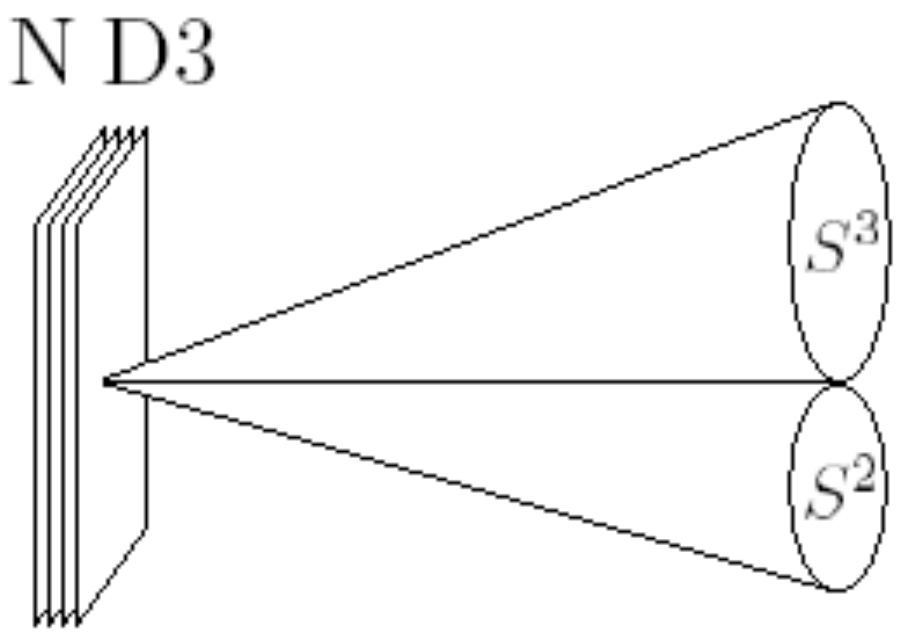}}
		\subfloat[M D5 branes wrapping the vanishing two cycle of the conifold]{\includegraphics[width=0.33\textwidth]{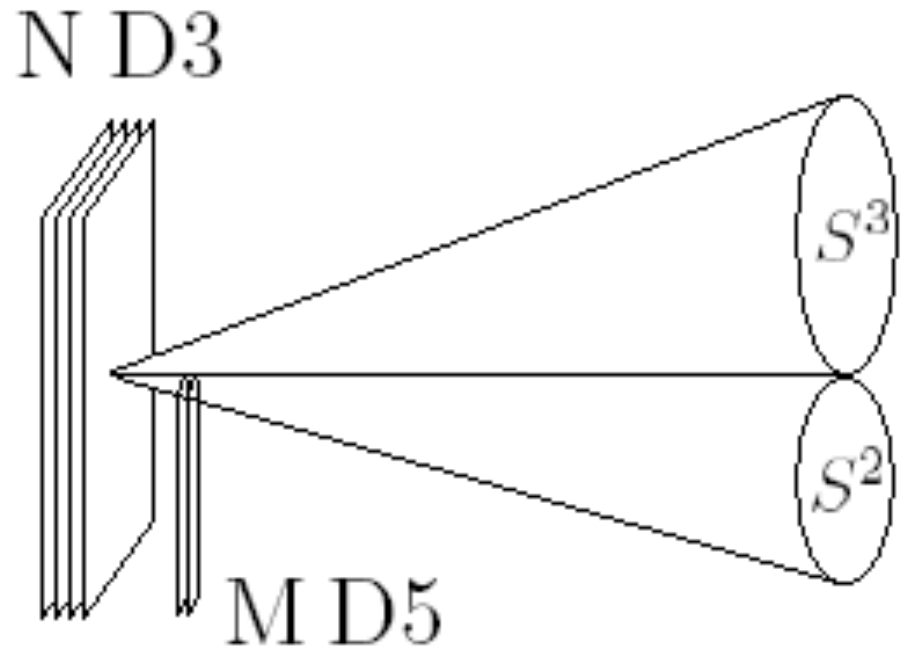}}
        \subfloat[M Anti-5 branes separated from the D5 and D3 branes but also located at the tip $r=0$.]{\includegraphics[width=0.33\textwidth]{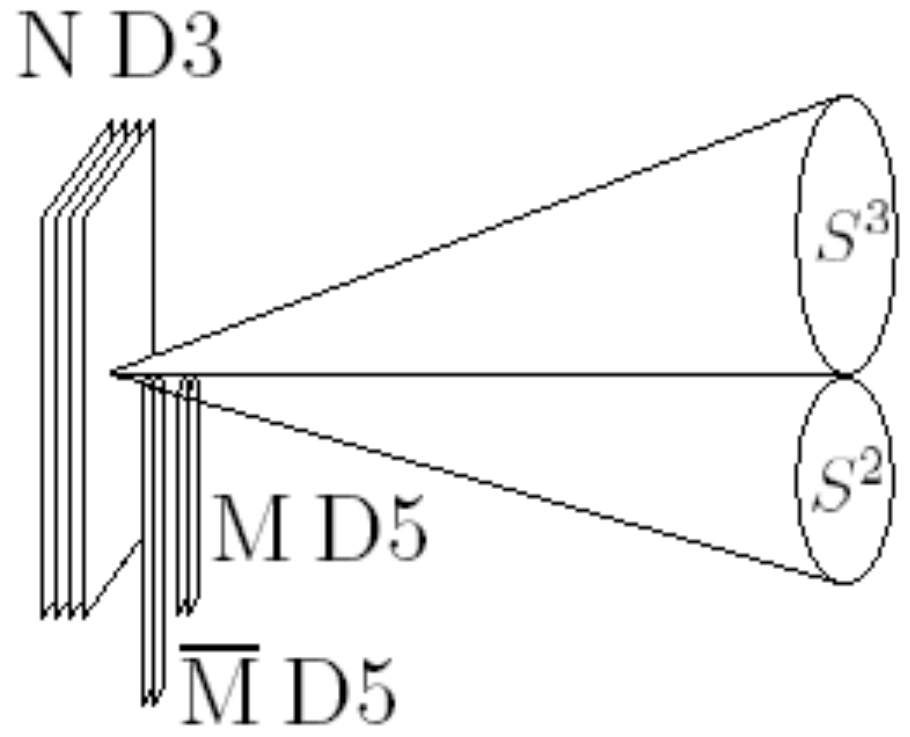}}
		\caption{{Brane construction of conformal field theory, non-conformal field theory and UV completed non-conformal field theory }}
\end{center}

\end{figure}

How exactly we expect this cancellation to be realized for the flux in the UV was further conjectured to take the form 
\bg\label{expansionrc}
\frac{1}{r^c}\approx 1-\ln(r)c+\frac{1}{2}\ln ^2(r)c^2+\ldots , ~~~~~~~{\text{for}}~c\ll 1
\nd
where the role of c would be played by one of the parameters in \eqref{limitesdeSUGRA}. Defining $c$ appropriately, this relation could easily be used to replace the logarithmic dependence of the fluxes on $r$. The 3-form fluxes should then take the form of \eqref{premierF3} and \eqref{H33}. This obviously allows these fluxes to decay at large radius. This form of the fluxes is further allowed as a solution in the far IR as we can always choose to cut the parameter expansion to first order in \eqref{limitesdeSUGRA}, where we recover our previous solution directly. Also, the decaying of the 3-form flux can be associated to a decrease in the number of 5-branes. This appears in the formulation of the flux in the interpolating region as $M(r)$. $M(r)$ being constant and non-zero corresponds to the IR region 1.

Now, once this is taken care of, this geometry can be capped by an asymptotically AdS geometry (region 3) where only the axion-dilaton $\tau$ is running and D7-branes live according to F-theory's law. 

To summarize, this work made progress towards defining the input to the equations of motion in the presence of a black hole and non-ISD fluxes, However, the form of their backreactions on the geometry were only conjectured and the equations not precisely solved. It was suggested that the non-extremality and the flavor branes would alter the geometry by making the resolution of the conifold squashed, {\it i.e.} running with the radial direction, thereby affecting the 3-form flux and making it non-ISD. Finding the precise way that the flux is affected is the object of the next chapter.

%%%%%%%%%%%%%%%%%%%%%%%%
%Ajouter arrive quoi ans le UV
%Landau Poles et un des 2 coupling explose dans le UV
%UV of OKS = runaway behavier pendant que QCD est assymptotically free
%UV de KS et QCD sont different mais IR sont semblable aprce les 2 ont SU(M) avec fundammental matter

%expansion r^-c=1-clogr+...

%in Region 2 there are anti five-brane sources that make $F_3$ non-closed

%M(r)

%pas calabi-yau

%flux diviser entre ouyang et backreactions

%metrique total et interne voir noms premiere ligne de 2e apragraphe

%flux de la forme 2.55

\section{Towards an Exact Solution to the Thermal Supergravity Equations}

In \cite{OurLastPaper}, we took a closer look at the full supergravity equations of motion associated to the setup presented in the previous section \cite{Mia:2009wj,Mia:2010zu}. Significant improvements towards a solution with all the necessary backreactions were made. Here, we will review the most important features of the results obtained as it is the introduction to the next chapter. 

The approach taken to solve type IIB supergravity equations of motion is to follow the same guidelines as in GKP \cite{GKP}, presented in section \ref{SectionGKP}. We take the same metric as in \eqref{bhmet1} with \eqref{inmate} as the squashed internal directions. The type IIB action \eqref{tubra} and equations of motions \eqref{ricci_T} still read the same as in GKP, but now with the above metric inserted in the Ricci tensor. Keep in mind that if we take the non-extremal factor to be $g(r)=1$, we recover the extremal case of chapter \ref{DtermPaper} where the flux is ISD and the resolution factor is constant. Let us now have a look at the equations of motion.

The 5-form is given by \eqref{5form}  
\bg  \label{5formv2}
\widetilde{F}_5=(1+\ast_{10})d\alpha\wedge dx^0\wedge dx^1\wedge dx^2\wedge dx^3 
\nd 
where we chose $\alpha=e^{4A}$. The Bianchi identity, coming from the $C_4$ equation, reads
\bg \label{bianchi5} 
d\widetilde{F}_5= H_3\wedge F_3 ~,
\nd
where we neglected the local sources. We took the warp factor to be of the form $h=h^0+h^1$, where $h^0$ is the first contribution to $h$ and given by \eqref{warpfacteurOuyang}, while $h^1$ is a correction of order \eqref{limitesdeSUGRA}. Using our metric ansatz, the Bianchi identity now reads
\bg \label{EQA}
&&\Bigg[\partial_r\partial_r h^1+ \frac{1}{g}\partial_{\theta_i}\left(\bar{g}_0^{\theta_i\theta_i}\partial_{\theta_i} h^1\right)
+ \frac{r_h^4/r^4}{g}\partial_{\theta_i}\left(\bar{g}_0^{\theta_i\theta_i}\partial_{\theta_i} h^0\right)\Bigg]
r^5 + 5r^4\partial_r h^1=4L^4 \partial_r F\nonumber\\
\nd
The presence of $L^4$ on the RHS tells us that the order of correction in $F$ is even smaller then in $h^1$.

Looking at the Einstein equations, a first relation we obtain is
\bg \label{BHfactorA1}
R_t^t-R_x^x=0
\nd
which leads to
\bg\label{BHfactorA}
\widetilde{\triangledown}^2 B-3\widetilde{g}^{mn} \partial_m B \partial_n B=-\frac{1}{3}e^{3B}\widetilde{\triangledown}^2e^{-3B}=0
\nd
To first order, this gives us the usual non-extremal factor \eqref{nonextrfact}. But the inclusion of the squashing will slightly modify it to 
\bg
e^{2B}=1-\frac{\bar{r}^4_h}{r^4}+G
\nd
where G is small and at least ${\cal O}(g_sM,g_sN_f)$. 

Combining the Bianchi identity with the trace of the Einstein equations along the spacetime directions, we get:
\begin{eqnarray}\label{GKP_BH}
\widetilde{\triangledown}^2(e^{4A+B}-\alpha)&=&\frac{e^{2A-B}}{6\textrm{Im}\tau}|{i}
G_3- \ast_6G_3|^2+e^{-6A-3B}|\partial (e^{4A+B}-\alpha)|^2\nonumber\\
&&+3e^{-2A-2B}\partial_mB\partial^m(e^{4A+B}-\alpha)+\textrm{local
source}
\end{eqnarray} 
This is interesting as it is a non-extremal generalization of \eqref{eqnGKPextremal} from GKP. 

Note that by choosing $\alpha=e^{4A}$, we no longer get a cancellation of every individual terms of \eqref{GKP_BH}. This includes the term for the 3-form flux, $G_3$, which will now have to be non-ISD in order to satifsy this equation. This is in complete accordance with \cite{KT-non-ex1,KT-non-ex2} which anticipated that a non-ISD 3-form flux was necessary at non-zero temperature. The choice $\alpha=e^{4A}$ also brings us to limit our result to non-compact manifolds. 

It was further demonstrated  in \cite{OurLastPaper} that the 3-form flux did not contribute to \eqref{GKP_BH} to first order in the squashing factor $F$. This means that the details of the deviation from the extremal ISD case (corresponding to the Ouyang embedding) are not crucial here since we have restricted ourselves to first order corrections. More precisely, the 3-form fluxes deviate from Ouyang's fluxes as 
\bg\label{onguflux}
F_3 ~\sim ~  M[1+ {\cal O}(F)],~~~~~~~~ H_3 ~ \sim ~ g_sM[1+ {\cal O}(F)]
\nd
This will contribute to the Bianchi identity \eqref{bianchi5} as
\bg
\frac{1}{N}H_3\wedge F_3=\frac{1}{N}H^{(0)}_3\wedge F^{(0)}_3+{\cal O}(g_sM^2/N\cdot F)
\nd
which is a second order correction to the Ouyang solution for $H^{(0)}_3$ and $F^{(0)}$. These will further backreact on the warp factor $e^{4A}$ when we will look for higher order solutions. This is another way that temperature will enter the geometry as the squashed resolution factor will depend on the black hole horizon, $F\sim a^2\sim a^2(r_h)$. These corrections are very interesting as they tell us how flux compactification gets modified by thermal effects. The method to compute these fluxes will be shown in the next chapter.

This greatly simplifies our equation \eqref{GKP_BH} as it shows that the $G_3$ flux does not contribute to first order. Surprisingly, calculations in \cite{OurLastPaper} further showed that the other terms of \eqref{GKP_BH} are also exact to first order. This means that the first order contribution of the two equations that combined into \eqref{GKP_BH} cancel each other. In order to solve both the Bianchi identity \eqref{EQA} and the trace of the Einstein equations, we can instead solve for the Bianchi identity and their combination \eqref{GKP_BH}. If we only look at first order, \eqref{GKP_BH} is satisfied automatically, as we just saw, leaving only the Bianchi identity \eqref{EQA} as our defining equation. 

The other Einstein equations to be solved are the ones along the internal directions. Since $G_3$ is not self-dual, it is not as obvious how to solve for each of them individually. However, limiting our calculations to the region
\begin{equation}\label{rrange} 
r ~ \ge ~ r_h \left({N\over M}\right)^{1/4} ~,
\end{equation}
these Einstein equations may be combined and simplified to 
\bg \label{TrRmnA}
&&\frac{\widetilde{R}}{6}+\frac{4}{3}\widetilde{g}^{mn}\partial_mA \partial_nA \left(e^{-2B}-1\right)+\frac{\widetilde{g}^{mn}}{6}\left(
3\widetilde{\triangledown}_m\partial_nB+\partial_mB\partial_nB\right)\nonumber\\
&& ~~~~~-\frac{\widetilde{g}^{mn}}{3} \partial_{(m} A \partial_{n)} B
=\frac{\widetilde{g}^{mn}\partial_m\tau \partial_n\bar\tau}{12|{\rm Im}\tau|^2}
\nd
where $\widetilde{R}=\widetilde{g}^{mn}\widetilde{R}_{mn}$ and we have ignored all local sources.

Interestingly, \eqref{EQA}, \eqref{BHfactorA} and \eqref{TrRmnA} are exactly the three equations we need to fully determine our three unknowns: $e^{4A}$, $e^{2B}$ and $F$. A numerical calculation has been performed, showing that our ansatz effectively allowed for a solution. See \cite{OurLastPaper} for the numerical results. 

Two results are to be noted. The first is that the resolution $F$ was computed to be negative. It might seem unnatural that we get $F\sim a^2 < 0$. However, since $F$ appears as $(1+F)$ in the metric \eqref{inmate} and that $F\ll 1$, this will not be a problem. As we will see in the next chapter, this will not be a problem for the 3-form fluxes either. The second interesting result is less obvious from the summary presented here, but it was shown that $a^2$, associated with the resolution $F$, is given by
\bg\label{respagi}
a^2 = && a_0^2 + \frac{5g_s M^2 p_{11} r_h^2}{32\pi N} + \frac{g_s M^2}{N}\frac{r_h^2}{4\pi}
\left[p_{12}{\rm log}~r + p_{13}{\rm log}^2 r\right] \nonumber\\
&& + {1\over 4\pi}\left({g_sM^2\over N}\right)\left(g_sN_f\right) r_h^4
\left(p_{14}{{\rm log}~r\over r^2} + {p_{15} \over r^2}\right)
{\rm log}\left(\rm{sin}\frac{\theta_1}{2}\rm{sin}\frac{\theta_2}{2}\right)
\nd
where $p_{ij}$ are constants determined in the details of the expansion and $a_0^2$ is the bare resolution parameter. This is actually a precise expansion of how the black hole factor $r_h$ enters the geometry through the resolution factor.

The first part of \cite{OurLastPaper} then provided a good geometric ansatz allowing to solve for all supergravity equations at first order and at high temperature.

~\newline\\
\begin{flushleft}
{\it Summary}
\end{flushleft}

In this chapter, we developed models that brought us several steps closer to a full gravitational dual to QCD. Building on top of the flavour that we added to the KS solution in chapter \ref{DtermPaper}, here we added temperature. We now have a gravity dual of a gauge theory with colours, flavours, temperature and which leads to confinement in the IR. However, our solution is still only valid at large temperature, large number of colours and we still do not have a precise formulation of the thermal fluxes. This is an issue that will be addressed in detail in the next chapter.

\chapter{The Backreacted Thermal 3-Form Fluxes \label{actfluxes}}
%%%%%%%%%%%%%%%%%%%%%%%%%%%%%%%%%%%%%%%%%%%%%%%%%%%%%%%%%%%%%%%%%%%%%%

In the earlier chapter, a detailed derivation of the geometry of the non-extremal limit of a Klebanov-Strassler type solution with a background warped resolved conifold was successfully derived. The ansatz used to solve for the fluxes of the form \eqref{onguflux} where the three-form fluxes was divided into two pieces: one leading term coming from the known Ouyang fluxes, and the subleading corrections from the various backreactions. For these corrections, both the RR and NS three-form fluxes were shown to receive contributions from the bare resolution parameter $a_0^2$ and the $g_sM^2/N$ terms in addition to the ${\cal O}(r_h)$ terms. In the following chapter, we will not only justify this but also provide the form of the three-form fluxes including the aforementioned corrections. Note that our analysis will not be affected by the constraint \eqref{rrange} that we had to impose to solve EOMs in the above subsection, allowing the radial coordinate to take all values above $r_h$. 

As we just mentioned, \cite{Mia:2009wj} and \cite{jpsi1} conjectured how, using the metric \eqref{bhmet1} and \eqref{inmate}, the Ouyang flux of chapter \ref{DtermPaper}, should naturally be extended to become non-extremal, non-ISD and satisfy the supergravity equations of motion. More precisely, the ansatz for the RR three-form flux $\widetilde{F}_3 \equiv F_3 - C_0 H_3$ takes the following form:
\begin{eqnarray}\label{premierF3}
&&{\widetilde F}_3  = \left(\widetilde{a}_o - {3 \over 2\pi r^{g_sN_f}} \right)
\sum_\alpha{2M(r)c_\alpha\over r^{\epsilon_{(\alpha)}}}
\left({\rm sin}~\theta_1~ d\theta_1 \wedge d\phi_1-
\sum_\alpha{f_\alpha \over r^{\gamma_{(\alpha)}}}~{\rm sin}~\theta_2~ d\theta_2 \wedge
d\phi_2\right)\nonumber\\
&&~~ \wedge~ {e_\psi\over 2}-\sum_\alpha{3g_s M(r)N_f d_\alpha\over 4\pi r^{\sigma_{(\alpha)}}}
~{dr}\wedge e_\psi \wedge \left({\rm cot}~{\theta_2 \over 2}~{\rm sin}~\theta_2 ~d\phi_2
- \sum_\alpha{g_\alpha \over r^{\rho_{(\alpha)}}}~
{\rm cot}~{\theta_1 \over 2}~{\rm sin}~\theta_1 ~d\phi_1\right)\nonumber \\
&& -\sum_\alpha{3g_s M(r) N_f e_\alpha\over 8\pi r^{\tau_{(\alpha)}}}
~{\rm sin}~\theta_1 ~{\rm sin}~\theta_2 \left({\rm cot}~{\theta_2 \over 2}~d\theta_1 +
\sum_\alpha{h_\alpha \over r^{\delta_{(\alpha)}}}~
{\rm cot}~{\theta_1 \over 2}~d\theta_2\right)\wedge d\phi_1 \wedge d\phi_2\label{brend1}
\end{eqnarray}
with $\widetilde{a}_o = 1 + {3\over 2\pi}$ and is defined in the intermediate region $r_{\rm min} < r < r_o$. The additional contributions to \eqref{brend1} are all proportional to powers of $r_h$, as they should vanish in the zero temperature ISD limit. Finally, we will refine the definition of the quantity $\epsilon_{(\alpha)}$  in the following way:
\bg\label{epdef}
\epsilon_{(\alpha)} ~ = ~ \alpha + \sum_n {b_{\alpha n}\over r^n}
\nd
with $b_{\alpha n}$ are functions of $g_sN_f, M$ and the horizon radius $r_h$. The coefficients $b_{\alpha n}$ are assumed to be small such that the expansion \eqref{expansionrc} is valid. The coefficients
$\rho_{(\alpha)}, \sigma_{(\alpha)}, \delta_{(\alpha)}$, etc are also defined in a similar fashion.
The other coefficients, for example $c_\alpha, ... h_\alpha$ would again be functions of $g_sN_f$ and $r_h$, but also of the resolution factor $a$ (including the internal angular coordinates).
The resolution factor appears from the gravity dual that was considered in \cite{Mia:2009wj}, i.e a resolved warped deformed conifold with the resolution factor, can be viewed as a function dependent on the horizon radius $r_h$. We will argue this in details below. The coefficients $b_{\alpha n}$ can be represented in terms of the following matrix:
\bg\label{za}
b_{\{\alpha n\}} ~ \equiv ~ \begin{pmatrix} b_{00}& ~b_{01} & ~b_{02}& ~b_{03} & ....\\
b_{10} & ~b_{11} & ~b_{12} & ~b_{13} & ....\\
b_{20} & ~b_{21} & ~b_{22} & ~b_{23} & ....\\
b_{30} & ~b_{31} & ~b_{32} & ~b_{33} & ....\\
... & ... & ... & ... & ....\\
b_{m0} & ~b_{m1} & ~b_{m2} & ~b_{m3} & ....
\end{pmatrix}
\nd
The elements of the matrix $b_{\alpha n}$ can be determined in terms of the $c_0, c_1, c_2, ..$ coefficients that
appears in the expansion ${c_\alpha\over r^{\epsilon_{(\alpha)}}}$. This is one reason for writing the various
powers of $r$ using different symbols. For example $\sigma_{(\alpha)}$ will have a similar expansion as \eqref{epdef}, but with a different matrix. The various elements of the matrix will now be determined in terms of  $d_0, d_1, d_2,$ etc as one would expect. For the first case, we have managed to determine $c_\alpha$ up to few terms by matching with the first order terms of chapter \ref{DtermPaper}. This was done by taking the small $b_{\alpha n}$ expansion of $\sum_\alpha{c_\alpha\over r^{\epsilon_{(\alpha)}}}$ and comparing order by order. The result gives us:
\bg\label{calpha}
c_0 ~ = ~ 1 + {\cal O}(r_h), ~~~ c_1 ~ = ~ {\cal O}(r_h), ~~~
c_2 ~ = ~ {9a^2 g_sN_f\over 2\pi \zeta^2} \left(1 - {3\over 2}{\rm log}~\zeta\right) + {\cal O}(r_h)
\nd
where $a$ is the resolution factor and $\zeta$ is a parameter whose importance will become apparent soon, but could also simply be set to 1. Once we know $c_\alpha$, it is not too difficult to get the relations between the various components of the matrix \eqref{za}. One may now show that the components $b_{\alpha n}$ satisfy:
\bg\label{kdev}
&&b_{00} ~ = ~ b_{01} ~ = ~ b_{10} ~ = ~ {\cal O}(r_h) \nonumber\\
&& c_0 b_{02}~ + ~ c_1 b_{11}~ + ~ c_2 b_{20}~ = ~ -{27 a^2 g_s N_f\over 4\pi \zeta^2} ~ + {\cal O}(r_h)\nonumber\\
&&2 c_0 b_{00} b_{01}~ + ~ c_1 b_{10}~ = ~ {\cal O}(r_h)\nonumber\\
&&c_0 b^2_{01}~ + ~ 2c_0b_{00} b_{02}~ + ~ 2c_1 b_{10}b_{11}~ + ~ c_2 b_{20}~ = ~ {\cal O}(r_h)
\nd
Putting these first order terms back together, one may show that:
\bg\label{relshow}
\sum_{\alpha} {c_\alpha\over r^{\epsilon_{(\alpha)}}} ~ &=& ~ 1 + {9 a^2 g_sN_f \over 2\pi \zeta^2 r^2}
\left(1 - {3\over 2} {\rm log}~\zeta\right) - {27 a^2 g_s N_f \over 4\pi \zeta^2} \cdot {{\rm log}~r\over r^2}
+ {{\cal O}(r_h) \over r^2} + {\cal O}(r_h)\nonumber\\
 & \equiv  & ~ 1 + {9g_sN_f\over 4\pi} \cdot {a^2(r_h, g_sN_f)\over
(\zeta r)^2}\cdot \left[2- 3 {\rm log}~(\zeta r)\right] + {\cal O}(r_h, g_s^2N_f^2)
\nd
which is consistent with what was discussed in \cite{Mia:2009wj}. Namely, the resolution parameter $a$ can be thought of as a function of ($r_h, M, g_sN_f$), including the radial and the angular directions, i.e
\bg\label{resfac}
a^2 ~ = ~ a^2(r_h, M, g_s N_f) ~ = ~ a^2_0 + \sum_{\alpha = 1}^\infty {b_\alpha g_s^\alpha[M(r)r_h]^{\alpha
+ 1}\over N^\alpha r^{\epsilon_{(\alpha)}}}
\nd
with $b_\alpha$ being functions of the angular directions so that this is consistent with \eqref{respagi}.
The relation between these two forms of $a^2$ can be determined by comparing \eqref{resfac} with \eqref{respagi} (note that \eqref{respagi} implies $b_0 = 0$). 

The other factors appearing in the flux \eqref{premierF3} are also given in terms of series expansions similar to \eqref{relshow} above:
\bg\label{monfuen}
&&\sum_{\alpha} {d_\alpha\over r^{\sigma_{(\alpha)}}} ~ = ~ 1
+ {18 a^2(r_h, M, g_sN_f) {\rm log}~(\zeta r)\over (\zeta r)^2} +
{\cal O}(r_h, M, g_sN_f)\nonumber\\
&&\sum_{\alpha} {e_\alpha\over r^{\tau_{(\alpha)}}} ~ = ~ 1
- {18 a^2(r_h, M, g_sN_f) {\rm log}~(\zeta r)\over (\zeta r)^2} +
{\cal O}(r_h, M, g_sN_f)
\nd

Looking back at the flux \ref{premierF3}, note that there are also {\it squashing} factors given by ${f_\alpha\over r^{\gamma_{(\alpha)}}}, {g_\alpha\over r^{\rho_{(\alpha)}}}$ and ${h_\alpha\over r^{\delta_{(\alpha)}}}$. These squashing factors distort the spheres and therefore affect the fluxes on them. These are exactly the factors we are interested in as they tell us how the fluxes become non-ISD once non-extremality and squashing are set in. These squashing factors can also be computed by the same procedure and are given at first order by:
\bg\label{squash}
\sum_{\alpha} {f_{\alpha}\over r^{\gamma_{(\alpha)}}} ~& = &~ 1 - {729\over 32\pi^2} \cdot
{g_s^2 N_f^2 a^4(r_h, M, g_sN_f)\over (\zeta r)^4} \cdot {\rm log}~(\zeta r)\left[2-3 {\rm log}~(\zeta r)\right] +
{\cal O}(r_h, M, g_s^2N_f^2)\nonumber\\
&& + {81\over 8\pi} \cdot
{g_sN_fa^2(r_h, M, g_sN_f) {\rm log}~(\zeta r)\over (\zeta r)^2}\\
& = & ~ 1 + {81\over 2} \cdot
{g_s N_f a^2(r_h, M, g_sN_f) {\rm log}~(\zeta r) \over 4\pi r^2 + 9 g_s N_f a^2(r_h, M, g_sN_f) [2 -
3~{\rm log}~(\zeta r)]}
+ {\cal O}(r_h, M, g_s^2 N_f^2)\nonumber\\
\sum_{\alpha} {g_{\alpha}\over r^{\rho_{(\alpha)}}} ~& = &~ 1
+ {36 a^2(r_h, M, g_sN_f) {\rm log}~(\zeta r)\over (\zeta r)^3}
- {648 a^4(r_h, M, g_sN_f) {\rm log}^2(\zeta r)\over (\zeta r)^5} + {\cal O}(r_h, M, g_s^2N_f^2)\nonumber\\
& = & ~ 1 + {36 a^2(r_h, M, g_sN_f)~{\rm log}~(\zeta r) \over (\zeta r)^3 +
18 a^2(r_h, M, g_sN_f) \zeta r ~{\rm log}~(\zeta r)} + {\cal O}(r_h, M, g_s^2 N_f^2)\nonumber\\
\sum_{\alpha} {h_{\alpha}\over r^{\delta_{(\alpha)}}} ~& = &~ 1
+ {36 a^2(r_h, M, g_sN_f) {\rm log}~(\zeta r)\over (\zeta r)^2}
+ {648 a^4(r_h, M, g_sN_f) {\rm log}^2(\zeta r)\over (\zeta r)^4} + {\cal O}(r_h, M, g_s^2N_f^2)\nonumber\\
& = & ~ 1 + {36 a^2(r_h, M, g_sN_f)~{\rm log}~(\zeta r) \over (\zeta r)^2 +
18 a^2(r_h, M, g_sN_f)~{\rm log}~(\zeta r)} + {\cal O}(r_h, M, g_s^2 N_f^2)\nonumber
\nd
The far IR physics is then determined from \eqref{premierF3} and the squashing factors \eqref{squash} by making the replacement $(\zeta r)\to r$ to the radial coordinate (or taking $\zeta=1$). Concerning the issue raised in section (\ref{sectionMia}) about the sign of $a^2$, note that all the flux components are expressed in terms of $a^2$ and therefore the sign of $a^2$ can be directly inserted here. 

Considering all the above, this gives us a more detailed formulation of a result obtained in \cite{Mia:2009wj}. Namely, the 3-form flux takes the general form
\begin{eqnarray}
{\widetilde F}_3 & = & 2M {\bf A_1} \left(1 + {3g_sN_f\over 2\pi}~{\rm log}~r\right) ~e_\psi \wedge
\frac{1}{2}\left({\rm sin}~\theta_1~ d\theta_1 \wedge d\phi_1-{\bf B_1}~{\rm sin}~\theta_2~ d\theta_2 \wedge
d\phi_2\right)\nonumber\\
&& -{3g_s MN_f\over 4\pi} {\bf A_2}~{dr\over r}\wedge e_\psi \wedge \left({\rm cot}~{\theta_2 \over 2}~{\rm sin}~
\theta_2 ~d\phi_2
- {\bf B_2}~ {\rm cot}~{\theta_1 \over 2}~{\rm sin}~\theta_1 ~d\phi_1\right)\nonumber \\
&& -{3g_s MN_f\over 8\pi}{\bf A_3} ~{\rm sin}~\theta_1 ~{\rm sin}~\theta_2 \left({\rm cot}~{\theta_2 \over 2}~d\theta_1 +
{\bf B_3}~ {\rm cot}~{\theta_1 \over 2}~d\theta_2\right)\wedge d\phi_1 \wedge d\phi_2\label{brend}
\end{eqnarray}
where we have taken $M(r) ~\to ~ M$ in the far IR, and
the various coefficients ${\bf A}_i, {\bf B}_i$ are related to \eqref{relshow} and \eqref{squash} as:
\bg\label{fiveclap}
&& \sum_\alpha {c_\alpha \over r^{\epsilon_{(\alpha)}}} ~ \equiv ~ {\bf A}_1, ~~~
\sum_\alpha {d_\alpha \over r^{\sigma_{(\alpha)}}} ~ \equiv ~ {\bf A}_2, ~~~
\sum_\alpha {e_\alpha \over r^{\tau_{(\alpha)}}} ~ \equiv ~ {\bf A}_3 \nonumber\\
&& \sum_\alpha {f_\alpha \over r^{\gamma_{(\alpha)}}} ~ \equiv ~ {\bf B}_1, ~~~
\sum_\alpha {g_\alpha \over r^{\rho_{(\alpha)}}} ~ \equiv ~ {\bf B}_2, ~~~
\sum_\alpha {h_\alpha \over r^{\delta_{(\alpha)}}} ~ \equiv ~ {\bf B}_3
\nd
As we mentioned before, the additional contributions to \eqref{fiveclap}, included as ${\bf A}_i,{\bf B}_i$ , are all proportional to powers of $r_h$ and others sources, as they vanish in the extremal ISD case. It is these asymmetry factors that incorporate the corrections into the fluxes. Needless to say, the functional form for ${\widetilde F}_3$ is consistent with \eqref{onguflux}.

The NS three-form flux $H_3$ is now more interesting. Unlike $\widetilde{F}_3$, it has to be closed. This is a crucial property of $H_3$ that we will exploit to help us define its exact form. When the resolution parameter $a$ and $M$ are just constants, it is easy to construct a closed $H_3$. In the presence of non-constant $a$ and $M(r)$, finding a closed $H_3$ is much less trivial. For our case
$H_3$ is given by\footnote{We corrected a minor typo in \cite{jpsi1}.}:

\begin{eqnarray}\label{H33}
H_3 &=&  \sum_\alpha {6g_s M(r) k_\alpha \over r^{\beta_{(\alpha)}}}\Bigg[1+\frac{1}{2\pi} -
\frac{\left({\rm cosec}~\frac{\theta_1}{2}~{\rm cosec}~\frac{\theta_2}{2}\right)^{g_sN_f}}{2\pi r^{{9g_sN_f\over 2}}}
\Bigg]~ \left[dr + \sum_i {\cal O}(r_h)d\sigma_i\right]\nonumber\\
&&\wedge \frac{1}{2}\Bigg({\rm sin}~\theta_1~ d\theta_1 \wedge d\phi_1
-\sum_\alpha{p_\alpha \over r^{\kappa_{(\alpha)}}} ~{\rm sin}~\theta_2~ d\theta_2 \wedge d\phi_2\Bigg)\nonumber\\
&+&\sum_\alpha \frac{3g^2_s M(r) N_f l_\alpha}{8\pi r^{\theta_{(\alpha)}}}
\Bigg(\frac{dr}{r}\wedge e_\psi -\frac{1}{2}de_\psi \Bigg) \wedge \Bigg({\rm cot}~\frac{\theta_2}{2}~d\theta_2
-\sum_\alpha{q_\alpha \over r^{\xi_{(\alpha)}}}~{\rm cot}~\frac{\theta_1}{2}
~d\theta_1\Bigg) \nonumber\\
&+&g_s {dM(r) \over dr}
\left(b_1(r)\cot\frac{\theta_1}{2}\,d\theta_1+b_2(r)\cot\frac{\theta_2}{2}\,d\theta_2\right)\wedge e_\psi
\wedge dr\nonumber\\
&+&{3g_s\over 4\pi} {dM(r) \over dr}\left(1+g_sN_f -{1\over r^{2g_sN_f}} + {9a^2g_sN_f\over r^2} + b_3(r)\right)\nonumber\\
&&\cdot\log\left(\sin\frac{\theta_1}{2}\sin\frac{\theta_2}{2}\right)
 \sin\theta_1\,d\theta_1\wedge d\phi_1\wedge dr\nonumber\\
&-&{g_s\over 12\pi}{dM(r) \over dr} \Bigg(2 -{27a^2g_sN_f\over r^2} + 9g_sN_f -{1\over r^{16g_sN_f}} -
{1\over r^{2g_sN_f}}\Bigg) \nonumber\\
&&\cdot\log\left(\sin\frac{\theta_1}{2}\sin\frac{\theta_2}{2}\right)\sin\theta_2\,d\theta_2\wedge d\phi_2\wedge dr \nonumber\\
&-&
{g_sb_4(r)\over 12\pi}{dM(r) \over dr}~ \log\left(\sin\frac{\theta_1}{2}\sin\frac{\theta_2}{2}\right)
\sin\theta_2\,d\theta_2\wedge d\phi_2\wedge dr
\label{brend2}
\end{eqnarray}
with ($k_\alpha, ..., q_\alpha$) being constants, $\sigma_i \equiv (\theta_i, \phi_i)$
 and $b_n = \sum_m {a_{nm}\over r^{m + \widetilde{\epsilon}_m}}$
where $a_{nm} \equiv a_{nm}(g_sN_f, M, r_h)$ and $\widetilde{\epsilon}_m \equiv \widetilde{\epsilon}_m(g_sN_f, M, r_h)$.

Note that in addition to the simplest ${\cal O}(r_h)$ term that we mentioned above, there would be more terms
of the same order that vanish in the ISD limit. We might also wonder about terms of the form $da/dr$ and
$da/d\sigma_i$. From the form of \eqref{resfac} we see that these terms themselves are of ${\cal O}(M)$ so to this
order, they could either be absorbed in the $dM(r)/dr$ or ${\cal O}(r_h)$ terms. The various squashing factors have been computed like for ${\widetilde F}_3$ and are now given by:
\bg\label{sqafac}
\sum_\alpha {k_\alpha\over r^{\beta_{(\alpha)}}} ~ &= & 1 - {3a^2(r_h, M, g_sN_f) \over (\zeta r)^2} +
{\cal O}(r_h, M, g^2_sN^2_f)\nonumber\\
\sum_\alpha {l_\alpha\over r^{\theta_{(\alpha)}}} ~ &= & 1 + {36 a^2(r_h, M, g_sN_f) ~{\rm log}~(\zeta r)
\over \zeta r} +
{\cal O}(r_h, M, g^2_sN^2_f)\nonumber\\
\sum_\alpha {p_\alpha\over r^{\kappa_{(\alpha)}}} ~ &= & 1 + {3 g_sa^2(r_h, M, g_sN_f) \over (\zeta r)^2} +
{9 g_sa^4(r_h, M, g_sN_f) \over (\zeta r)^4}
+ {\cal O}(r_h, M, g^2_sN^2_f)\nonumber\\
& = & 1 + {3g_s a^2(r_h, M, g_sN_f) \over (\zeta r)^2 - 3 a^2(r_h, M, g_sN_f)} + {\cal O}(r_h, M, g_s^2 N_f^2)\nonumber\\
\sum_\alpha {q_\alpha\over r^{\xi_{(\alpha)}}} ~ &= & 1 + {72 a^2(r_h, M, g_sN_f)~{\rm log}~(\zeta r)
 \over \zeta r} -
{2592 a^4(r_h, M, g_sN_f)~{\rm log}^2~(\zeta r) \over (\zeta r)^2}
+{\cal O}(r_h, M, g^2_sN^2_f)\nonumber\\
& = & 1 + {72 a^2(r_h, M, g_sN_f)~{\rm log}~(\zeta r) \over \zeta r + 36 a^2(r_h, M, g_sN_f)~ {\rm log}~(\zeta r)}
+ {\cal O}(r_h, M, g_s^2 N_f^2)
\nd
Note again that the resolution parameter in all the coefficients appear as $a^2$. To study the far IR physics, we consider the limit $M(r) \to M$ and the expansions of \eqref{sqafac} are again respectively related to the ${\bf A_4, A_5, B_4}$ and ${\bf B_5}$ defined below. This then reproduces again the far IR result of \cite{jpsi1} as well as the expected ansatze \eqref{onguflux}. Namely:
\begin{eqnarray}
H_3 &=&  {6g_s {\bf A_4} M}\Bigg(1+\frac{9g_s N_f}{4\pi}~{\rm log}~r+\frac{g_s N_f}{2\pi}
~{\rm log}~{\rm sin}\frac{\theta_1}{2}~
{\rm sin}\frac{\theta_2}{2}\Bigg)\frac{dr}{r}\nonumber \\
&& \wedge \frac{1}{2}\Bigg({\rm sin}~\theta_1~ d\theta_1 \wedge d\phi_1
- {\bf B_4}~{\rm sin}~\theta_2~ d\theta_2 \wedge d\phi_2\Bigg)
+ \frac{3g^2_s M N_f}{8\pi} {\bf A_5} \Bigg(\frac{dr}{r}\wedge e_\psi -\frac{1}{2}de_\psi \Bigg)\nonumber  \\
&& \hspace*{1.5cm} \wedge \Bigg({\rm cot}~\frac{\theta_2}{2}~d\theta_2
-{\bf B_5}~{\rm cot}~\frac{\theta_1}{2} ~d\theta_1\Bigg)
\end{eqnarray}
with the necessary ${\cal O}(r_h)$ terms that vanish when the horizon radius vanishes.

In the far IR the closure of $H_3$ is again non-trivial because the resolution parameter $a$ is no longer a constant like in \cite{DtermPaper}, although $M(r) \to M$. All the information about the running of $a$ is captured in the coefficients ${\bf A_{4, 5}}$ and the squashing factors ${\bf B_{4,5}}$. In the following section we will determine the resolution parameter to ${\cal O}(M^2)$ in the IR. This means that up to this order, the closure of each component of $dH_3$ implies the following three conditions:
\bg\label{3cond} 
&& (i)~~\alpha_2 {\rm cot}~{\theta_2\over 2}
~{\partial {\bf A_5}\over \partial \theta_1} + \alpha_2 ~{\bf
A_5}{\rm cot}~{\theta_1\over 2} ~{\partial {\bf B_5}\over \partial
\theta_1}
+ {\cal O}(r_h, M^3, g_sN_f) = 0\\
&& (ii)~~\alpha_1 {\rm sin}~{\theta_1} ~{\partial {\bf A_4}\over \partial \theta_2} +
\alpha_2 {\rm cos}~{\theta_1}~{\rm cot}~{\theta_2\over 2} ~{\partial {\bf A_5}\over \partial \theta_1}
- \alpha_3 {\rm sin}~{\theta_1}~{\rm cot}~{\theta_2\over 2} ~{\partial {\bf A_5}\over \partial r} \nonumber\\
&& ~~~~~~~~~~~~~~~ +\alpha_2 {\bf A_5}
{\rm cos}~{\theta_1}~{\rm cot}~{\theta_1\over 2} ~{\partial {\bf B_5}\over \partial \theta_2} +
{\cal O}(r_h, M^3, g_sN_f) = 0\nonumber\\
&&(iii)~~\alpha_1~{\bf B_4}~{\rm sin}~{\theta_2}~{\rm cot}~{\theta_1\over 2}~{\partial {\bf A_4}\over \partial \theta_1}
- \alpha_1 ~{\bf A_4}~{\rm sin}~{\theta_2}~{\partial {\bf B_4}\over \partial \theta_1}
-\alpha_2 ~{\bf A_5}~{\rm cos}~{\theta_2}~{\rm cot}~{\theta_1\over 2} ~{\partial {\bf B_5}\over
\partial \theta_2}\nonumber\\
&& ~~~~~~~~~~~~~~~- \alpha_3 ~{\bf A_5}~{\rm sin}~{\theta_2}~{\rm cot}~{\theta_1\over 2}
~{\partial {\bf B_5}\over \partial r}
+ \alpha_2 ~{\rm cos}~{\theta_2}~{\rm cot}~{\theta_2\over 2} ~{\partial {\bf A_5}\over \partial \theta_1}\nonumber\\
&&~~~~~~~~~~~~~~~~ -\alpha_3 ~{\bf B_5}~{\rm sin}~{\theta_2}~{\rm cot}~{\theta_1\over 2}
~{\partial {\bf A_5}\over \partial r} + {\cal O}(r_h, M^3, g_sN_f) = 0\nonumber
\nd
where $\alpha_1, \alpha_2$ and $\alpha_3$ are defined as:
\bg\label{alpde}
&&\alpha_1 = {3g_s M \over 2}\Bigg(1+\frac{9g_s N_f}{4\pi}~{\rm log}~r+\frac{g_s N_f}{2\pi}
~{\rm log}~{\rm sin}\frac{\theta_1}{2}~
{\rm sin}\frac{\theta_2}{2}\Bigg)\nonumber\\
&&\alpha_2 = {3g_s^2 MN_f\over 8\pi r}, ~~~~~~~~ \alpha_3 = -{3g_s^2 MN_f\over 16\pi} = - {r\alpha_2\over 2}
\nd
The RR three-form flux ${\widetilde F}_3 \equiv F_3 - C_0 H_3$ is not closed, but it satisfies the condition $d{\widetilde F}_3 = -dC_0 \wedge H_3$, which is equivalent to the statement that $F_3$ is closed. As described in \cite{jpsi1}, the closure of $F_3$ is only in Region 1, as in Region 2 there are anti five-brane sources that make $F_3$ non-closed. 
Since $C_0$ is given by the Ouyang embedding \eqref{7brembDterm}, we have that 
\bg
dC_0=k(d\psi-d\phi_1-d\phi_2)
\nd
where $k=\frac{N_f}{4\pi}$. Implementing this with the squashing and non-extremality, the closure of $F_3$ in Region 1 implies the following nine conditions on the various coefficients of the three-form fluxes:
\bg\label{condon3}
&& (iv) ~~\beta_1 {\partial {\bf A_1}\over \partial r}~{\rm sin}~\theta_1
- \beta_2 ~ {\partial {\bf A_2}\over \partial \theta_1}~{\rm cot}~{\theta_1\over 2}~{\rm sin}~\theta_1
- \beta_2 ~{\bf A_2}~ {\partial {\bf B_2}\over \partial \theta_1}~{\rm cot}~{\theta_1\over 2}~{\rm sin}~\theta_1
\nonumber\\
&& + k\alpha_1 ~{\bf A_4}~{\rm sin}~\theta_1
+ k\alpha_2 ~{\bf A_5}~ {{\bf B_5}}~{\rm cot}~{\theta_1\over 2}~{\rm cos}~\theta_1
+ k\alpha_2 ~{\bf A_5}~ {{\bf B_5}}~{\rm cot}~{\theta_1\over 2} + {\cal O}(r_h, M^3, g_sN_f)= 0\nonumber\\
&& (v)~~ - \beta_1 ~{\bf B_1}~{\partial {\bf A_1}\over \partial r}~{\rm sin}~\theta_2
+ \beta_1 ~{\bf A_1}~ {\partial {\bf B_1}\over \partial r}~{\rm sin}~\theta_2
+ \beta_2 ~{\partial {\bf A_2}\over \partial \theta_2}~{\rm cot}~{\theta_2\over 2}~{\rm sin}~\theta_2\nonumber\\
&& - k\alpha_1 ~{\bf A_4}~{\bf B_4}~{\rm sin}~\theta_2
- k\alpha_2 ~{\bf A_5} ~{\rm cot}~{\theta_2\over 2}~{\rm cos}~\theta_2
+ k\alpha_2 ~{\bf A_5}~{\rm cot}~{\theta_2\over 2} + {\cal O}(r_h, M^3, g_sN_f) = 0\nonumber\\
&& (vi)~~ \beta_1 ~{\partial {\bf B_2}\over \partial r}~{\rm cos}~{\theta_2}~{\rm sin}~\theta_1
- \beta_2 ~{\partial {\bf A_2}\over \partial \theta_1}~{\rm cot}~{\theta_2\over 2}~{\rm sin}~\theta_1 ~{\rm cos}~\theta_1
+ \beta_2 ~{\bf B_2}~ {\partial {\bf A_2}\over \partial \theta_1}~{\rm cot}~{\theta_1\over 2}~{\rm sin}~\theta_1 ~
{\rm cos}~\theta_2\nonumber\\
&&- \beta_2 ~{\bf A_2}~ {\partial {\bf B_2}\over \partial \theta_1}~{\rm cot}~{\theta_1\over 2}~{\rm sin}~\theta_1~{\rm cos}~\theta_2
+ \beta_3 ~{\partial {\bf A_3}\over \partial r}~{\rm cot}~{\theta_2\over 2}
- k\alpha_1 ~{\bf A_4}~{\rm sin}~\theta_1\nonumber\\
&&~~~~~~ - k\alpha_2 ~{\bf A_5} ~{\bf B_5}~{\rm cot}~{\theta_1\over 2}~{\rm cos}~\theta_1
+ k\alpha_2 ~{\bf A_5}~{\bf B_5}~{\rm cot}~{\theta_1\over 2}~{\rm cos}~\theta_2 + {\cal O}(r_h, M^3, g_sN_f)= 0
\nonumber\\
&& (vii)~~
- \beta_1 ~{\bf B_1}~ {\partial {\bf A_1}\over \partial r}~{\rm sin}~\theta_2~{\rm cos}~\theta_1
- \beta_1 ~{\bf A_1}~ {\partial {\bf B_1}\over \partial r}~{\rm sin}~\theta_2~{\rm cos}~\theta_1
- \beta_2 ~{\partial {\bf A_2}\over \partial \theta_2}~{\rm cot}~{\theta_2\over 2}~{\rm sin}~\theta_2~{\rm cos}~\theta_1
\nonumber\\
&& - \beta_2 ~{\bf B_2}~ {\partial {\bf A_2}\over \partial \theta_2}~{\rm cot}~{\theta_1\over 2}~{\rm sin}~\theta_1~{\rm cos}~\theta_2
- \beta_2 ~{\bf A_2}~ {\partial {\bf B_2}\over \partial \theta_2}~{\rm cot}~{\theta_1\over 2}~{\rm sin}~\theta_1~{\rm cos}~\theta_2
+ \beta_3 ~{\bf B_3}~ {\partial {\bf A_3}\over \partial r}~{\rm cot}~{\theta_1\over 2}\nonumber\\
&& ~~~~~~~~+ \beta_3 ~{\bf A_3}~ {\partial {\bf B_3}\over \partial r}~{\rm cot}~{\theta_1\over 2}
 + k\alpha_2 ~{\bf A_5}~{\rm cos}~\theta_1~{\rm cot}~{\theta_2\over 2}
- k\alpha_1 ~{\bf A_4} ~{\bf B_4}~{\rm sin}~\theta_2 \nonumber\\
&&~~~~~~~~~~~~~~~~~~~- k\alpha_2 ~{\bf A_5}~{\rm cot}~{\theta_2\over 2}~{\rm cos}~\theta_2
+ {\cal O}(r_h, M^3, g_sN_f) = 0 \nonumber\\
&& (viii)~~
\beta_1 ~{\partial {\bf A_1}\over \partial \theta_2}~{\rm cos}~\theta_2
+ \beta_1 ~{\bf A_1}~ {\partial {\bf B_1}\over \partial \theta_1}~{\rm sin}~\theta_2~{\rm cos}~\theta_1
- \beta_3 ~{\bf B_3}~ {\partial {\bf A_3}\over \partial \theta_1}~{\rm cot}~{\theta_1\over 2}
+\beta_3 ~{\partial {\bf A_3}\over \partial \theta_2}~{\rm cot}~{\theta_2\over 2}\nonumber\\
&&- \beta_3 ~{\bf A_3}~ {\partial {\bf B_3}\over \partial \theta_1}~{\rm cot}~{\theta_1\over 2}
- k\alpha_3 ~{\bf A_5}~{\rm cot}~{\theta_2\over 2}~{\rm sin}~\theta_1
+ k\alpha_3 ~{\bf A_5}~{\bf B_5}~{\rm cot}~{\theta_1\over 2}~{\rm sin}~\theta_2 +
{\cal O}(r_h, M^3, g_sN_f) = 0 \nonumber\\
&& (ix)~~
\beta_1 ~{\partial {\bf A_1}\over \partial \theta_2}~{\rm sin}~\theta_1
+k \alpha_3 ~{\bf A_5}~{\rm cot}~{\theta_2\over 2}~{\rm sin}~\theta_1 + {\cal O}(r_h, M^3, g_sN_f) = 0 \nonumber\\
&& (x)~~\beta_1 ~{\bf A_1}~{\partial {\bf B_1}\over \partial \theta_1}~{\rm sin}~\theta_2
+k \alpha_3 ~{\bf A_5}~{\bf B_5}~{\rm cot}~{\theta_1\over 2}~{\rm sin}~\theta_2 + {\cal O}(r_h, M^3, g_sN_f) = 0
\nonumber\\
&& (xi)~~
- \beta_2 ~{\partial {\bf A_2}\over \partial \theta_1}~{\rm cot}~{\theta_2\over 2}~{\rm sin}~\theta_2
-k \alpha_2 ~{\bf A_5}~{\bf B_5}~{\rm cot}~{\theta_1\over 2}~{\rm cos}~\theta_2
+k \alpha_2 ~{\bf A_5}~{\bf B_5}~{\rm cot}~{\theta_1\over 2}\nonumber\\
&&~~~~~~~~~~~~~~~~~~~~~~~~~~~~~~~~~~~~+ {\cal O}(r_h, M^3, g_sN_f) = 0 \nonumber\\
&& (xii)~~
- \beta_2 ~{\bf A_2}~ {\partial {\bf B_2}\over \partial \theta_2}~{\rm cot}~{\theta_1\over 2}~{\rm sin}~\theta_1
+ \beta_2 ~{\partial {\bf A_2}\over \partial \theta_2}~{\rm cot}~{\theta_2\over 2}~{\rm sin}~\theta_2
-k \alpha_2 ~{\bf A_5}~{\rm cot}~{\theta_2\over 2}\nonumber\\
&&~~~~~~~~~~~~~~~~~~
-k \alpha_2 ~{\bf A_5}~{\rm cot}~{\theta_2\over 2}~{\rm cos}~\theta_1 + {\cal O}(r_h, M^3, g_sN_f) = 0\nonumber\\
\nd
where we have already defined $\alpha_k, {\bf A_n}$ and ${\bf B_m}$. The $\beta_i$ are now defined as:

\bg\label{adevin}
\beta_1 = M\left(1+ {3g_sN_f\over 2\pi}{\rm log}~r\right), ~~~~~~ \beta_2 = -{3g_sMN_f\over 4\pi r} = {2\beta_3\over r}
\nd

Notice that in \eqref{3cond} and \eqref{condon3} we have separated the ${\cal O}(r_h, M^3, g_sN_f)$ corrections from the resolution parameter $a^2$ in the fluxes. We may also absorb these corrections to the resolution parameter and write the three-form fluxes completely in terms of ${\cal O}(M^3)$ corrections to the ${\cal O}(M)$ terms in the original Ouyang solution. This may also be interpreted as though every flux components sees a different resolution parameter $a_k^2$.

Note also that we don't have an ${\cal O}(M^2)$ corrections to the Ouyang three-form fluxes. From \eqref{resfac} we could have expected an ${\cal O}(M)$ term for $a^2$. Fortunately, \eqref{respagi} showed us that $b_0=0$. Also, the scenario with $a^2_0$ being of ${\cal O}(M/N)$ is more likely, as in the absence of wrapped D5-branes the gauge theory is conformal with the gravity dual given by $AdS_5 \times T^{1,1}$ \cite{klebwit}. Only in the presence of wrapped D5-branes does the gravity dual become a resolved warped-deformed conifold so the resolution parameter $a_0$ should depend on $M$. It is then clear that the fluxes we computed will contribute to ${\cal O}(M^3)$ terms to the original Ouyang fluxes. As we anticipated, this helped us to get a consistent background in the presence of fluxes and a black hole.

Before we end this section, let us also see how the squashing factors in the three-form fluxes behave in the light of the result \eqref{respagi}. To ${\cal O}(g_sM^2/N)$ the resolution parameter $a^2$ is only a function of the radial coordinate $r$. This means that ${\bf A_n}, {\bf B_m}$ can be written as functions of $r$ to this order satisfying the closure conditions \eqref{3cond} and \eqref{condon3}. For example, combining \eqref{respagi},  \eqref{3cond} and the condition ($ii$), ${\bf A_5}$ takes the following integral form:
\bg\label{intfora5}
{\bf A_5} = {1\over {\rm sin}~\theta_1~{\rm cot}~{\theta_2\over 2}}\int dr {a_5\over \alpha_3}
\nd
where $a_5$ is a function of the angular $\theta_i$ and the radial $r$ variable such that ${\partial {\bf A_5}\over \partial \theta_i} = 0$. Similarly, using the condition ($iii$),
\bg\label{b5what}
{\bf B_5} = {{\rm sin}~\theta_1~{\rm cot}~{\theta_2\over 2} \over {\rm sin}~\theta_2~{\rm cot}~{\theta_1\over 2}}
~{\int dr ~ c_5 \alpha_3^{-1}(r) \over \int dr' ~ a_5 \alpha_3^{-1}(r')}
\nd
where again $c_5$ is like $a_5$ discussed above. Once ($a_5, c_5$) are determined the two integral forms
\eqref{intfora5} and \eqref{b5what} not only satisfy the closure conditions \eqref{3cond} but also the
necessary equations of motion. These two integral forms are also consistent with the conditions ($ix$) and ($x$) of \eqref{condon3} because $\alpha_3 \equiv -{3g_s^2MN_f\over 16\pi}$ is a constant. Note however that, to this order, the integral form for (${\bf A_4}, {\bf B_4}$) cannot be determined by this method, although we will know (${\bf A_4}, {\bf B_4}$) in terms of the resolution parameter $a^2$ up to the ${\cal O}(r_h, M^3, g_sN_f)$ corrections.

On the other hand, if we {\it assume} that (${\bf A_4}, {\bf B_4}$) have some integral representation, both (${\bf A_1}, {\bf B_1}$) will also have an integral representation. If this is the case, then:
\bg\label{a1b1}
&&{\bf A_1} = \int {dr\over \beta_1} ~\left({a_1\over {\rm sin}~\theta_1} + k\alpha_1 ~{\bf A_4}
+ k\alpha_2 ~{\bf A_5}~ {{\bf B_5}}~{\rm cot}~{\theta_1\over 2}~{\rm cot}~\theta_1
+ k\alpha_2 ~{\bf A_5}~ {{\bf B_5}}~{\rm csc}^2~{\theta_1\over 2}\right) \nonumber\\
&& {\bf B_1} = \int {dr \over \beta_1 {\bf A_1}}~\left({b_1 \over {\rm sin}~\theta_2}
+ k\alpha_1 ~{\bf A_4}~{\bf B_4}
+ k\alpha_2 ~{\bf A_5} ~{\rm cot}~{\theta_2\over 2}~{\rm cot}~\theta_2
- k\alpha_2 ~{\bf A_5}~{\rm csc}^2~{\theta_2\over 2}\right)\nonumber\\
\nd
where ($a_1, b_1$) are functions of $r$ and $\theta_i$ such that ${\partial{\bf A_1}\over \partial\theta_i}
= {\partial{\bf B_1}\over \partial\theta_i} = 0$ in the same sense as mentioned earlier for the other cases. The other squashing factors in \eqref{condon3} do not however have such simpler integral forms.

Another interesting thing to note is that from condition ($xii$) of \eqref{condon3} we might get a simpler
form for ${\bf A_5}$, namely:
\bg\label{a5now}
{\bf A_5} = - {ra_7\over 2k\alpha_3 {\rm cot}~{\theta_2\over 2}(1+ {\rm cos}~\theta_1)}
\nd
This may seem to be different from \eqref{intfora5} that we derived earlier. This is however not the case because the coefficient $a_7$ can be related to $a_5$ by the following relation:
\bg\label{a7a5}
a_5 = {1\over k}\left({\partial a_7\over \partial \theta_1} + r {\partial^2 a_7\over \partial \theta_1 \partial r}\right)
\nd
One may also compute a similar relation for ${b_5}$ and $b_7$ from condition ($xi$) of \eqref{condon3} as above. The final result will again be consistent with what we got earlier, establishing the fact that the system is well defined with the given set of boundary conditions. Therefore these analyses now complete the non-extremal flux side of the story expected in \cite{Mia:2009wj, jpsi1} in a satisfactory manner.

\chapter{Conclusion}

Over the course of this thesis, motivated by cosmological and particle physics theories, we expanded the spectrum of Klebanov-Strassler type solutions of String Theory. In the first part, we have applied the Ouyang embedding of D7-branes \cite{ouyang} to the warped resolved conifold background of Pando--Zayas and Tseytlin \cite{pt}. We found a supergravity background that breaks supersymmetry spontaneously due to fluxes of type (1,2) without generating a bulk cosmological constant. The pullback of the NS B--field onto the D7--worldvolume gives rise to D-terms, which vanish in the limit of vanishing resolution parameter $a\to 0$, i.e. when we approach the original singular background of \cite{ouyang}. We also find that the worldvolume flux in the original embedding is non-primitive and should therefore break supersymmetry. A cancellation of this effect by adding gauge fluxes would be worth further study. We should then also re-examine the D-terms we find on the resolved conifold. In the case we studied, the D7 gauge fluxes were zero and the D-terms entirely due to the non--primitive NS B--field. In general we would also expect F--terms from the D7 worldvolume theory. 

In parallel to the IIB discussion we have also studied the F-theory lift of our background. We showed how the non--primitive ISD $G_3$--flux lifts to non--primitive self--dual $G_4$ flux, which should be proportional to $J\wedge J$. We gave an explicit construction of the normalisable harmonic forms that correspond to the D7--branes. These harmonic forms would appear as second cohomologies of the compactified fourfold. We showed how a compact non-K\"ahler threefold base could be constructed which would have the required local features of a resolved conifold background that we studied for the type IIB scenario.

In the second part of the thesis, we used these results and the work of \cite{Mia:2009wj,jpsi1,OurLastPaper} on thermal solutions to explore further String Theory solutions with a particle physics perspective. Using the supergravity equations of motion in a GKP fashion, we obtained a much clearer picture of the backreactions from the black hole, branes and fluxes on the non-extremal geometry and its UV completion. More particularly, we gave a detailed analysis of the backreaction on the 3-form fluxes from the branes and non-extremal geometry, which induced flavor and temperature respectively. As was anticipated in \cite{KT-non-ex1,KT-non-ex2}, we showed that these backreactions tend to make the 3-from flux non-ISD. Interestingly, the appearance of the black hole radius $r_h$ on the fluxes is, to the order that we present here, implicit: it enters the fluxes only via the squashing factor F(r) (or $a^2$). This thereby confirms a result that was anticipated in \cite{Mia:2009wj,jpsi1}.

%conn w real world?
%QCD

%fit into big pict of ST?
%new sol

%what can be computed?
%QGP
%phase transitions

%trouver un vrai microphysical mechanism pour inflation
%trouver un F-term

%As soon as we want to apply our model to inflationary model building, we would want to add D3--branes into the picture. This gives rise to new degrees of freedom and further influences the F-terms. Related to our flux choice is another issue that deserves mentioning. The (1, 2) flux that we choose is ISD and solves equations of motion. One may also choose AISD flux if one changes the ans\"atze for the background geometry, i.e. if one ventures beyond conformal Calabi--Yau compactifications. Typically one can show that a compact conformally Calabi-Yau background only allows ISD fluxes that are also primitive. As we saw above, non-primitive ISD fluxes are allowed on a compact non-K\"ahler background or on a non-compact Calabi-Yau background. However AISD fluxes are generically part of the solution to the equations of motion on non-K\"ahler backgrounds. Some recent papers dealing with this are \cite{uranga1, uranga2, kachrunew}.

%small temperature dual
%full 7-brane backreaction

\newpage
%%%%%%%%%%%%%%%%%%%%%%%%%%%%%%%%%%%%%%%%%%%%%%%%%%%%%%%%%%%%%%%%%%%%%%%%%%%%%%%%%%%%%%%%%%%%%
%\begin{appendix}
%\def\theequation{\Alph{section}.\arabic{equation}}
%%%%%%%%%%%%%%%%%%%%%%%%%%%%%%%%%%%%%%%%%%%%%%%%%%%%%%%%%%%%%%%%%%%%%%%%%%%%%%%%%%%%%%%%%%%%%%

\appendix

\chapter{Ouyang embedding of D7--branes on the resolved conifold}\label{embed}

In this appendix we describe how D7--branes can be embedded in the PT background. We use the Ouyang \cite{ouyang} embedding 
\begin{equation}
  z \,=\, \mu^2\,,
\end{equation}
where $z$ is one of the holomorphic coordinates defined in \eqref{holocoord}. While this choice was orginally made for the singular conifold, it continues to give a consistent holomorphic embedding on both patches. From \eqref{overlap}, it is clear that selecting $z=\mu^2$ on $H_-$ implies that $-u \lambda=\mu^2$ on the intersection with $H_+$, which consistently gives $z=\mu^2$ on all of $H_+$. 

While the case $\mu \neq 0$, where the D7-brane does not extend to the tip of the throat, is of primary interest for inflationary models, we set $\mu=0$ for simplicity of calculation. As a consistency check we should always be able to recover a supersymmetric solution in the limit $a\to 0$.
The D7--brane induces a non--trivial axion--dilaton
\begin{equation}\label{dilbehav}
  \tau \,=\,  \frac{i}{g_s}+\frac{N}{2\pi i} \log z\,,
\end{equation} 
where $N$ is the number of embedded D7-branes. 
Our goal is to determine the change the dilaton induces in the other fluxes and the warp factor. We will closely follow the method laid out in \cite{ouyang} and solve the SuGra equation of motion only in linear order $g_sN$. That said, we neglect any backreaction onto the geometry beyond a change in the warp factor, i.e. we will assume the manifold remains a conformal resolved conifold.

Consider first the Bianchi identity, which in leading order becomes ($H_3$ indicates the unmodified NS flux from \eqref{fluxres}, whereas the hat indicates the corrected flux at leading order)
\begin{eqnarray}
  d\hat G_3 & =& d\hat F_3 - d \tau \wedge \hat H_3 - \tau \wedge d\hat H_3 = -d \tau \wedge H_3 +\mathcal{O}((g_s N)^2)  \\ \nonumber
  & = & -\bigg( \frac{N}{2 \pi i } \frac{dz}{z} \bigg) \wedge \big( df_1(\rho) \wedge  d\theta_1 \wedge \sin\theta_1\,d\phi_1 
    + df_2(\rho) \wedge  d\theta_2 \wedge \sin\theta_2\,d\phi_2 \big) +\mathcal{O}((g_s N)^2)\,.
\end{eqnarray}
In order to find a 3--form flux that obeys this Bianchi identity, we make an ansatz
\begin{equation}
  \hat G_3 \,=\, \sum \alpha_i\,\eta_i
\end{equation}
where $\{\eta_i\}$ is a basis of imaginary self--dual (ISD) 3--forms on the resolved conifold given in \eqref{eta}.
We find a particular solution in terms of only four of above eight 3--forms
\begin{eqnarray}\label{particular1}
  P_3 &=& \alpha_1(\rho)\,\eta_1 + e^{-i\psi/2}\alpha_3(\rho,\theta_1)\,\eta_3 + e^{-i\psi/2}\alpha_4(\rho,\theta_2)\,\eta_4 
    +\alpha_8(\rho)\,\eta_8\,,
\end{eqnarray}
with 
\begin{eqnarray}\nonumber
  \alpha_3 &=& -3\sqrt{6} g_sNP\,\frac{72 a^4-3\rho^4+a^2\rho^2(\log(\rho^2+9 a^2)-56\log \rho)}
    {8\pi\rho^3(\rho^2+6 a^2)^2}\,\cot\frac{\theta_1}{2}\\[1ex]
  \alpha_4 &=& -9\sqrt{6} g_sNP\,\frac{\rho^2-9a^2\log(\rho^2+9 a^2)}{8\pi\rho^4\sqrt{\rho^2+6 a^2}}\,
    \cot\frac{\theta_2}{2}\\[1ex] \nonumber
  \alpha_8 &=& \frac{3a^2}{\rho^2+3a^2}\left[3g_sNP\frac{-9(\rho^2+4a^2)+28\rho^2\log\rho+(81a^2+13\rho^2)\log(\rho^2+9a^2)}
    {8\pi\rho^3\sqrt{\rho^2+6a^2}\sqrt{\rho^2+9a^2}}+\alpha_1(\rho)\right]
\end{eqnarray}
Note that $a_8$ is implicitly given by $\alpha_1$, which in turn is determined via the first order ODE
\begin{eqnarray}\nonumber
  \alpha_1'(\rho) &=& \frac{-3}{\rho(\rho^2+3a^2)(\rho^2+9a^2)\sqrt{\rho^2+6a^2}}\,\left[\frac{(162a^6+78a^4\rho^2+15a^2\rho^4+\rho^6)}
    {\sqrt{\rho^2+6a^2}}\,\alpha_1(\rho)\right.\\[1ex]
  & & \left.\; + 3g_sNP\frac{-162a^6+99a^4\rho^2+63a^2\rho^4+6\rho^6 +14a^2\rho^2(\rho^2+9a^2)\log\frac{\rho^2}{\rho^2+9a^2}}{4\pi\rho^3\sqrt{\rho^2+9a^2}}\right]\,.
\end{eqnarray}
Letting $a\to 0$ in above equations, we do indeed recover the singular conifold solution of \cite{ouyang}. Keeping the resolution parameter $a$ finite instead, we can solve for $\alpha_1(\rho)$ 
\begin{equation}
   \alpha_1(\rho) \,=\, \frac{3g_sNP}{8\pi\rho^3}\frac{ \left[18 a^2 - 36(\rho^2+3a^2)\log\left(\frac{\rho}{a}\right) 
    + (10\rho^2+72a^2)\log\left(\frac{\rho^2}{\rho^2+9a^2}\right)\right]} {\sqrt{\rho^2+6a^2}\sqrt{\rho^2+9a^2}}
\end{equation}
Furthermore, we find a homogeneous solution
\begin{eqnarray}
  G_3^{hom} &=& \beta_1(z,\rho)\,\eta_1 + e^{-i\psi/2}\beta_3(\rho,\theta_1)\,\eta_3 
    + e^{-i\psi/2}\beta_4(\rho,\theta_2)\,\eta_4 \\ \nonumber 
  & &  + e^{-i\psi}\beta_5(\rho,\theta_1,\theta_2)\,\eta_5 +\beta_8(z,\rho)\,\eta_8\,,
\end{eqnarray}
with
\begin{eqnarray}\label{betas}\nonumber 
  \beta_1 &=& \frac{-3i}{8\rho^3\sqrt{\rho^2+6a^2}\sqrt{\rho^2+9a^2}}\,\big[12(\rho^2+3a^2)\log z+18a^2+10(\rho^2-9a^2)\log \rho\\ \nonumber
    & & \phantom{8\rho^3\sqrt{\rho^2+6a^2}\sqrt{\rho^2+9a^2}} +(13\rho^2+99a^2)\log(\rho^2+9a^2)\big]\\[1ex] \nonumber
  \beta_3 &=& 3i\sqrt{6}\,\left(\frac{-36a^4+3\rho^4+2a^2\rho^2\big(20\log\rho-\log(\rho^2+9a^2)\big)}{4\rho^3(\rho^2+6a^2)^2}\right)\,
    \cot\frac{\theta_1}{2}\\[1ex]
  \beta_4 &=& -9i\sqrt{6}\,\left(\frac{\rho^2-6a^2\log(\rho^2+9a^2)}{4\rho^4\sqrt{\rho^2+6a^2}}\right)\,\cot\frac{\theta_2}{2}\\[1ex] \nonumber
  \beta_5 &=& \frac{-9i\,(\cot\frac{\theta_1}{2}\,\cos\theta_2+\cot\theta_1)}{2\rho^2\sqrt{\rho^2+9a^2}\sin\theta_2}\\[1ex]
    \nonumber
  \beta_8 &=& \frac{-27ia^2}{8\rho^3\sqrt{\rho^2+6a^2}\sqrt{\rho^2+9a^2}}\,\big[4\log z+6-10\log \rho-\log(\rho^2+9a^2)\big]
\end{eqnarray}
This solution has the right singularity structure at $z=0$ and $\rho=0$, but it does not transform correctly under $SL(2,\mathbb{Z})$; only the particular solution does. We therefore conclude that the correction to the 3--form flux, which is in general a linear combination of $P_3$ and $G_3^{hom}$, is given by \eqref{particular1} only
\begin{equation}
  \hat{G}_3 \,=\, G_3+P_3\,.
\end{equation}
We can now determine the change in the remaining fluxes and the warp factor, at least to linear order in $(g_sN)$. 
We find the corrected fluxes from the equations
\begin{equation}
  \hat H_3 \,=\, \frac{ \overline{\hat{G}}_3 - \hat{G}_3 }{\tau - \bar{\tau}}\qquad\mbox{and}\qquad 
    \widetilde{F}_3 \,=\, \frac{\hat{G}_3 + \overline{\hat{G}}_3}{2}\,,
\end{equation}
which evaluates to
\begin{eqnarray}\nonumber
  \hat H_3 &=& d\rho\wedge e_\psi\wedge(c_1\,d\theta_1+c_2\,d\theta_2) + d\rho\wedge(c_3\sin\theta_1\,d\theta_1\wedge d\phi_1-c_4\sin\theta_2\,d\theta_2\wedge d\phi_2)\\
  & & +\left(\frac{\rho^2+6a^2}{2\rho}\,c_1\sin\theta_1\,d\phi_1 
    -\frac{\rho}{2}\,c_2\sin\theta_2\,d\phi_2\right)\wedge d\theta_1\wedge d\theta_2\,,\\ \nonumber
  \widetilde{F}_3 &=& -\frac{1}{g_s}\,d\rho\wedge e_\psi\wedge(c_1\sin\theta_1\,d\phi_1+c_2\sin\theta_2\,d\phi_2)\\ \nonumber
  & &  +\frac{1}{g_s}\,e_\psi\wedge(c_5\sin\theta_1\,d\theta_1\wedge d\phi_1-c_6\sin\theta_2\,d\theta_2\wedge d\phi_2)\\ 
  & & -\frac{1}{g_s}\,\sin\theta_1\sin\theta_2\left(\frac{\rho}{2}\,c_2 \,d\theta_1-\frac{\rho^2+6a^2}{2\rho}\,c_1\,d\theta_2 \right)
    \wedge d\phi_1\wedge d\phi_2\,.
\end{eqnarray}
We have introduced the coefficients
\begin{eqnarray}\label{defc}\nonumber
  c_1 &=& \frac{g_s^2PN}{4\pi\rho(\rho^2+6a^2)^2}\,\big(72 a^4-3\rho^4-56a^2\rho^2\log\rho+a^2\rho^2\log(\rho^2+9a^2)\big)\,
    \cos\frac{\theta_1}{2}\\ 
  c_2 &=& \frac{3g_s^2PN}{4\pi\rho^3}\,\big(\rho^2-9a^2\log(\rho^2+9a^2)\big)\, \cos\frac{\theta_2}{2}\\ \nonumber
  c_3 &=& \frac{3g_sP\rho}{\rho^2+9a^2}+\frac{g_s^2PN}{8\pi\rho(\rho^2+9a^2)}\,\Big[-36a^2-36\rho^2\log a+34\rho^2\log\rho\\ \nonumber
    & & \qquad\qquad\qquad\qquad +(10\rho^2+81a^2)\log(\rho^2+9a^2)+12\rho^2\log\left(\sin\frac{\theta_1}{2}\sin\frac{\theta_2}{2}\right)\Big]\\  
    \nonumber
  c_4 &=& \frac{3g_sP(\rho^2+6a^2)}{\kappa\rho^3}+\frac{g_s^2NP}{8\pi\kappa\rho^3}\,\Big[18a^2-36(\rho^2+6a^2)\log a+(34\rho^2+36a^2) 
    \log\rho\\ \nonumber 
  & & \qquad\qquad + (10\rho^2+63a^2)\log(\rho^2+9a^2)+(12\rho^2+72a^2)\log\left(\sin\frac{\theta_1}{2}\sin\frac{\theta_2}{2}\right)\Big]
\end{eqnarray}
\begin{eqnarray}\nonumber
  c_5 &=& g_sP+\frac{g_s^2PN}{24\pi(\rho^2+6a^2)}\,\Big[18a^2-36(\rho^2+6a^2)\log a+8(2\rho^2-9a^2)\log\rho\\ \nonumber
  & & \qquad\qquad\qquad\qquad\qquad\qquad\qquad\qquad\qquad\qquad\;\;+(10\rho^2+63a^2)\log(\rho^2+9a^2)
    \Big]\\ \nonumber
  c_6 &=& g_sP+\frac{g_s^2PN}{24\pi\rho^2}\,\Big[-36a^2-36\rho^2\log a+16\rho^2\log\rho+(10\rho^2+81a^2)\log(\rho^2+9a^2)\Big]
\end{eqnarray}
This allows us to write the NS 2--form potential
\begin{eqnarray}
  B_2 &=& \left(b_1(\rho)\cot\frac{\theta_1}{2}\,d\theta_1+b_2(\rho)\cot\frac{\theta_2}{2}\,d\theta_2\right)\wedge e_\psi\\ \nonumber
  & + &\left[\frac{3g_s^2NP}{4\pi}\,\left(1+\log(\rho^2+9a^2)\right)\log\left(\sin\frac{\theta_1}{2}\sin\frac{\theta_2}{2}\right)
    +b_3(\rho)\right]\sin\theta_1\,d\theta_1\wedge d\phi_1\\ \nonumber 
  & - & \left[\frac{g_s^2NP}{12\pi\rho^2}\left(-36a^2+9\rho^2+16\rho^2\log\rho+\rho^2\log(\rho^2+9a^2)\right)
    \log\left(\sin\frac{\theta_1}{2}\sin\frac{\theta_2}{2}\right)+b_4(\rho)\right]\\ \nonumber
  & & \qquad\qquad \times \sin\theta_2\,d\theta_2\wedge d\phi_2
\end{eqnarray}
with the $\rho$-dependent functions 
\begin{eqnarray}\label{defb}\nonumber
  b_1(\rho) &=& \frac{g_S^2NP}{24\pi(\rho^2+6a^2)}\big(18a^2+(16\rho^2-72a^2)\log\rho+(\rho^2+9a^2)\log(\rho^2+9a^2)\big)\\
  b_2(\rho) &=& -\frac{3g_s^2NP}{8\pi\rho^2}\big(\rho^2+9a^2\big)\log(\rho^2+9a^2)
\end{eqnarray}
and $b_3(\rho)$ and $b_a(\rho)$ are given by the first order differential equations
\begin{eqnarray}\nonumber
  b_3'(\rho) &=& \frac{3g_sP\rho}{\rho^2+9a^2} + \frac{g_s^2NP}{8\pi\rho(\rho^2+9a^2)}\Big[-36a^2-36a^2\log a+34\rho^2\log\rho\\ \nonumber
  & & \qquad\qquad\qquad\qquad\qquad\qquad+(10\rho^2+81a^2)
    \log(\rho^2+9a^2)\Big]\\ 
  b_4'(\rho) &=& -\frac{3g_sP(\rho^2+6a^2)}{\kappa\rho^3} - \frac{g_s^2NP}{8\pi\kappa\rho^3}\Big[18a^2-36(\rho^2+6a^2)\log a\\ \nonumber 
  & & \qquad\qquad\qquad+(34\rho^2+36a^2)\log\rho +(10\rho^2+63a^2)\log(\rho^2+9a^2)\Big]
\end{eqnarray}
The five--form flux is as usual given by 
\begin{equation}
  \hat{F}_5 \,=\, (1+\tilde{*}_{10})(d\hat h^{-1}\wedge d^4x)\,.
\end{equation}
In order to solve the supergravity equations of motion, the warp factor has to fulfill
\begin{equation}
  \hat h^2\,\Delta \hat h^{-1}-2 \hat h^3\,\partial_m \hat h^{-1}\,\partial_n \hat h^{-1} g^{mn}\,=\,-\Delta\hat{h}
    \,=\,*_6 \left(\frac{\hat G_3\wedge \overline{\hat G}_3}{6\left(\overline{\tau}-\tau\right)}\right)\,=\,\frac{1}{6}*_6 d\hat{F}_5\,,
\end{equation}
where $\Delta$ is the Laplacian on the unwarped resolved conifold and all indices are raised and lowered with the unwarped metric.
This should be evaluated in linear order in N, since we solved the SuGra eom for the fluxes also in linear order. However, we were unable to find an analytic solution to this problem, so we need to employ some simplification. We can take the limit $\rho\gg a$, i.e. we restrict ourselves to be far from the tip. As in the limit $a\to 0$ we recover the singular conifold setup \cite{ouyang}, we know our solution takes the form 
\begin{eqnarray}
  \hat h(\rho,\theta_1,\theta_2) &=& 1+\frac{L^4}{r^4}\,\left\{1+\frac{24g_sP^2}{\pi\alpha'Q}\,\log\rho\left[1+\frac{3g_sN}{2\pi\alpha'}
    \left(\log\rho+\frac{1}{2}\right)\right.\right.\\ \nonumber
  && \left.\left. \qquad\qquad\qquad\qquad +\frac{g_sN}{2\pi\alpha'}\,\log\left(\sin\frac{\theta_1}{2}\sin\frac{\theta_2}{2}\right)\right]\right\} 
    +\mathcal{O}\left(\frac{a^2}{\rho^2}\right)
\end{eqnarray}
with $L^4=27\pi g_s\alpha'Q/4$. Unfortunately, we cannot give an explicit expression for the $a^2/\rho^2$ corrections. However, above result is already an improvement over using the simple Klebanov--Tseytlin warp factor (which is strictly only valid for the singular solution, but is often employed in the deformed KS geometry).

%\end{appendix}

\chapter{List of Acronyms}
\begin{center}
\begin{longtable}{l l}
% \begin{table}
 \hline
\textbf{Acronym} & \textbf{Definition} \\
\hline
KW & Klebanov-Witten\cite{klebwit}\\
KS & Klebanov-Strassler\cite{ks}\\
KT & Klebanov-Tseytlin\cite{kt}\\
PT & Pando Zayas-Tseytlin\cite{pt}\\
GKP & Giddings-Kachru-Polchinski\cite{GKP}\\
AdS & Anti de Sitter\\
ADS & Affleck-Dine-Seiberg\\
CFT & conformal Field Theory\\
ISD & Imaginary Self-Dual\\
AISD & Anti Imaginary Self-Dual\\
RG & Renormalisation Group\\
QCD & Quantum Chromodynamics\\
SuGra & Supergravity\\
CY & Calabi-Yau\\
KKLMMT & Kachru-Kalosh-Linde-Maldacena-McAllister-Trivedi\cite{KKLMMT}\\
TN & Taub-NUT\\
RHIC & Relativistic Heavy Ion Collider

% \end{table}
\end{longtable}
\end{center}

%\bibliographystyle{utphys}

%\bibliography{Thesisbib}

\bibHeading{References}
\bibliographystyle{utphys}      %{unsrt}
\bibliography{MaTheseArxivV2.bbl}

%\bibliographystyle{utphys}

%\bibliography{draft}

\end{document}